\documentclass{article}

\PassOptionsToPackage{numbers, compress}{natbib}
\usepackage{xspace}
\usepackage{soul}
\usepackage{url}
\usepackage{tipa}
\usepackage{graphicx}
\usepackage{amsmath}
\usepackage{amsfonts}
\usepackage{multirow}
\usepackage{amsthm}
\usepackage{booktabs}
\usepackage{epstopdf}
\usepackage{xcolor}
\usepackage{bbm}
\usepackage{dsfont}
\usepackage{setspace}
\usepackage{braket}
\usepackage{tablefootnote}
\usepackage[normalem]{ulem}
\usepackage{hyperref}
\usepackage[nameinlink]{cleveref}
\usepackage{url}
\usepackage{makecell}
\usepackage{wrapfig}
\usepackage{sansmath}
\usepackage{subcaption}
\usepackage{float}
\usepackage{pifont}
\useunder{\uline}{\ul}{}
\usepackage{tcolorbox}
\usepackage{fancyvrb}
\usepackage{listings}
\useunder{\uline}{\ul}{}
\usepackage[textsize=tiny]{todonotes}
\usepackage[toc,page,header]{appendix}
\usepackage{minitoc}
\graphicspath{{images/}}

\usepackage{tikz}
\usetikzlibrary{tikzmark}
\tikzset{mycircled/.style={circle,draw,inner sep=0.05em,line width=0.1em, scale=0.8}}
\usepackage{xcolor}
\definecolor{mypurple}{HTML}{6600CC}

\usepackage[preprint]{neurips_2026}

\usepackage[utf8]{inputenc} % allow utf-8 input
\usepackage[T1]{fontenc}    % use 8-bit T1 fonts
\usepackage{hyperref}       % hyperlinks
\usepackage{url}            % simple URL typesetting
\usepackage{booktabs}       % professional-quality tables
\usepackage{amsfonts}       % blackboard math symbols
\usepackage{nicefrac}       % compact symbols for 1/2, etc.
\usepackage{microtype}      % microtypography
\usepackage{xcolor}         % colors
\usepackage{float}

\title{Quantum Spectral Model: Data Reuploading with Input-Conditioned Frequency Support}

\author{
Peiyong Wang\thanks{Corresponding author. Also available at \texttt{addwater0315 at gmail.com}} \\ CSIRO Technology \\ Research Way, Clayton VIC 3168, Australia \\ \texttt{Peiyong.Wang at csiro.au}  \\
\And
Udaya Parampalli \\ School of Computing and Information Systems\\ The University of Melbourne\\ Grattan Street, Parkville, VIC 3010, Australia \\ \texttt{udaya at unimelb.edu.au}
\And
Casey R. Myers\thanks{Also with Pawsey Supercomputing Centre, 1 Bryce Avenue, Kensington WA, 6151, Australia} \\ School of Physics, Mathematics and Computing\\ The University of Western Australia \\ 35 Stirling Hwy, Crawley WA, 6009, Australia  \\ \texttt{casey.myers at uwa.edu.au}
}

\begin{document}

\maketitle

\begin{abstract}
A central design principle in modern machine learning and artificial intelligence is to align a model's inductive bias with the structure of its input data.
For matrix-valued inputs, relevant matrix-level relationships can be characterised through spectral values and spectral subspaces; however, common coordinate-wise rotation-gate data-encoding unitaries used in most quantum machine learning models do not explicitly construct such a matrix-level representation.
We introduce Quantum Spectral Models (QSMs), in which we construct the generator of the data-encoding unitary directly from each input matrix.
We study three QSM variants based on symmetric, global block, and non-overlapping patch-local block Hamiltonians.
Their outputs admit truncated Fourier representations in which input-dependent spectral gaps supply candidate phase carriers, while spectral subspaces help determine their coefficients.
We evaluate the QSMs and comparison quantum models on two matrix representations of Pendigits and two controlled synthetic tasks defined by spectral statistics.
At the largest evaluated circuit depth, QSM variants lead the tested quantum models in mean test accuracy across all four benchmarks.
The patch-local QSM leads on Pendigits, whereas the global block-Hamiltonian QSM leads on the controlled spectral tasks.
Ablations show a task-dependent reversal: subspace-preserving controls perform better on Pendigits, whereas spectral-value-only controls lead among the tested ablations on the synthetic tasks.
Together, these results shed new light on quantum machine-learning model design by showing how input-conditioned spectral representations can provide an analysable inductive bias, while offering a broader perspective on structure-aware model design in machine learning and artificial intelligence.
  
\end{abstract}

\section{Introduction}\label{sec:introduction}

\begin{figure}[htbp]
    \centering
    \includegraphics[width=1\linewidth]{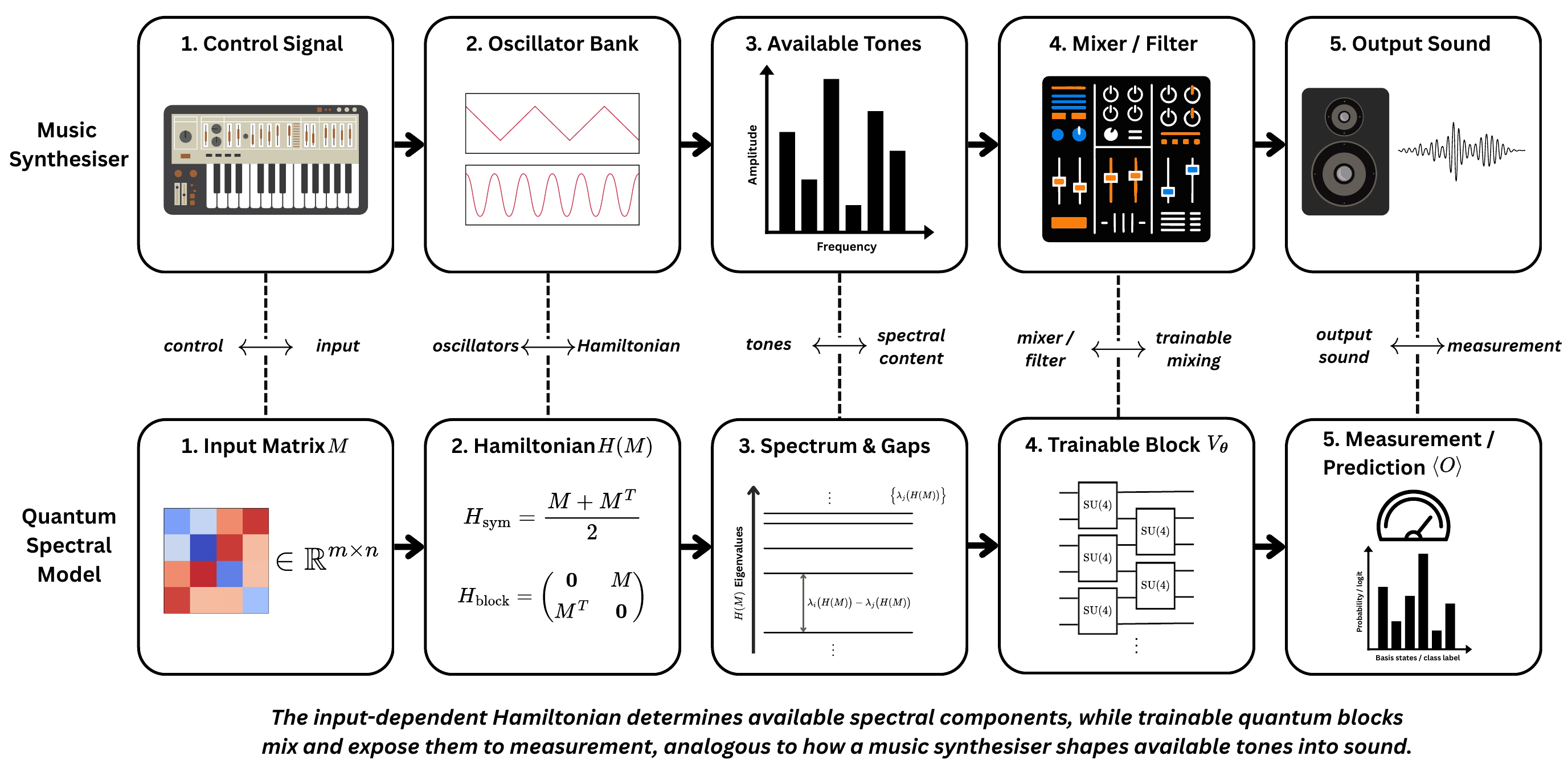}
    \caption{
    \textbf{Synthesiser view of the quantum spectral model. Top, conceptual signal flow in a music synthesiser; bottom, the corresponding flow in a quantum spectral model.} Reading the panels from left to right, the control signal corresponds to the input matrix, the oscillator bank to the input-derived Hamiltonian, the available tones to its spectrum and spectral gaps, the mixer and filter to the trainable quantum block, and the output sound to quantum measurement and prediction. A matrix input $M$ defines a Hamiltonian $H(M)$ whose spectrum supplies sample-conditioned candidate phase carriers and whose spectral projectors help determine their coefficients through the trainable mixer and quantum measurement. Unlike a coordinate-wise rotation-gate data-encoding unitary with globally determined candidate frequency support, the quantum spectral model conditions both its candidate carriers and coefficient structure on each input matrix.
    }
    \label{fig:main-overview-fig-one}
\end{figure}

The design of modern deep neural networks often reflects inductive biases about the structure of their input data. Convolutional neural networks encode spatial locality \cite{Lecun1998-qe}, while graph neural networks encode relational structure \cite{Battaglia2018-ir}. Decoder-only Transformers impose an autoregressive ordering through masked self-attention, preventing each token from accessing future tokens during language modelling \cite{Radford2018-bq}. Operator-learning architectures, meanwhile, learn maps between function spaces rather than between fixed-dimensional vectors \cite{neuraloperatorlearning}. The same representational decision arises in quantum machine learning before training begins: classical data must first be mapped to data-dependent quantum operations \cite{Schuld2021-kv}. This encoding specifies the quantum representation on which the trainable circuit acts.

Quantum models represent data through unitary evolution, whose dependence on a Hermitian generator can be understood through its spectral decomposition. Data reuploading circuits alternate data-encoding unitaries with trainable quantum layers \cite{Perez-Salinas2020-gi}. A quantum Fourier model (QFM) is a data-reuploading model whose observable outputs can be analysed as truncated Fourier series: the spectra of the data-encoding generators determine the candidate frequencies, while the trainable circuit controls their coefficients and interference at measurement \cite{Schuld2021-kv}. From this perspective, encoder design determines which frequencies, modes or basis functions are made available to the model.

Common data-encoding unitaries in reuploading circuits inject individual scalar features through single-qubit rotation gates \cite{Schuld2021-kv}. Trainable-frequency rotation-gate data-encoding unitaries can adjust their globally shared candidate frequency support during optimisation \cite{Jaderberg2024-ye}. Once trained, that support remains shared across samples. These encoders process matrix entries through coordinate-wise rotations rather than explicitly constructing a generator from matrix-level spectral values, spectral subspaces, trajectory order or local organisation.

In this work, we instead construct the generator of the data-encoding unitary directly from each input matrix. We call the resulting model class a \textbf{Quantum Spectral Model} (QSM). A QSM maps a matrix-structured input to a Hamiltonian and uses the corresponding time evolution as the data-encoding unitary. The resulting candidate frequency support is therefore conditioned on the input matrix rather than determined solely by the generator family or globally learned frequency parameters. In the synthesiser analogy in \Cref{fig:main-overview-fig-one}, a conventional rotation-gate data-encoding unitary provides a fixed oscillator bank, whereas a QSM allows each input to determine the available components; the trainable mixer then combines these components before measurement.

Hamiltonian embedding changes both the candidate support and the coefficients of the resulting truncated Fourier series. We refer to each application of a data-encoding unitary as an upload. The one-upload candidate support comprises eigengaps of the symmetrised input matrix for the symmetric-Hamiltonian QSM and signed singular-value gaps, including gaps involving zero modes, for the block-Hamiltonian QSM. In both cases, the candidate support varies across samples, while the Fourier coefficients depend on input-derived spectral projectors or subspaces together with the initial state, trainable mixer and quantum measurement observable (see \Cref{eqn:sym-ham-coeff,eqn:sym-ham-freq,eqn:block-ham-freq,eqn:block-ham-coeff}). A particular observable output may realise only a subset of the candidate support when coefficients vanish or cancel. A QSM therefore does more than pass eigenvalues or singular values to a classifier: its distinguishing feature is the coupling of sample-conditioned candidate phase carriers and sample-conditioned spectral subspaces through the trainable mixer and quantum measurement.

This construction makes quantum data encoding a choice of inductive bias, because the input-derived operator determines which spectral information is made available to the trainable circuit. We therefore focus on the representational consequences of this operator choice within a common upload--mixer framework. Frequency-related learning phenomena have been studied in classical neural networks, where spectral bias interacts with data-manifold geometry \cite{Xu2019-ad,Rahaman2018-uo}. In quantum neural networks, trainable-frequency QFMs make the candidate frequency support itself optimisable \cite{Jaderberg2024-ye}. Spectral representations also depend on the choice of transform, operator and basis \cite{Belis2026-hl}. Classical architectures obtain spectral structure from fixed Fourier transforms \cite{lee-thorp-etal-2022-fnet}, spectral parameterisations of convolutional networks \cite{rippel2015spectralrepresentationsconvolutionalneural}, graph-Laplacian bases \cite{bruna2014spectralnetworkslocallyconnected} or Fourier-space operators learned across a training distribution \cite{li2021fourier,guibas2022efficient}. In a QSM, the current input matrix instead defines the Hamiltonian, whose spectral decomposition supplies sample-conditioned candidate phase carriers and coefficient-forming subspaces for unitary evolution.

We study two global QSM variants. The symmetric-Hamiltonian QSM follows our earlier Hamiltonian embedding, $H_{\mathrm{sym}}(M)=(M+M^T)/2$ \cite{Wang2025-hx}. We additionally study the block-Hamiltonian QSM, which uses $H_{\mathrm{block}}(M)=\begin{pmatrix}0&M\\M^T&0\end{pmatrix}$ (see \Cref{eqn:sym-ham-emb,eqn:block-ham-emb}). We also evaluate fixed and trainable patch-$SU(4)$ data-encoding unitaries and a quantum spectral model with a non-overlapping patch-local block-Hamiltonian-based data-encoding unitary to test the effect of local structure; their definitions are given in Appendix \Cref{sec:appendix-patch-and-mixer}.

We investigate how sample-conditioned candidate frequency support changes representation, optimisation and final-state geometry within a common upload--mixer modelling framework. We compare the three QSM variants with fixed rotation-gate, trainable-frequency rotation-gate, fixed patch-$SU(4)$ and trainable patch-$SU(4)$ data-encoding unitaries on two matrix representations of Pendigits \cite{pen-based_recognition_of_handwritten_digits_81} and two controlled spectral tasks. At depth 32, a QSM variant attains the largest mean among the tested quantum models in all four cases: the patch-local block-Hamiltonian QSM leads on both Pendigits representations, while the block-Hamiltonian QSM leads on both synthetic tasks (see \Cref{fig:main-res,tab:main-test-accuracy}). The gradient diagnostics show that the coordinate-wise rotation-gate data-encoding unitaries have much more strongly suppressed mixer-gradient variance at high-depth initialisation, whereas several high-performing QSMs retain larger final gradient scales (see \Cref{fig:main-grad-analysis}). Their final-state fidelity kernels also show more label-aligned geometry in the fixed diagnostic batch, while the patch-$SU(4)$ controls demonstrate that gradient magnitude or fidelity-kernel spectral spread alone does not determine accuracy (see \Cref{fig:main-latent-summary}).

We further use ablations to distinguish the roles of spectral values and subspaces. On Pendigits, spectrum-only controls perform poorly, whereas controls that preserve the input-dependent spectral subspaces retain substantially higher accuracy.
The ordering reverses on the synthetic tasks, whose labels are defined by eigenvalue or singular-value statistics: spectral-value-only controls are strong, while vector-only controls are weaker (\Cref{tab:core-ablation-summary,tab:ablation-summary}). This reversal establishes a task-dependent division of labour within the sample-conditioned representation: Pendigits relies more strongly on input-dependent subspace geometry and its interaction with the trainable mixer, whereas the controlled tasks reward the relevant spectral values directly.

We make four contributions that connect the QSM construction, its Fourier analysis and the empirical comparisons. First, we formulate QSMs as matrix-conditioned models with Hamiltonian-based data-encoding unitaries and identify their representational distinction from encoders with globally shared frequency support. Second, we derive the eigengap and signed-singular-gap candidate supports of the symmetric- and block-Hamiltonian QSMs, together with the dependence of their coefficients on input-derived spectral projectors or subspaces. Third, controlled quantum-model comparisons across reuploading depths, complemented by gradient and fidelity diagnostics, characterise depth-dependent performance, optimisation behaviour and final-state geometry across four benchmark cases. Fourth, value-versus-subspace and structure-destroying ablations identify which component of the sample-conditioned QSM representation is informative for each task.

\section{Related Work}\label{sec:related-work}

\noindent\textbf{Variational quantum learning.}
Variational quantum circuits (VQCs), trained with classical optimisation algorithms, 
are widely studied as models for near-term quantum applications \cite{Cerezo2021-xl}. In quantum machine learning, they have been applied to classification \cite{farhi2018classificationquantumneuralnetworks} and generative modelling \cite{Dallaire-Demers2018-gv,Huang_2021} tasks. However, their optimisation can be impeded by barren plateaus \cite{McClean2018-pf,Larocca2025-ev} and by poor local minima even in shallow regimes without barren plateaus \cite{Anschuetz2022-ey}.

\noindent\textbf{Quantum data encoding.}
When learning from classical data, the encoding map determines the quantum representation \cite{Schuld2021-ak}. Common schemes represent features through rotation angles, state amplitudes or computational-basis states \cite{Schuld2021-ak}. Hamiltonian embedding constructs an evolution generator from a matrix-valued sample \cite{Wang2025-hx}. Block encodings provide a related but distinct matrix representation: they embed a scaled matrix as a subblock of a larger unitary used as an algorithmic primitive \cite{Camps_2022,camps2023explicitquantumcircuitsblock}. Our block-Hamiltonian construction instead uses the matrix to define the generator of the data-encoding evolution. In reuploading models, the encoding generator constrains the frequency set available to the trainable circuit \cite{Schuld2021-kv}.

\noindent\textbf{Hamiltonian learning and recognition.}
Hamiltonian learning infers an unknown physical generator from controlled measurements, using either structural assumptions or ansatz-free black-box protocols \cite{Haah2024-pn,Gu2024-yg,Hu2025-dw}. Hamiltonian recognition instead selects a generator from a finite candidate set, including through quantum-signal-processing and variational protocols \cite{Zhu2026-sw}. Both address inverse problems for a fixed unknown generator, whereas a QSM constructs $H(M)$ from each input matrix for forward prediction, making its candidate gap support and projector-dependent coefficients sample-conditioned.

\noindent\textbf{Data reuploading and quantum Fourier models.}
Data reuploading circuits interleave data-encoding and trainable unitaries and were introduced as universal quantum classifiers \cite{Perez-Salinas2020-gi}. Their outputs admit truncated Fourier-series representations whose frequencies are determined by the spectrum of the encoding generator. In particular, Pauli-generated rotation gates provide a finite grid of integer-spaced coordinate frequencies, with the spacing set by the coefficient multiplying each input feature, such as $\alpha$ in $R_y(\alpha x)$ \cite{Schuld2021-kv}. When we analyse their observable outputs through this finite Fourier representation, we refer to these data-reuploading models as quantum Fourier models (QFMs). Trainable-frequency models optimise the generator spectrum \cite{Jaderberg2024-ye}, while recent theory quantifies the depth needed for fixed-frequency encodings to approximate tunable ones \cite{Liu2026-kr}. Linear-combination-of-unitaries-inspired architectures can instead select prescribed Fourier components \cite{Tang2026-mx}. These constructions therefore provide architecture-fixed or globally learned candidate support shared across samples. Other analyses identify coefficient constraints and vanishing expressivity in QFMs \cite{Mhiri2025-bn}. Moreover, for high-dimensional inputs, limited-qubit reuploading models can approach random-guessing performance as the encoding depth grows, motivating wider architectures rather than deeper and narrower ones \cite{Wang2025-jk}. QSMs instead construct the data-encoding generator from each input, so their candidate frequency support varies across samples.

\noindent\textbf{Spectral methods in quantum machine learning.}
Classical neural networks often learn low-frequency components before high-frequency ones, although the geometry of the data manifold can modify this bias \cite{Xu2019-ad,Rahaman2018-uo}. A recent framework derives an analogous low-frequency bias for quantum neural networks \cite{Lu2026-gw}. Model spectra have also been proposed as objects that quantum routines can learn, regularise or manipulate \cite{Belis2026-hl}. QSMs address an earlier representational choice: the input determines which candidate phase carriers are available to the model.

\noindent\textbf{Classical spectral architectures.}
Classical spectral architectures can obtain spectral structure from fixed transforms, Fourier-space maps learned across a training distribution, or domain- and graph-derived operators. Fourier neural operators parameterise integral kernels directly in Fourier space to learn maps between function spaces \cite{li2021fourier}. Spectral representations support pooling, regularisation and filter parameterisation in convolutional networks \cite{rippel2015spectralrepresentationsconvolutionalneural}. FNet uses fixed, unparameterised Fourier transforms for token mixing \cite{lee-thorp-etal-2022-fnet}, whereas adaptive Fourier neural operators learn token mixing in the Fourier domain \cite{guibas2022efficient}. Graph networks derive spectral bases from graph shift operators, commonly a Laplacian; because the operator is tied to graph structure, the basis can change when the graph structure changes, while filters are learned over its spectrum \cite{bruna2014spectralnetworkslocallyconnected,3157382.3157527,10.1109/TSP.2018.2879624,lin2024equivariant}. Within this broader landscape, QSMs take an input-adaptive route: each matrix sample defines a Hermitian generator whose spectral subspaces and gap-derived candidate phase carriers enter the unitary representation.

\section{Preliminaries}\label{sec:preliminary}

\noindent\textbf{Circuit architecture and projector readout.}
A depth-$L$ data-reuploading circuit alternates data-encoding unitaries with trainable mixers \cite{Perez-Salinas2020-gi}:
\begin{equation}\label{eqn:qfm}
    \ket{\psi(\boldsymbol{x};\boldsymbol{\Theta})}
    = V_L(\boldsymbol{\theta}_L)S_L(\boldsymbol{x};\boldsymbol{\phi}_L)
    \cdots
    V_1(\boldsymbol{\theta}_1)S_1(\boldsymbol{x};\boldsymbol{\phi}_1)
    \ket{\psi_0}.
\end{equation}
Here $\boldsymbol{\Theta}=\{(\boldsymbol{\theta}_\ell,\boldsymbol{\phi}_\ell)\}_{\ell=1}^L$ collects the mixer and encoder parameters, with $\boldsymbol{\phi}_\ell$ omitted for fixed encoders. We use $\ket{\psi_0}=\ket{+}^{\otimes n}$ on $n$ qubits. For $N$ classes, let $m=\lceil\log_2N\rceil$ and associate class $j$ with the projector \cite{Wang2025-hx}
\begin{equation}
    P_j=\ket{j}\bra{j}\otimes\mathbf{I}^{\otimes(n-m)},
    \qquad j\in\{0,\ldots,N-1\},
\end{equation}
where $\ket{j}$ is an $m$-qubit computational-basis state. The raw projector mass $q_j$ and the conditional class probability $p_j$ used in the experiments are
\begin{equation}\label{eqn:logits}
    \begin{aligned}
        q_j(\boldsymbol{x};\boldsymbol{\Theta})
        &=\bra{\psi(\boldsymbol{x};\boldsymbol{\Theta})}P_j
          \ket{\psi(\boldsymbol{x};\boldsymbol{\Theta})},\\
        p_j(\boldsymbol{x};\boldsymbol{\Theta})
        &=\frac{q_j(\boldsymbol{x};\boldsymbol{\Theta})}
        {\sum_{k=0}^{N-1}q_k(\boldsymbol{x};\boldsymbol{\Theta})},
        \qquad \sum_{k=0}^{N-1}q_k>0.
    \end{aligned}
\end{equation}
If $N=2^m$, the projectors exhaust the label register and $p_j=q_j$. Otherwise, $p_j$ conditions the measurement distribution on the named class outcomes and leaves the predicted class unchanged. The Fourier analysis below concerns the observable outputs $q_j$: the nonlinear normalisation in \Cref{eqn:logits} need not preserve a finite Fourier representation.

\noindent\textbf{Fourier representation of observable outputs.}
For fixed generator parameters, an observable output of a data-reuploading circuit admits a truncated Fourier expansion \cite{Schuld2021-kv}
\begin{equation}
    f_{\boldsymbol{\theta}}(\boldsymbol{x})
    =\sum_{\boldsymbol{\omega}\in\Omega}
    c_{\boldsymbol{\omega}}(\boldsymbol{\theta})
    e^{-i\boldsymbol{\omega}\cdot\boldsymbol{x}},
    \qquad \Omega\subseteq\mathbb{R}^d.
\end{equation}
The generator eigenvalues and upload pattern determine the maximal candidate support $\Omega$; the corresponding spectral projectors, initial state, trainable mixers and quantum measurement observable determine the coefficients. Equal candidate support therefore need not imply equal observable outputs. Consequently, the realised Fourier support of a particular output may be smaller than $\Omega$, because candidate frequencies with zero coefficients do not appear in the output \cite{Mhiri2025-bn}.

For one upload of a single scalar input feature, let $S(x)=e^{-ixK}$, where $K$ is Hermitian with distinct eigenvalues $\kappa_a$ and eigenspace projectors $\Pi_a$. Then
\begin{equation}
    K=\sum_a\kappa_a\Pi_a,
    \qquad
    S(x)=\sum_a e^{-ix\kappa_a}\Pi_a,
    \qquad
    \sum_a\Pi_a=\mathbf{I},\quad \Pi_a\Pi_b=\delta_{ab}\Pi_a.
\end{equation}
Defining $O_{\boldsymbol{\theta}}=V(\boldsymbol{\theta})^\dagger O V(\boldsymbol{\theta})$, the one-upload output is
\begin{equation}
    \begin{aligned}
        f_1(x;\boldsymbol{\theta})
        &=\bra{\psi_0}S(x)^\dagger O_{\boldsymbol{\theta}}S(x)\ket{\psi_0}\\
        &=\sum_{a,b}e^{-ix(\kappa_b-\kappa_a)}
          \bra{\psi_0}\Pi_aO_{\boldsymbol{\theta}}\Pi_b\ket{\psi_0}\\
        &=\sum_{a,b}c_{ab}(\boldsymbol{\theta})e^{-ix(\kappa_b-\kappa_a)}.
    \end{aligned}
\end{equation}
The maximal one-upload support is therefore
\begin{equation}
    \Omega_1=\{\kappa_b-\kappa_a:a,b\}.
\end{equation}
For frequency sets $A,B\subseteq\mathbb{R}^d$, we denote their Minkowski sum by $A\oplus B=\{\boldsymbol{a}+\boldsymbol{b}:\boldsymbol{a}\in A,\boldsymbol{b}\in B\}$ and use $\bigoplus_{\ell=1}^{L}A_\ell$ for its repeated application.
For layer-dependent generators, let $\Omega_{1,\ell}$ denote the corresponding gap set at upload $\ell$. Because the phase factors multiply across uploads, their frequencies add; hence, the maximal $L$-upload candidate support and the Fourier support of a particular output satisfy \cite{Schuld2021-kv}
\begin{equation}
    \Omega_L=\bigoplus_{\ell=1}^L\Omega_{1,\ell},
    \qquad
    \operatorname{supp}_{\mathrm{F}}(f)\subseteq\Omega_L.
\end{equation}
If every upload uses the same generator, this reduces to
\begin{equation}
    \Omega_L=\underbrace{\Omega_1\oplus\cdots\oplus\Omega_1}_{L\text{ uploads}}.
\end{equation}

\noindent\textbf{Frequency support of rotation encoders.}
We summarise the maximal one-upload supports of the three rotation encoders used in the experiments for $d$ scalar features $\boldsymbol{x}=(x_1,\ldots,x_d)$. Appendix~\ref{sec:rot-gate-freq-derivation} gives the derivations.

For $S_{R_y}(\boldsymbol{x})=\bigotimes_{j=1}^dR_y(\alpha x_j)$, the support is
\begin{equation}\label{eqn:ry-freq-support}
    \Omega_1^{R_y}=\Big\{\alpha(n_1,\ldots,n_d):n_j\in\{-1,0,1\}\Big\}.
\end{equation}
For $S_{R_yR_z}(\boldsymbol{x})=\bigotimes_{j=1}^dR_z(\beta x_j)R_y(\alpha x_j)$, the one-feature support is
\begin{equation}\label{eqn:ryrz-local-freq-support}
    \Omega_{\mathrm{local}}^{R_yR_z}
    =\Big\{n\alpha+m\beta:n,m\in\{-1,0,1\}\Big\},
\end{equation}
and the full vector-valued support is
\begin{equation}\label{eqn:ryrz-freq-support}
    \Omega_1^{R_yR_z}
    =\Big\{(\omega_1,\ldots,\omega_d):
    \omega_j=n_j\alpha+m_j\beta,\ n_j,m_j\in\{-1,0,1\}\Big\}.
\end{equation}
For the component-wise trainable-frequency encoder, upload scale $\gamma_j$ enters the rotation $R_y(\gamma_jx_j)$ for feature $j$ \cite{Jaderberg2024-ye}. Its one-upload support is
\begin{equation}\label{eqn:trainable-freq-support}
    \Omega_1^{\mathrm{TF}}(\boldsymbol{\gamma})
    =\Big\{(n_1\gamma_1,\ldots,n_d\gamma_d):n_j\in\{-1,0,1\}\Big\}.
\end{equation}
The experiments use a separate scale vector $\boldsymbol{\gamma}_\ell$ at each upload. Once $\alpha$, $\beta$ or the $\boldsymbol{\gamma}_\ell$ are fixed, these maximal supports are shared across samples: changing $\boldsymbol{x}$ changes the evaluation point but not the available frequency set. This sample-independent support provides the baseline for the Hamiltonian embeddings introduced in \Cref{sec:hamembed}.

\section{Hamiltonian Embedding Produces Sample-Conditioned Frequency Support}\label{sec:hamembed}

Let $M=M_{\boldsymbol{x}}\in\mathbb{R}^{p\times q}$ denote the matrix representation of an input $\boldsymbol{x}$, with $p\geq q$ for the layouts considered here. In our previous work, we adopted the symmetric Hamiltonian embedding as the generator of a data-encoding unitary for classification \cite{Wang2025-hx}:
\begin{equation}\label{eqn:sym-ham-emb}
    H_{\mathrm{sym}}(M)=\frac{\widetilde{M}+\widetilde{M}^{T}}{2},
\end{equation}
where $\widetilde{M}$ is the square matrix obtained by zero-column padding when $M$ is rectangular. Symmetrisation removes the skew-symmetric component of $\widetilde{M}$. We therefore also consider the block-Hamiltonian embedding
\begin{equation}\label{eqn:block-ham-emb}
    H_{\mathrm{block}}(M)=
    \begin{pmatrix}
        \boldsymbol{0}_{p\times p} & M \\
        M^{T} & \boldsymbol{0}_{q\times q}
    \end{pmatrix},
\end{equation}
which retains $M$ as an off-diagonal block. This construction has the same Hermitian block form used in the quantum polar decomposition algorithm, with $A=M^{T}$ for real $M$ \cite{lloyd2020quantumpolardecompositionalgorithm}. Here it serves as a sample-derived generator of a data-encoding unitary rather than as a subroutine for implementing the polar factors.

When required by the circuit dimension or projector readout, either Hamiltonian is promoted by direct-sum padding, $H\mapsto H\oplus\boldsymbol{0}$; we retain the same symbol for the padded operator. This convention preserves its nonzero spectrum and introduces additional zero modes. Within a QSM, the two global data-encoding unitaries are
\begin{equation}
    S_{H_{\mathrm{sym}}}(t;M)=e^{-iH_{\mathrm{sym}}(M)t/2},
    \qquad
    S_{H_{\mathrm{block}}}(t;M)=e^{-iH_{\mathrm{block}}(M)t/2}.
\end{equation}
We also consider a QSM with a non-overlapping patch-local block-Hamiltonian-based data-encoding unitary. This variant partitions $M$ into $R$ patches $P_1,\ldots,P_R$ and applies $H_{\mathrm{block}}(P_r)$ on disjoint registers. The experiments use four $2\times2$ patches, defined in Appendix~\ref{sec:appendix-patch-and-mixer}.

For each fixed $M$, the analysis below treats the upload time $t$, or the vector of patch times, as the Fourier variable. Accordingly, $\Omega_1(M)$ denotes sample-conditioned candidate support in the upload-time variables; the entries of $M$ condition the generator and remain fixed throughout this expansion.
Appendix~\ref{sec:H-block-derivation-details} gives the full derivations, including padding, degeneracies and layer-specific upload times.

\noindent\textbf{Symmetric embedding produces eigengap carriers.}
Let $\mathcal{A}_{\mathrm{sym}}(M)$ denote the index set of distinct eigenvalues of $H_{\mathrm{sym}}(M)$. For each $a\in\mathcal{A}_{\mathrm{sym}}(M)$, let $\lambda_a(M)$ be the corresponding eigenvalue and let $\Pi_a^{\mathrm{sym}}(M)$ project onto its full eigenspace. Then
\begin{equation}
    H_{\mathrm{sym}}(M)
    =\sum_{a\in\mathcal{A}_{\mathrm{sym}}(M)}
    \lambda_a(M)\Pi_a^{\mathrm{sym}}(M),
    \qquad
    S_{H_{\mathrm{sym}}}(t;M)
    =\sum_a e^{-i\lambda_a(M)t/2}\Pi_a^{\mathrm{sym}}(M).
\end{equation}
For a quantum measurement observable $O$, with the trainable mixer absorbed into $O_{\boldsymbol{\theta}}$, the resulting one-upload observable output is
\begin{equation}\label{eqn:sym-ham-one-layer}
    f_1^{H_{\mathrm{sym}}}(M;t,\boldsymbol{\theta})
    =\sum_{a,b}A_{ab}^{H_{\mathrm{sym}}}(M,\boldsymbol{\theta})
    e^{-i[\lambda_b(M)-\lambda_a(M)]t/2},
\end{equation}
where the pair amplitudes and the coefficient at frequency $\omega$ are
\begin{equation}\label{eqn:sym-ham-coeff}
    \begin{aligned}
        A_{ab}^{H_{\mathrm{sym}}}(M,\boldsymbol{\theta})
        &=\bra{\psi_0}\Pi_a^{\mathrm{sym}}(M)O_{\boldsymbol{\theta}}
          \Pi_b^{\mathrm{sym}}(M)\ket{\psi_0},\\
        C_{\omega}^{H_{\mathrm{sym}}}(M,\boldsymbol{\theta})
        &=\sum_{\substack{a,b:\\ (\lambda_b(M)-\lambda_a(M))/2=\omega}}
          A_{ab}^{H_{\mathrm{sym}}}(M,\boldsymbol{\theta}).
    \end{aligned}
\end{equation}
The maximal one-upload candidate support therefore satisfies
\begin{equation}\label{eqn:sym-ham-freq}
    \begin{aligned}
        \Omega_1^{H_{\mathrm{sym}}}(M)
        &=\left\{\frac{\lambda_b(M)-\lambda_a(M)}{2}:
          a,b\in\mathcal{A}_{\mathrm{sym}}(M)\right\},\\
        \operatorname{supp}_{\mathrm{F}}(f_1^{H_{\mathrm{sym}}})
        &\subseteq\Omega_1^{H_{\mathrm{sym}}}(M).
    \end{aligned}
\end{equation}
Thus, for the symmetric-Hamiltonian QSM, the candidate frequencies depend on the eigengaps of the sample-derived Hamiltonian, while the realised coefficients also depend on its eigenspaces, the initial state, the trainable mixer and the readout.

\noindent\textbf{Block-Hamiltonian embedding produces signed singular-value gaps.}
If the nonzero singular values of $M$ are $\sigma_j(M)$, then the nonzero eigenvalues of $H_{\mathrm{block}}(M)$ are $\pm\sigma_j(M)$. Let $\mathcal{A}_{\mathrm{block}}(M)$ denote the index set of distinct eigenvalues of $H_{\mathrm{block}}(M)$. For each $a\in\mathcal{A}_{\mathrm{block}}(M)$, let $\mu_a(M)$ be the corresponding eigenvalue, including zero when present, and let $\Pi_a^{\mathrm{block}}(M)$ project onto its full eigenspace:
\begin{equation}
    H_{\mathrm{block}}(M)
    =\sum_{a\in\mathcal{A}_{\mathrm{block}}(M)}
    \mu_a(M)\Pi_a^{\mathrm{block}}(M),
    \qquad
    \mu_a(M)\in\{+\sigma_j(M),-\sigma_j(M),0\}.
\end{equation}
The corresponding observable output has the expansion
\begin{equation}
    f_1^{H_{\mathrm{block}}}(M;t,\boldsymbol{\theta})
    =\sum_{a,b}A_{ab}^{H_{\mathrm{block}}}(M,\boldsymbol{\theta})
    e^{-i[\mu_b(M)-\mu_a(M)]t/2},
\end{equation}
with
\begin{equation}\label{eqn:block-ham-coeff}
    \begin{aligned}
        A_{ab}^{H_{\mathrm{block}}}(M,\boldsymbol{\theta})
        &=\bra{\psi_0}\Pi_a^{\mathrm{block}}(M)O_{\boldsymbol{\theta}}
          \Pi_b^{\mathrm{block}}(M)\ket{\psi_0},\\
        C_{\omega}^{H_{\mathrm{block}}}(M,\boldsymbol{\theta})
        &=\sum_{\substack{a,b:\\ (\mu_b(M)-\mu_a(M))/2=\omega}}
          A_{ab}^{H_{\mathrm{block}}}(M,\boldsymbol{\theta}).
    \end{aligned}
\end{equation}
Its maximal one-upload candidate support is
\begin{equation}\label{eqn:block-ham-freq}
    \begin{aligned}
        \Omega_1^{H_{\mathrm{block}}}(M)
        &=\left\{\frac{\mu_b(M)-\mu_a(M)}{2}:
          a,b\in\mathcal{A}_{\mathrm{block}}(M)\right\},\\
        \operatorname{supp}_{\mathrm{F}}(f_1^{H_{\mathrm{block}}})
        &\subseteq\Omega_1^{H_{\mathrm{block}}}(M).
    \end{aligned}
\end{equation}
For the block-Hamiltonian QSM, these gaps include half-differences and half-sums of singular values; when a zero eigenspace is present, they also include $\pm\sigma_j(M)/2$. The coefficients depend on the associated singular subspaces together with the initial state, mixer and readout.

\noindent\textbf{Patch-local embedding factorises support across patches.}
For patch $P_r$, let $\Omega_1^{H_{\mathrm{block}}}(P_r)$ be the candidate set in \Cref{eqn:block-ham-freq}. Because the patch uploads act on disjoint registers with independent times $\boldsymbol{t}=(t_1,\ldots,t_R)$, their maximal one-upload support is
\begin{equation}\label{eqn:patch-block-freq}
    \Omega_1^{\mathrm{patch}\text{-}H_{\mathrm{block}}}(M)
    =\Omega_1^{H_{\mathrm{block}}}(P_1)\times\cdots\times
     \Omega_1^{H_{\mathrm{block}}}(P_R).
\end{equation}
This Cartesian product arises because the patches have independent upload-time variables. If they instead shared a single scalar time, the corresponding scalar support would be the Minkowski sum of the patch-level candidate sets.
The patch-local carriers are therefore determined by the singular spectra of individual patches rather than that of the full matrix. Their amplitudes depend on the corresponding local singular subspaces, while the trainable mixer and readout can couple modes from different patches. The joint data-encoding unitary, its coefficients and the extension to layer- and patch-specific times $t_{\ell,r}$ are derived in \Cref{eqn:patch-block-joint-upload,eqn:patch-block-one-layer,eqn:patch-block-coeff}.

These three QSM data-encoding variants determine two coupled parts of the representation. Spectral values determine the maximal candidate phase carriers, whereas spectral projectors or singular subspaces participate in forming the realised coefficients through their interaction with the initial state, trainable mixer and readout. The global variants condition these quantities on the full matrix, while the patch variant conditions them separately on non-overlapping patches. The choice of data-encoding unitary therefore determines both the available sample-conditioned phase carriers and the subspaces through which the circuit forms their realised coefficients.

\section{Experiments}\label{sec:experimentsetting}

\begin{figure}[ht]
    \centering
    \begin{subfigure}[t]{1\textwidth}
        \includegraphics[width=1\linewidth]{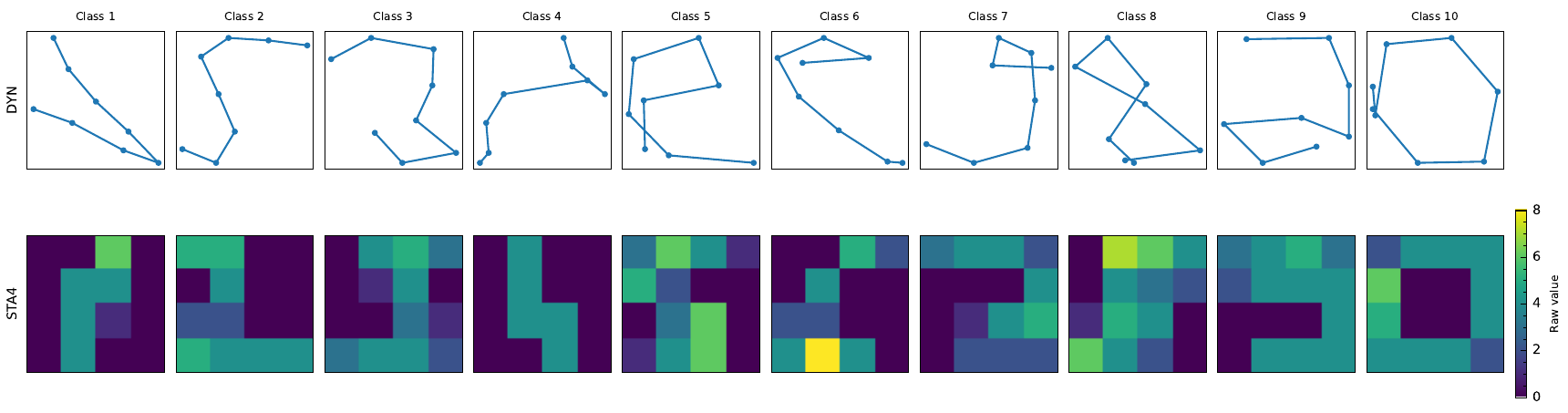}
        \caption{ }\label{fig:pendigit-data-sample}
    \end{subfigure}
    \vfill
    \begin{subfigure}[t]{0.5\textwidth}
        \includegraphics[width=1\linewidth]{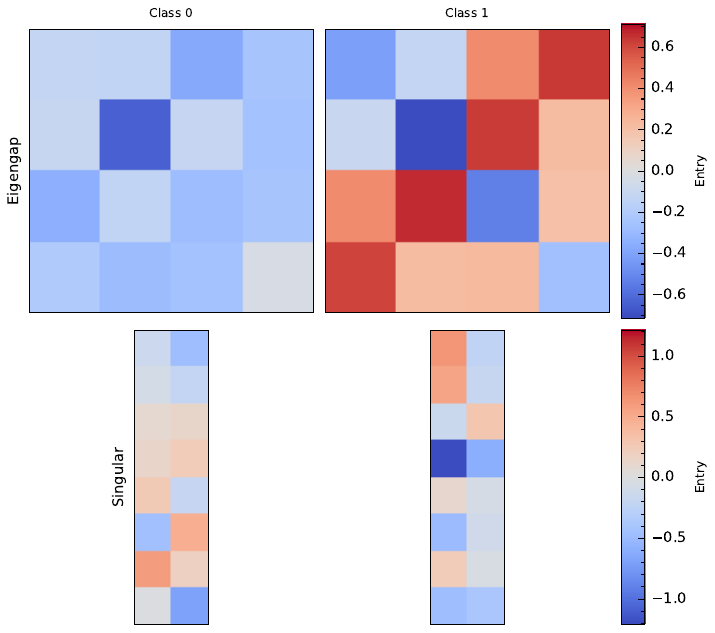}
        \caption{ }\label{fig:synthetic-data-sample}
    \end{subfigure}
    \caption{
    \textbf{Representative training samples from the real-world and controlled synthetic benchmarks.}
    \textbf{(a) Pendigits.} One example from each of the ten original classes is shown using two paired representations. The upper row shows the \texttt{DYN} representation, in which eight ordered two-dimensional pen coordinates are connected in their recorded sequence. The lower row shows the corresponding \texttt{STA4} representation as a $4\times4$ bitmap-like matrix. Raw feature values are displayed for visualisation, whereas the model inputs are standardised using statistics computed from the training split; the colour bar reports the raw \texttt{STA4} entry values.
    \textbf{(b) Synthetic spectral benchmarks.} One noisy training input from each class is shown for the \texttt{SYNTHETIC EIGENGAP} task (top, $4\times4$ observations generated from latent symmetric matrices) and the \texttt{SYNTHETIC SINGULAR} task (bottom, $8\times2$ rectangular observations). For \texttt{SYNTHETIC EIGENGAP}, class $1$ indicates that the gap between the two largest eigenvalues of the latent clean matrix exceeds $\tau_{\mathrm{eig}}=0.75$. For \texttt{SYNTHETIC SINGULAR}, class $1$ indicates that the sum of the two leading latent singular values exceeds $\tau_{\mathrm{sv}}=2$; class $0$ denotes the complementary condition in each task. The displayed inputs include relative Gaussian noise with $\epsilon=0.05$, while the labels are computed from the corresponding clean latent spectra. Colours encode the noisy matrix entries using a separate zero-centred scale for each task.
    }\label{fig:data-sample}
\end{figure}

\noindent\textbf{Benchmark datasets.} We use Pendigits \cite{pen-based_recognition_of_handwritten_digits_81} as a compact real-world benchmark. Each digit is represented either by a \texttt{DYN} $8\times2$ pen-trajectory matrix or by a \texttt{STA4} $4\times4$ bitmap-like matrix, providing two structurally different views of the same ten-class task (\Cref{fig:pendigit-data-sample}). We also consider two binary synthetic tasks: \texttt{SYNTHETIC EIGENGAP}, based on noisy observations of latent $4\times4$ symmetric matrices, and \texttt{SYNTHETIC SINGULAR}, based on noisy $8\times2$ rectangular matrices (\Cref{fig:synthetic-data-sample}). Their labels are determined from the clean latent eigenvalues or singular values, making them controlled tests of sensitivity to spectral structure. Preprocessing, data splits and data-generation settings are given in Appendix \Cref{sec:additional-dataset-info}.

\noindent\textbf{Encoders and training protocol.} All models follow the upload--mixer template in \Cref{eqn:qfm}, with $\ket{\psi_0}=\ket{+}^{\otimes n}$. We compare eight data-encoding constructions: fixed-$R_y$, fixed-$R_yR_z$, trainable-frequency $R_y$, fixed patch-$SU(4)$, trainable patch-$SU(4)$, patch-local block-Hamiltonian, symmetric Hamiltonian and block Hamiltonian. They are defined in \Cref{sec:preliminary,sec:hamembed} and Appendix \Cref{sec:appendix-patch-and-mixer}. For each construction, we evaluate reuploading depths $L\in\{1,2,4,8,16,32\}$ over 20 random parameter-initialisation seeds. Block $\ell$ applies the upload $S_\ell(\boldsymbol{x})$ followed by the trainable mixer $V_\ell(\boldsymbol{\theta}_\ell)$.

The comparison controls the dataset splits, upload--mixer template, optimiser, training budget and number of random initialisation seeds, allowing us to examine how different data-encoding constructions behave under a common experimental protocol. However, the constructions differ in both qubit and parameter count: rotation-gate data-encoding unitaries use 16 qubits, patch-local data-encoding unitaries use eight, and the global QSMs use two to four qubits before any label-register expansion, with the Pendigits global QSMs promoted to four qubits. The mixer therefore contributes $15(n-1)$ parameters per layer, while the trainable data-encoding unitaries introduce different additional parameter groups. These resource differences follow from the register layouts and parameterisations adopted by the respective data-encoding constructions. We therefore compare the predictive performance and inductive biases of the complete encoder--mixer configurations under a common experimental protocol, treating data encoding as a central architectural choice. The results compare the models as implemented, rather than isolating the encoder from all resource differences or measuring performance per qubit or trainable parameter. Appendix \Cref{sec:appendix-patch-and-mixer} gives the padding rules and exact depth-32 trainable-parameter counts.

For the two global QSMs, the data-encoding unitary at layer $\ell$ is
\begin{equation}
    S_{\mathrm{sym}, \ell} (M) = \exp \Bigg[ -\frac{i}{2} H_{\mathrm{sym}} (M) t_{\ell}  \Bigg], \quad S_{\mathrm{block}, \ell} (M) = \exp \Bigg[ -\frac{i}{2} H_{\mathrm{block}} (M) t_{\ell}  \Bigg],
\end{equation}
where $H_{\mathrm{sym}}(M)$ and $H_{\mathrm{block}}(M)$ are defined in \Cref{eqn:sym-ham-emb,eqn:block-ham-emb}. The trainable upload time $t_\ell$ scales the input-derived Hamiltonian and is initialised to $1/L$ unless a fixed time is specified.

The patch-local data-encoding constructions divide each input into four non-overlapping $2\times2$ patches and assign each patch to a disjoint qubit pair. For fixed patch-$SU(4)$, a patch $P_r\in\mathbb{R}^{2\times2}$ is flattened and mapped linearly to 15 coefficients of the two-qubit unitary
\begin{equation}
    U_{\mathrm{patch}}(P_r) = \exp \Bigg[ i \sum_{a=1}^{15} \theta_a (P_r) G_a  \Bigg],
\end{equation}
where $G_a$ are the 15 non-identity two-qubit Pauli generators. The fixed map is shared across layers and patches; the trainable variant uses a separate linear map for each layer and patch. The patch-local block-Hamiltonian QSM instead applies \Cref{eqn:block-ham-emb} to each patch, with trainable times $t_{\ell,r}$ initialised to $1/L$ unless fixed.

After each upload, $V_\ell$ applies nearest-neighbour $SU(4)$ blocks in a brick-wall pattern. The even pairs $(1,2),(3,4),\ldots$ precede the odd pairs $(2,3),(4,5),\ldots$. Each block has 15 trainable parameters, so an $n$-qubit mixer layer contains $n-1$ blocks. Mixer parameters are initialised from a zero-mean Gaussian with scale $0.01$.

All trainable parameters are optimised jointly. We use Adam \cite{Kingma2014-qs} with learning rate $0.01$, no weight decay and 2000 optimisation steps. Pendigits runs use training and evaluation batch sizes of 32 and 128, respectively; the corresponding synthetic-task sizes are 64 and 256.

\section{Results and Analysis}\label{sec:results}

We organise the analysis in four stages: we first compare final test accuracy across reuploading depth, then examine gradient behaviour, next analyse the final-state fidelity geometry of the trained models, and finally use value--subspace and structure-destruction ablations to identify which components of the sample-conditioned spectral representation are informative for each task.

\begin{figure}[ht]
    \centering
    \begin{subfigure}[t]{1\textwidth}
        \includegraphics[width=1\linewidth]{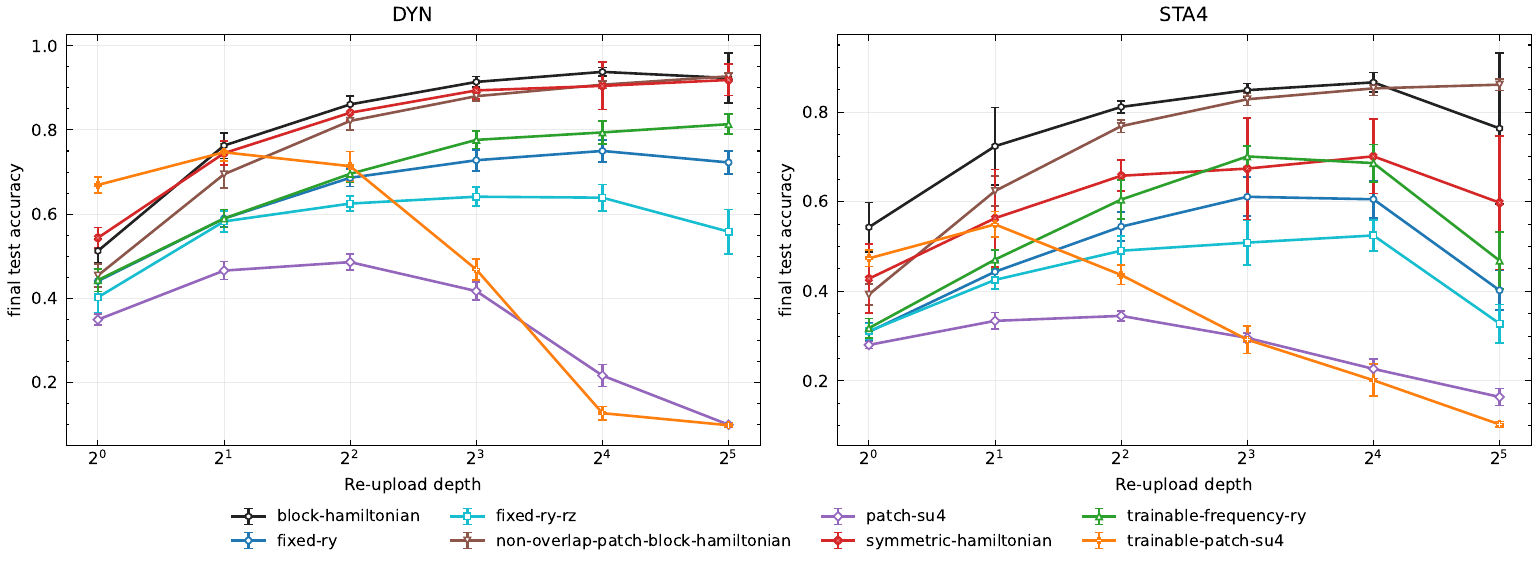}
        \caption{ }\label{fig:pendigit-res}
    \end{subfigure}
    \vfill
    \begin{subfigure}[t]{1\textwidth}
        \includegraphics[width=1\linewidth]{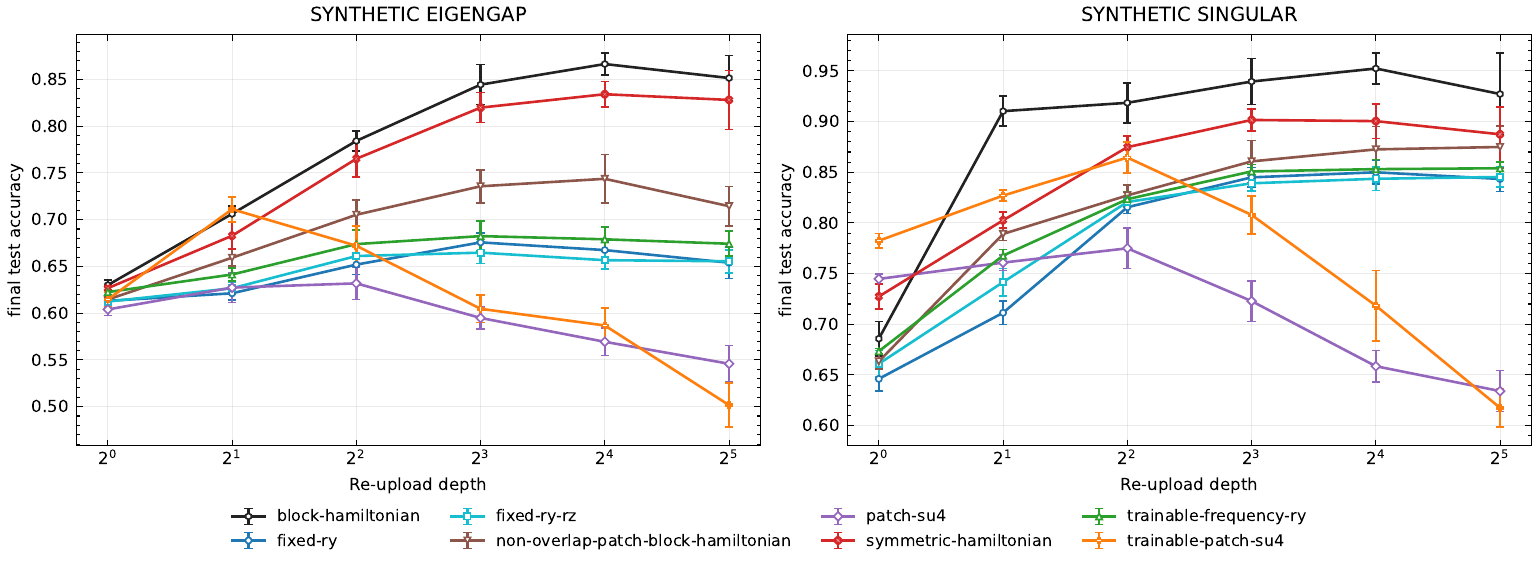}
        \caption{ }\label{fig:synthetic-res}
    \end{subfigure}
    \caption{
    \textbf{(a) Final test accuracy on the Pendigits benchmark as a function of reuploading depth.} The two panels show the \texttt{DYN} $8\times 2$ trajectory representation and the \texttt{STA4} $4\times 4$ matrix representation. Each point is the mean over 20 parameter-initialisation seeds, and error bars show one standard deviation across seeds. QSMs lead the tested depth-32 comparisons. The global block-Hamiltonian QSM has the largest observed mean across the six depths on both representations, while the patch-local block-Hamiltonian QSM has the largest mean at depth 32.
    \textbf{(b) Final test accuracy on the two synthetic spectral benchmarks as a function of reuploading depth.} The \texttt{SYNTHETIC EIGENGAP} task uses noisy $4\times 4$ symmetric matrices with labels determined by the largest latent eigengap, while the \texttt{SYNTHETIC SINGULAR} task uses noisy $8\times 2$ matrices with labels determined by the sum of the two leading latent singular values. Each point is the mean over 20 seeds, and error bars show one standard deviation across seeds. The global block-Hamiltonian QSM has the largest mean at depth 32 and the largest observed mean across the six depths on both tasks.
    }\label{fig:main-res}
\end{figure}

\noindent\textbf{QSMs attain the largest depth-32 mean test accuracy among the tested quantum models.} At depth 32, the patch-local block-Hamiltonian QSM has the largest mean accuracy on Pendigits \texttt{DYN} ($92.75\%\pm0.84\%$) and \texttt{STA4} ($86.14\%\pm1.33\%$), while the global block-Hamiltonian QSM has the largest mean on \texttt{SYNTHETIC EIGENGAP} ($85.15\%\pm2.40\%$) and \texttt{SYNTHETIC SINGULAR} ($92.72\%\pm4.04\%$). Thus, a QSM attains the largest depth-32 mean among the tested quantum models in each of the four benchmark cases (\Cref{fig:main-res} and \Cref{tab:main-test-accuracy}).

Across the six test-accuracy-versus-depth curves, the global block-Hamiltonian QSM has the largest observed mean for all four cases, occurring at depth 16 with $93.81\%\pm1.06\%$, $86.66\%\pm2.17\%$, $86.65\%\pm1.18\%$ and $95.24\%\pm1.53\%$, respectively. These values summarise the observed depth sweeps rather than performance at a validation-selected depth; the full curves and depth-32 results provide the corresponding depth-specific comparisons. Appendix \Cref{sec:appendix-additional-main-results} reports the validation summaries and validation-accuracy AUCs.

\begin{table*}[htbp!]
  \caption{\textbf{Final test accuracy for the main Pendigits and synthetic benchmark runs.}
  Values are mean $\pm$ standard deviation over 20 random seeds, reported in percentages.
  The largest-observed-mean columns report the depth and value of the maximum
  mean test accuracy across the six evaluated depths for each dataset--encoder
  pair, providing a descriptive summary of the depth sweep. The depth-32 column
  reports the same metric at the largest reuploading depth used in the
  experiments. Boldface marks the largest
  mean within each dataset block and accuracy column.}
  \label{tab:main-test-accuracy}
  \centering
  \small
  \setlength{\tabcolsep}{3pt}
  \begin{tabular}{llccc}
    \toprule
    Dataset & Encoder &
    \shortstack{Depth of largest\\observed mean} &
    \shortstack{Largest observed mean\\test acc. (\%)} &
    \shortstack{Depth-32 test\\acc. (\%)} \\
    \midrule
    Pendigits DYN & fixed-$R_y$ & 16 & $75.02\pm2.63$ & $72.29\pm2.77$ \\
     & fixed-$R_yR_z$ & 8 & $64.14\pm2.26$ & $55.84\pm5.30$ \\
     & trainable-frequency(TF)-$R_y$ & 32 & $81.37\pm2.36$ & $81.37\pm2.36$ \\
     & patch-$SU(4)$ & 4 & $48.60\pm1.87$ & $9.96\pm0.44$ \\
     & trainable patch-$SU(4)$ & 2 & $74.70\pm2.04$ & $9.84\pm0.47$ \\
     & patch $H_{\rm block}$ & 32 & $92.75\pm0.84$ & $\boldsymbol{92.75\pm0.84}$ \\
     & $H_{\rm sym}$ & 32 & $91.89\pm3.71$ & $91.89\pm3.71$ \\
     & $H_{\rm block}$ & 16 & $\boldsymbol{93.81\pm1.06}$ & $92.35\pm5.95$ \\
    \midrule
    Pendigits STA4 & fixed-$R_y$ & 8 & $61.10\pm4.35$ & $40.22\pm4.40$ \\
     & fixed-$R_yR_z$ & 16 & $52.47\pm3.49$ & $32.79\pm4.28$ \\
     & TF-$R_y$ & 8 & $70.11\pm2.35$ & $46.85\pm6.39$ \\
     & patch-$SU(4)$ & 4 & $34.51\pm1.08$ & $16.41\pm1.88$ \\
     & trainable patch-$SU(4)$ & 2 & $54.94\pm2.88$ & $10.27\pm0.62$ \\
     & patch $H_{\rm block}$ & 32 & $86.14\pm1.33$ & $\boldsymbol{86.14\pm1.33}$ \\
     & $H_{\rm sym}$ & 16 & $70.14\pm8.29$ & $59.79\pm14.90$ \\
     & $H_{\rm block}$ & 16 & $\boldsymbol{86.66\pm2.17}$ & $76.38\pm16.84$ \\
    \midrule
    Synthetic eigengap & fixed-$R_y$ & 8 & $67.57\pm0.98$ & $65.40\pm1.69$ \\
     & fixed-$R_yR_z$ & 8 & $66.47\pm1.18$ & $65.55\pm1.22$ \\
     & TF-$R_y$ & 8 & $68.24\pm1.62$ & $67.42\pm1.33$ \\
     & patch-$SU(4)$ & 4 & $63.19\pm1.72$ & $54.58\pm1.97$ \\
     & trainable patch-$SU(4)$ & 2 & $71.12\pm1.33$ & $50.16\pm2.36$ \\
     & patch $H_{\rm block}$ & 16 & $74.37\pm2.59$ & $71.44\pm2.11$ \\
     & $H_{\rm sym}$ & 16 & $83.42\pm1.37$ & $82.81\pm3.16$ \\
     & $H_{\rm block}$ & 16 & $\boldsymbol{86.65\pm1.18}$ & $\boldsymbol{85.15\pm2.40}$ \\
    \midrule
    Synthetic singular & fixed-$R_y$ & 16 & $85.00\pm1.15$ & $84.34\pm1.24$ \\
     & fixed-$R_yR_z$ & 32 & $84.54\pm1.01$ & $84.54\pm1.01$ \\
     & TF-$R_y$ & 32 & $85.40\pm0.63$ & $85.40\pm0.63$ \\
     & patch-$SU(4)$ & 4 & $77.50\pm2.00$ & $63.41\pm2.03$ \\
     & trainable patch-$SU(4)$ & 4 & $86.45\pm1.52$ & $61.75\pm1.91$ \\
     & patch $H_{\rm block}$ & 32 & $87.50\pm2.06$ & $87.50\pm2.06$ \\
     & $H_{\rm sym}$ & 8 & $90.16\pm1.08$ & $88.75\pm2.69$ \\
     & $H_{\rm block}$ & 16 & $\boldsymbol{95.24\pm1.53}$ & $\boldsymbol{92.72\pm4.04}$ \\
    \bottomrule
  \end{tabular}
\end{table*}

The depth-32 comparisons show that the largest mean test accuracy is obtained by a QSM across structurally different inputs and spectral rules. The \texttt{DYN} input is an ordered $8\times2$ trajectory, and its four non-overlapping $2\times2$ patches correspond to consecutive trajectory segments. By contrast, \texttt{STA4} is a $4\times4$ bitmap-like representation whose patches preserve local two-dimensional neighbourhoods, while the two synthetic labels depend on full-matrix eigengap or singular-value statistics. These contrasting input structures and task definitions provide the context for the encoder-specific performance patterns discussed below.

The rotation-gate baselines attain larger means on \texttt{DYN} than on \texttt{STA4}, both at depth 32 and at their highest observed means. This difference may reflect how \texttt{DYN} exposes coordinate values in their recorded sequence, whereas \texttt{STA4} organises them within a two-dimensional bitmap neighbourhood. The strongest rotation-gate result on \texttt{DYN}, $81.37\%\pm2.36\%$ for trainable-frequency $R_y$ at depth 32, remains below the depth-32 means of all three QSMs. On \texttt{STA4}, the explicit $2\times2$ locality of the patch-local block-Hamiltonian QSM offers a possible interpretation of its performance, paralleling the spatial locality encoded by convolutional networks \cite{Lecun1998-qe}. Taken together, these patterns suggest an interaction between data-encoding construction and input layout; the specific contribution of locality remains unresolved.

\begin{figure}[htbp]
    \centering
    \begin{subfigure}[t]{1\textwidth}
        \centering
        \includegraphics[width=0.79\linewidth]{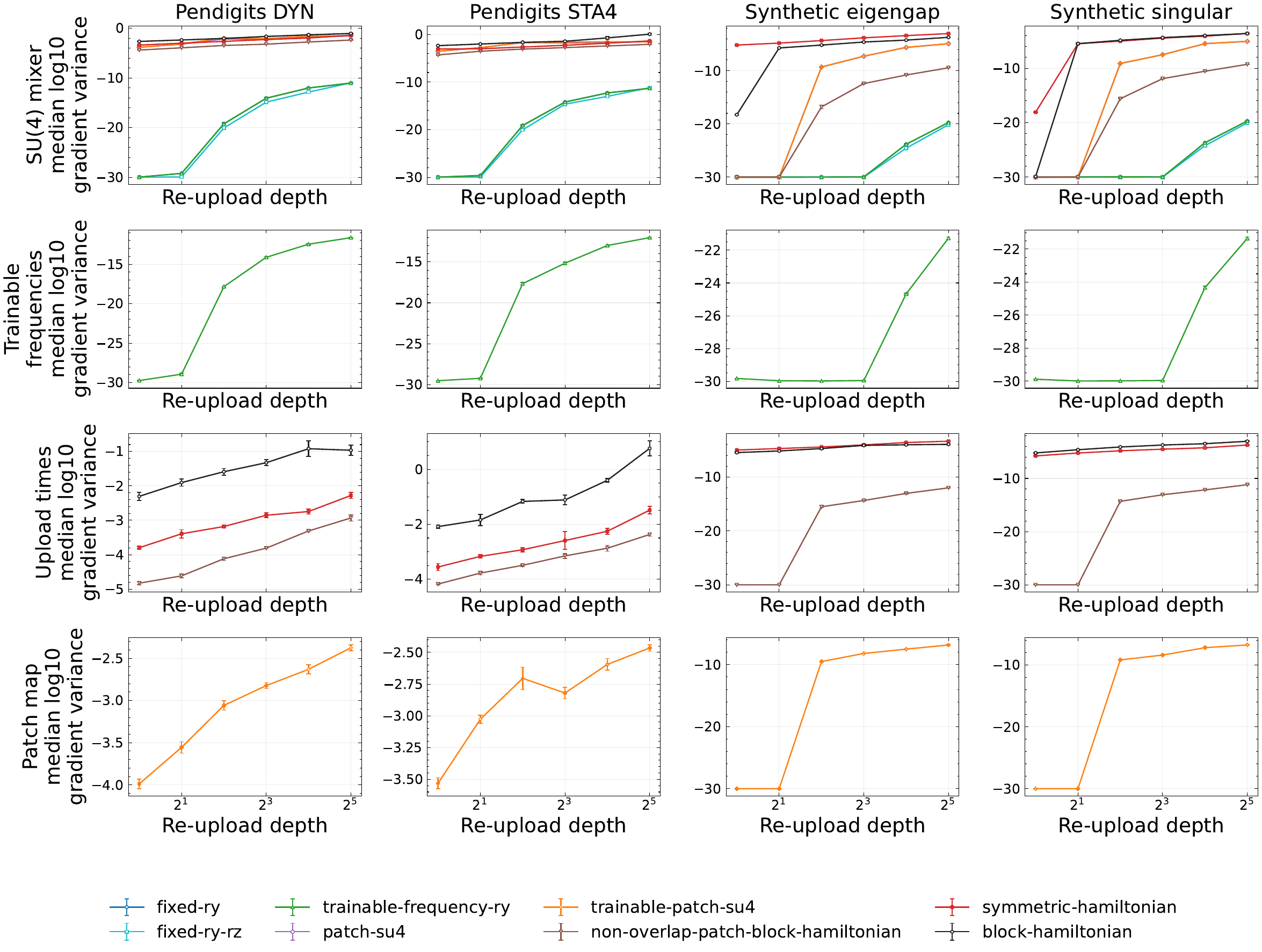}
        \caption{ }\label{fig:gradient-init-var}
    \end{subfigure}
    \vfill
    \begin{subfigure}[t]{1\textwidth}
        \centering
        \includegraphics[width=0.79\linewidth]{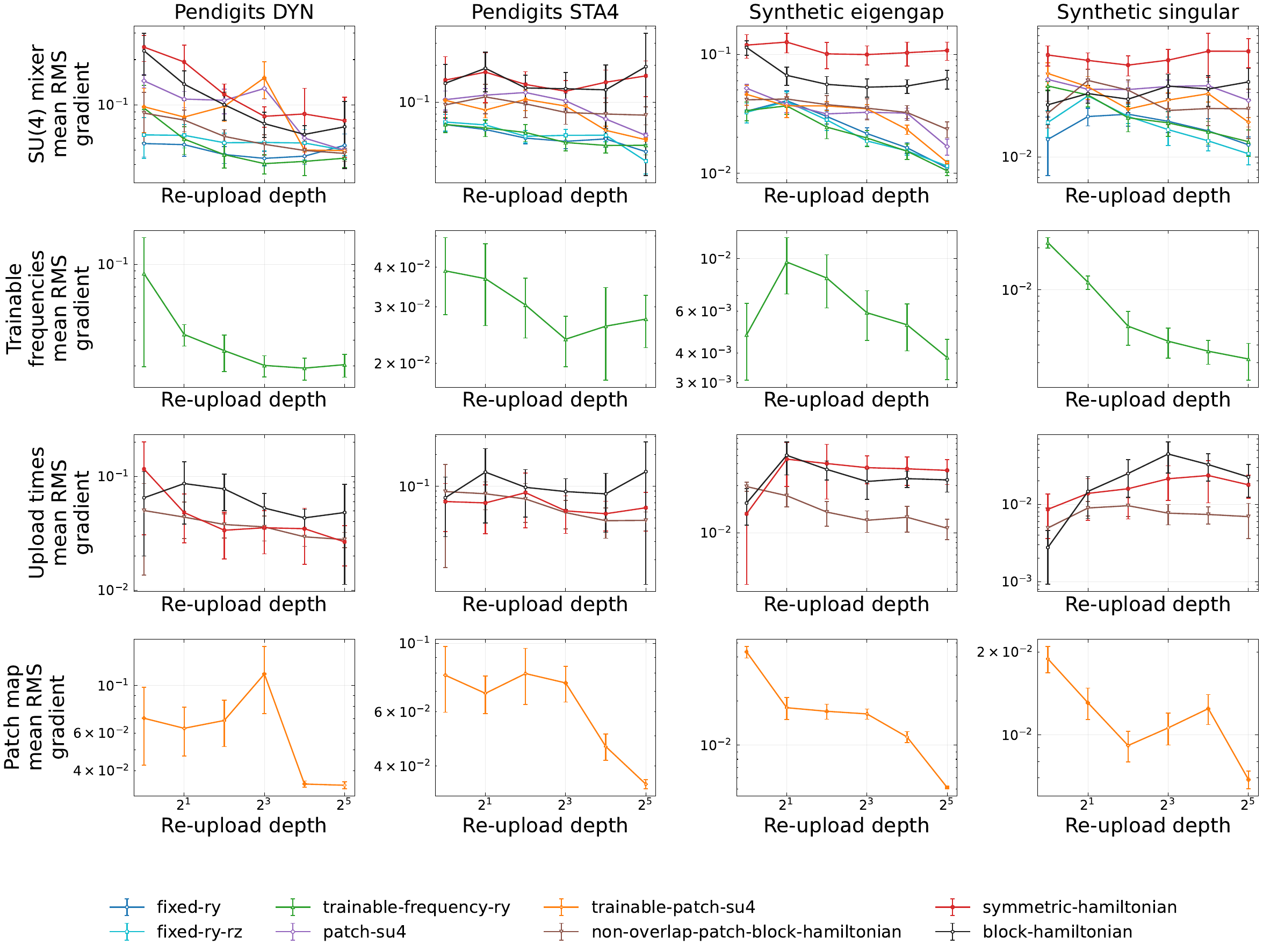}
        \caption{ }\label{fig:gradient-final-rms}
    \end{subfigure}
    \caption{
    \textbf{(a) Gradient variance at initialisation as a function of reuploading depth.} For each run configuration, gradients are computed on a fixed diagnostic training batch of size 32 for 20 random initialisations. The plotted quantity is the median $\log_{10}$ variance across parameters within each parameter group, averaged over 20 run seeds; error bars show one standard deviation across run seeds. The coordinate-wise rotation-gate data-encoding unitaries exhibit strongly suppressed mixer-gradient variance at large depth, whereas the QSMs retain substantially larger initialisation variance.
    \textbf{(b) Final-checkpoint RMS gradient as a function of reuploading depth.} Gradients are computed at the final trained checkpoint using the same diagnostic-batch procedure as in the initialisation diagnostics. The RMS is computed within each parameter group, and plotted values are means over 20 run seeds with one-standard-deviation error bars. The QSMs retain finite gradients at high depth, while the rotation-gate data-encoding unitaries generally show smaller final mixer-gradient scales, especially on the synthetic spectral tasks.
    }\label{fig:main-grad-analysis}
\end{figure}

\noindent\textbf{QSMs with Hamiltonian-based data-encoding unitaries retain stronger high-depth gradient signals, while gradient scale alone does not determine accuracy.} At $L=32$, the median $\log_{10}$ mixer-gradient variance \cite{McClean2018-pf} of the fixed and trainable-frequency rotation-gate data-encoding unitaries is approximately $-11$ on Pendigits and $-20$ on the synthetic tasks. For $H_{\mathrm{block}}$, the corresponding values are $-1.05\pm0.05$ on Pendigits \texttt{DYN}, $0.05\pm0.20$ on \texttt{STA4}, $-3.72\pm0.05$ on \texttt{SYNTHETIC EIGENGAP} and $-3.64\pm0.04$ on \texttt{SYNTHETIC SINGULAR}.
We compute these training-loss gradients on a fixed diagnostic batch of 32 examples and separate the shared $SU(4)$ mixer parameters from encoder-specific parameters. Within each run seed, the initialisation statistic measures the coordinate-wise variance of each parameter gradient across 20 parameter initialisations; we take the median log variance across a parameter group and then aggregate the run-level values over 20 run seeds, with sample-standard-deviation error bars. The coordinate-wise rotation encoders therefore show much more strongly suppressed parameter-wise variation across initialisations at high depth. Appendix \Cref{sec:appendix-grad-analysis} defines this statistic and the remaining gradient diagnostics.

The final-checkpoint root-mean-square (RMS) gradient gives a separate snapshot after optimisation (\Cref{fig:gradient-final-rms}). At depth 32, the mixer-gradient RMS values for $H_{\mathrm{sym}}$ and $H_{\mathrm{block}}$ are $0.079\pm0.034$ and $0.071\pm0.034$ on Pendigits \texttt{DYN}, $0.164\pm0.053$ and $0.194\pm0.169$ on \texttt{STA4}, $0.108\pm0.019$ and $0.062\pm0.011$ on \texttt{SYNTHETIC EIGENGAP}, and $0.061\pm0.015$ and $0.036\pm0.009$ on \texttt{SYNTHETIC SINGULAR}. On the synthetic tasks these values exceed the rotation-gate baseline values, which are near $10^{-2}$, showing that several high-performing QSMs retain larger mixer-gradient scales at the final checkpoint.

The empirical-Fisher diagnostic is the Gram matrix of per-example, ground-truth-label loss gradients, following the common machine-learning convention \cite{Kunstner2019-ef}. We use its effective rank as a finite-batch summary of how the sampled gradient energy is distributed across parameter-space directions.
At depth 32 on Pendigits \texttt{DYN}, patch-$SU(4)$ has empirical-Fisher effective rank $29.25\pm0.52$ but test accuracy $9.96\%\pm0.44\%$. The corresponding values for $H_{\mathrm{block}}$ are $9.39\pm2.72$ and $92.35\%\pm5.95\%$ (Appendix \Cref{sec:appendix-grad-analysis} and \Cref{tab:gradient-depth32-summary}). In this comparison, the empirical-Fisher effective-rank ordering does not match the test-accuracy ordering.

The gradient results establish two complementary observations: extreme initialisation-variance suppression accompanies the lower high-depth accuracy of the rotation-gate data-encoding unitaries, while the patch-$SU(4)$ controls show that larger gradient variance or empirical-Fisher rank is insufficient for high accuracy. These diagnostics describe the optimisation geometry of the tested runs but do not establish a causal mechanism or a necessary gradient condition for the QSM results. Appendix \Cref{sec:appendix-grad-analysis} gives the complete metric definitions, while \Cref{fig:appendix-grad-init-rms,fig:appendix-grad-init-fisher} reports additional initialisation-RMS and empirical-Fisher results.

\begin{figure}[htbp!]
    \centering
    \begin{subfigure}[t]{0.73\textwidth}
        \includegraphics[width=\linewidth]{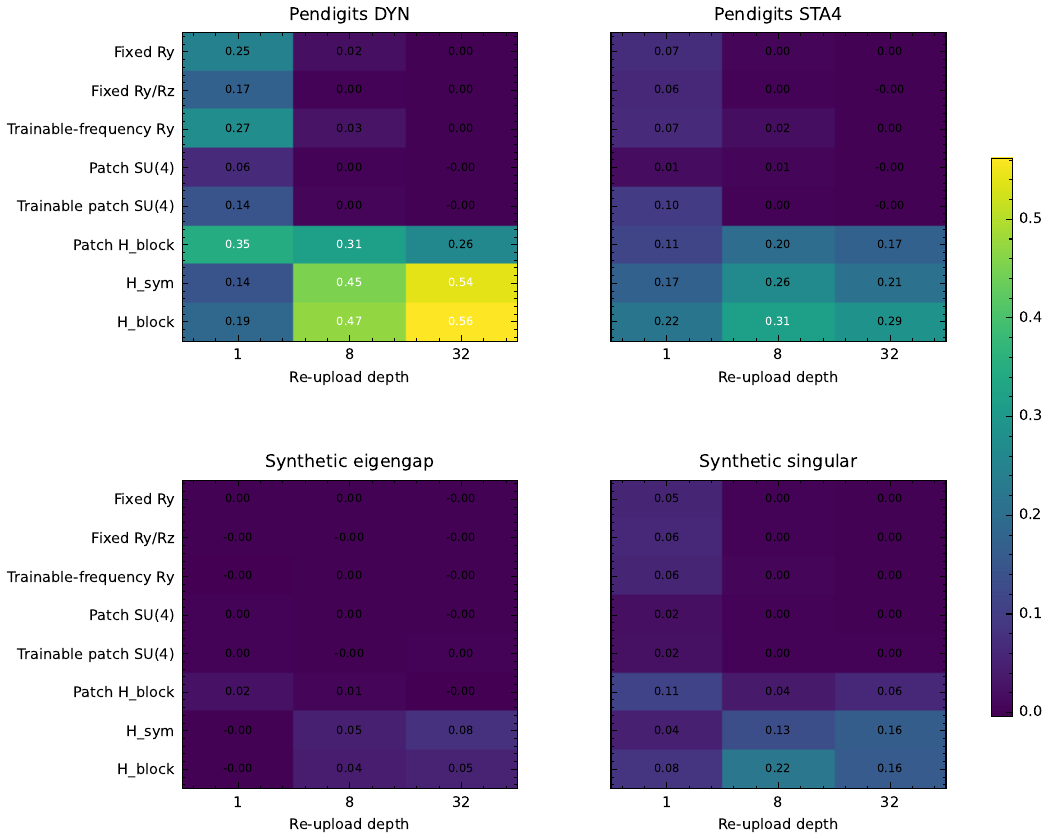}
        \caption{ }\label{fig:latent-summary-kernel-gap}
    \end{subfigure}
    \begin{subfigure}[t]{0.73\textwidth}
        \includegraphics[width=\linewidth]{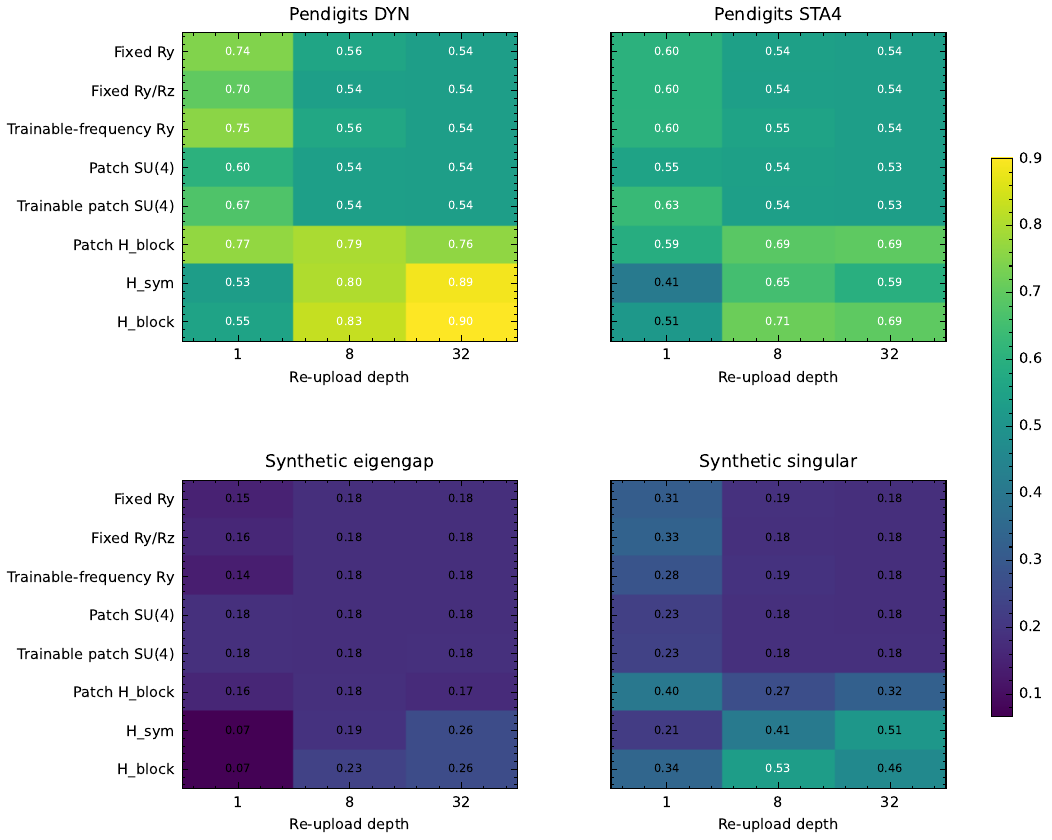}
        \caption{ }\label{fig:latent-summary-kernel-target-alignment}
    \end{subfigure}
    \caption{
    \textbf{(a) Within-minus-between fidelity gap of the final quantum states.} For each trained model, we compute the pure-state fidelity kernel $K_{ij} = |\braket{\psi_i | \psi_j}|^2$ on a fixed 32-example validation diagnostic batch and report the mean same-label off-diagonal fidelity minus the mean different-label off-diagonal fidelity. Larger values indicate stronger class separation in quantum-state geometry. At high depth, the global QSMs and the patch-local block-Hamiltonian QSM produce substantially larger gaps than the coordinate-wise rotation-gate and patch-$SU(4)$ baselines.
    \textbf{(b) Kernel-target alignment of the final quantum-state fidelity kernel.} The plotted metric is the centred alignment between the final fidelity kernel and the label-equality kernel on the same batch. Larger values indicate that the geometry of the trained quantum states is more aligned with class membership. Several high-performing QSMs show stronger alignment in the high-depth regime; this is a descriptive association rather than evidence that the data-encoding unitary causes the accuracy ordering. Both panels use the final checkpoint from seed 0 and contain no across-seed uncertainty estimate.
    }\label{fig:main-latent-summary}
\end{figure}

\noindent\textbf{Several high-performing quantum spectral models develop more label-aligned fidelity geometry.} At high depth, several of these models show larger \emph{within-minus-between fidelity gap}s and stronger kernel-target alignment than the coordinate-wise rotation and patch-$SU(4)$ baselines (see \Cref{fig:main-latent-summary}). We form the pure-state fidelity kernel $K_{ij}=|\braket{\psi_i|\psi_j}|^2$ from the final post-mixer states of a fixed 32-example validation batch. Centred kernel-target alignment \cite{on-kernel-target-alignment,centered-alignement-paper,Kornblith2019-vz} compares this kernel with the label-equality kernel, while the \emph{within-minus-between fidelity gap}, motivated by class separation in quantum metric learning \cite{lloyd2020quantumembeddingsmachinelearning}, is the mean same-label fidelity minus the mean different-label fidelity after excluding self-pairs (\Cref{eqn:within-minus-between-fidelity-gap}). Positive gaps indicate greater average similarity within classes on the diagnostic batch. Except for the 20-seed test accuracies in \Cref{tab:latent-depth32-summary}, all latent diagnostics use seed 0 and have no cross-seed uncertainty estimates; the appendix expands this provenance and scope.

At depth 32 on Pendigits \texttt{DYN}, $H_{\mathrm{block}}$ has diagnostic-batch accuracy 1.000, target alignment 0.900 and fidelity gap 0.561; $H_{\mathrm{sym}}$ gives 1.000, 0.887 and 0.538, respectively (\Cref{tab:latent-depth32-summary}). On \texttt{STA4}, patch $H_{\mathrm{block}}$ has diagnostic-batch accuracy 0.938, target alignment 0.694 and gap 0.173. On \texttt{SYNTHETIC SINGULAR}, $H_{\mathrm{sym}}$ and $H_{\mathrm{block}}$ give target alignments of 0.508 and 0.461 and gaps of 0.163 and 0.159. These observations show that several high-performing QSMs form visibly class-separated fidelity geometry within the diagnostic batch.

The fidelity gap and participation-ratio effective rank describe complementary aspects of the kernel geometry: the former measures label separation, while the latter measures how broadly the kernel eigenvalue mass is distributed across eigenmodes. At depth 32, the rotation-gate data-encoding unitaries have kernel effective rank near the batch-size maximum of 32 while their gaps are approximately zero (see \Cref{fig:appendix-latent-summary-kernel-effective-rank}). The patch-$SU(4)$ data-encoding unitaries show the same combination of high effective rank, near-zero gap and low diagnostic-batch accuracy. Near-maximal effective rank in a fidelity kernel is consistent with an identity-like Gram matrix, a limiting geometry that can arise through exponential concentration and impair generalisation in quantum kernel methods \cite{Thanasilp2024-ip}. The positive gaps of several QSMs instead indicate class-dependent off-diagonal structure within this diagnostic batch.

Appendix \Cref{sec-appendix-latent} gives the corresponding layerwise and checkpoint trajectories, including the fidelity gap, target alignment and adjacent-layer CKA.

\noindent\textbf{Classical baselines probe which representations make each task accessible.}
The synthetic labels are defined by spectral statistics of the clean latent matrices. On \texttt{SYNTHETIC EIGENGAP}, the $H_{\mathrm{sym}}$-eigenvalue MLP reaches $98.01\%\pm0.35\%$, compared with the largest observed QSM mean of $86.65\%\pm1.18\%$. On \texttt{SYNTHETIC SINGULAR}, the $H_{\mathrm{block}}$-singular-value MLP reaches $98.50\%\pm0.20\%$, compared with $95.24\%\pm1.53\%$. These near-ceiling accuracies provide an empirical check that the noisy observations retain the intended label-defining spectral signal, supporting the use of these benchmarks as controlled tests of eigenvalue- and singular-value-dependent learning.

On Pendigits, one-hidden-layer MLPs trained on raw entries reach $96.88\%\pm0.24\%$ on \texttt{DYN} and $91.32\%\pm0.46\%$ on \texttt{STA4} (see \Cref{tab:classical-baselines}). By contrast, MLPs trained only on $H_{\mathrm{sym}}$ eigenvalues reach $59.98\%\pm0.60\%$ and $39.80\%\pm0.50\%$, while those trained on $H_{\mathrm{block}}$ singular values reach $39.67\%\pm0.48\%$ and $30.62\%\pm0.53\%$. Under this protocol, reducing each input to its spectral values therefore removes substantial task-relevant information available in the raw matrices. The classical baselines reveal a task-dependent contrast: spectral values are almost sufficient for the controlled synthetic tasks, but not for Pendigits. The value--subspace ablations below examine how this contrast appears within the QSM representations.

\noindent\textbf{Ablations reveal task-dependent roles for spectral values and subspaces.} The central ablation result is a reversal between the real-world and controlled tasks (see \Cref{tab:core-ablation-summary}). The spectral-value controls retain each sample's values while using fixed reference bases derived from the training split. The spectral-subspace controls instead retain each sample's eigenvectors or singular vectors while replacing its values by coordinate-wise training-set medians. On Pendigits, the subspace-retaining controls preserve much more accuracy than the spectral-value controls. On the synthetic tasks, where the labels are defined by the corresponding spectral statistics, the spectral-value controls become strongest and the subspace-retaining controls are much weaker. This reversal shows that the useful component of sample-conditioned spectral geometry depends on the structure of the learning problem. To summarise performance across depth, we compare the largest mean test accuracy observed for each ablation over the six evaluated depths.

\begin{table*}[htbp!]
  \caption{\textbf{Value--subspace ablation summary.} Entries give the depth and
  largest observed mean final test accuracy (mean $\pm$ standard deviation over
  20 random seeds, in percentage) for the original encoder and the corresponding
  value-only and subspace-only controls. Every entry is a descriptive maximum
  across the tested depths rather than performance at a depth selected by
  independent validation; each control is retrained on its transformed
  dataset. Pendigits depends primarily on sample-dependent subspaces, whereas
  the controlled
  spectral tasks reverse the ordering. The complete ablation and permutation
  results are reported in Appendix \Cref{tab:ablation-summary}.}
  \label{tab:core-ablation-summary}
  \centering
  \small
  \setlength{\tabcolsep}{3.5pt}
  \begin{tabular}{lcccc}
    \toprule
    \shortstack{Task\\(reference encoder)} &
    \shortstack{Original\\(depth, acc.)} &
    \shortstack{Spectral values only\\(depth, acc.)} &
    \shortstack{Spectral subspaces only\\(depth, acc.)} &
    \shortstack{Primary observed\\dependence} \\
    \midrule
    \shortstack[l]{Pendigits DYN\\($H_{\rm block}$)} &
    L16, $93.81\pm1.06$ & L4, $31.65\pm2.44$ &
    L16, $93.56\pm0.84$ & subspaces \\
    \shortstack[l]{Pendigits STA4\\($H_{\rm block}$)} &
    L16, $86.66\pm2.17$ & L4, $25.25\pm2.02$ &
    L16, $84.74\pm4.51$ & subspaces \\
    \shortstack[l]{Synthetic eigengap\\($H_{\rm sym}$)} &
    L16, $83.42\pm1.37$ & L8, $97.92\pm0.38$ &
    L1, $61.28\pm0.07$ & values \\
    \shortstack[l]{Synthetic singular\\($H_{\rm block}$)} &
    L16, $95.24\pm1.53$ & L32, $98.42\pm0.30$ &
    L16, $49.88\pm1.96$ & values \\
    \bottomrule
  \end{tabular}
\end{table*}

On Pendigits \texttt{DYN} and \texttt{STA4}, the $H_{\mathrm{block}}$ spectral-value controls reach $31.65\%\pm2.44\%$ and $25.25\%\pm2.02\%$, whereas the spectral-subspace controls retain $93.56\%\pm0.84\%$ and $84.74\%\pm4.51\%$. The $H_{\mathrm{sym}}$ controls show the same ordering: spectral-value accuracies are $48.87\%\pm3.90\%$ and $28.35\%\pm2.11\%$, compared with $89.80\%\pm4.56\%$ and $64.30\%\pm2.02\%$ when sample-dependent eigenvectors are retained. Within these ablations, the Pendigits results therefore depend much more strongly on sample-dependent spectral subspaces than on spectral values alone.

The synthetic controls reverse this performance ordering. The $H_{\mathrm{sym}}$ spectral-value control reaches $97.92\%\pm0.38\%$ on \texttt{SYNTHETIC EIGENGAP}, whereas its spectral-subspace counterpart reaches $61.28\%\pm0.07\%$. On \texttt{SYNTHETIC SINGULAR}, the corresponding $H_{\mathrm{block}}$ values are $98.42\%\pm0.30\%$ and $49.88\%\pm1.96\%$. This reversal agrees with the data-generating rules and demonstrates that spectral values and subspaces can play different roles under the same upload--mixer architecture.

Fixed row and column permutations preserve singular values, whereas a general entry permutation does not. On \texttt{SYNTHETIC SINGULAR}, the largest observed mean for global $H_{\mathrm{block}}$ is $95.06\%\pm2.30\%$ after row/column permutation, 0.18 percentage points below the original, and $90.55\%\pm0.58\%$ after entry permutation, 4.69 points below. For patch $H_{\mathrm{block}}$ on \texttt{STA4}, entry and row/column permutations reduce the largest observed means by 3.04 and 2.22 points. This sensitivity is compatible with the use of fixed $2\times2$ patches, but does not isolate locality as its cause. The complete value/subspace and permutation results are given in Appendix \Cref{tab:ablation-summary}.

\section{Discussion and Outlook}\label{sec:conclusion}

Quantum Spectral Models (QSMs) operate in the spectral domain of each matrix input: we construct an input-dependent Hermitian generator from the matrix and use its time-evolution operator as the data-encoding unitary. For the symmetric- and block-Hamiltonian data-encoding unitaries, eigengaps or signed singular-value gaps determine the sample-conditioned candidate phase carriers. The associated spectral projectors represent the input-dependent subspaces that enter the corresponding Fourier coefficients. Candidate support specifies the carriers that are available in principle, whereas realised support contains only those whose coefficients remain nonzero after the initial state, trainable mixing and measurement observable are taken into account. Common coordinate-wise quantum Fourier models (QFMs) instead use architecture-defined candidate support, while trainable-frequency rotation-gate QFMs learn support that remains globally shared across samples after training \cite{Schuld2021-kv,Jaderberg2024-ye}. The key difference is that, in a QSM, both the candidate phase carriers and the spectral subspaces entering the corresponding Fourier coefficients depend on the individual input matrix.

At the largest evaluated depth, a QSM variant attains the highest mean test accuracy among the tested quantum models on each of the four benchmarks. The patch-local block-Hamiltonian QSM leads on the two Pendigits representations, whereas the global block-Hamiltonian QSM leads on the two controlled spectral tasks (see \Cref{fig:main-res} and \Cref{tab:main-test-accuracy}). Together with the differing trends across the full depth sweeps, this task-dependent ordering shows that the effect of circuit depth is specific to the encoder and dataset.

One possible, non-causal interpretation of the change in the depth-32 winner concerns the scale of structure exposed by each encoder. The two synthetic labels are defined by spectral statistics of the complete clean latent matrix, whereas each Pendigits input contains an explicit local organisation. For \texttt{DYN}, the four non-overlapping $2\times2$ patches are four consecutive two-point segments of the ordered pen trajectory; for \texttt{STA4}, they are four contiguous $2\times2$ quadrants of the bitmap-like matrix. The patch-local QSM therefore preserves these trajectory segments or spatial neighbourhoods, while the global block-Hamiltonian QSM retains relations across the full matrix. This correspondence is consistent with a locality-based inductive-bias interpretation and motivates future controlled comparisons that isolate locality from other encoder differences and clarify its contribution to the observed accuracy ordering.

The gradient and fidelity diagnostics add complementary descriptive evidence. At large depth, we observe extreme suppression of mixer-gradient variance across initialisations for the coordinate-wise rotation-gate encoders, whereas the QSMs retain greater variance and several high-performing variants retain larger final gradient scales (see \Cref{fig:main-grad-analysis}). In the fixed diagnostic batch, several QSMs also develop larger fidelity gaps and stronger kernel-target alignment (see \Cref{fig:main-latent-summary}). The patch-$SU(4)$ controls further show that the gradient and fidelity diagnostics capture different aspects of the learned model geometry. Future work could investigate how these measures interact throughout training and whether their joint behaviour provides a more complete account of predictive performance.

The value-versus-subspace ablations provide the most direct representational evidence. On Pendigits, value-only controls lose most of the original accuracy, while controls that preserve sample-dependent eigenvectors or singular vectors retain much more performance (\Cref{tab:core-ablation-summary,tab:ablation-summary}). The ordering reverses on the controlled spectral tasks, where the labels are defined by an eigengap or a leading-singular-value sum and the value-only controls become strongest. This task-dependent division of labour shows that spectral values make the synthetic rules directly accessible, whereas spectral subspaces and trainable mixing carry much of the useful information for Pendigits.

These results point to a general representation-design question: should a spectral representation be shared across a dataset or adapted to each input? Coordinate-wise QFMs use architecture-defined carriers, trainable-frequency QFMs learn support shared across samples, and several classical spectral architectures use fixed or domain-derived operators or Fourier-space maps learned across a training distribution \cite{bruna2014spectralnetworkslocallyconnected,li2021fourier}. A QSM takes a different approach by constructing the data-encoding generator from each input matrix, so that both the gap-derived candidate phase carriers and the spectral subspaces entering the Fourier coefficients adapt to the input. Because the time evolution of a Hermitian generator is governed by its eigenvalues and spectral projectors, using the input matrix as the generator makes spectral structure intrinsic to the data-encoding dynamics. This provides a particularly direct route to input-adaptive spectral representation learning in quantum models and motivates further investigation of when such representations are useful across quantum and classical models.

Building on these results, future work can test how broadly the observed representation effects persist across model scales, datasets and resource constraints. Resource-matched comparisons could hold qubit count, parameter count and locality fixed, select depth through independent validation, retain the task-appropriate classical baselines that lead on the present benchmarks, and extend the analysis to larger matrix-structured datasets. The nonzero fidelity gaps of several QSMs at the studied sizes also motivate qubit-scaling and finite-shot analyses of whether Hamiltonian-based data encodings resist exponential kernel concentration \cite{Thanasilp2024-ip}. Multi-seed latent-state diagnostics across independent batches, together with controlled global-versus-patch-local and structure-destroying ablations, could further clarify how spectral subspaces, locality and trainability interact. Together, these directions provide a path towards hardware-relevant, resource-normalised evaluations of whether the observed representation effects can support end-to-end quantum advantage.

A complementary direction for future work is to develop scalable implementations of dense input-dependent matrix operations. This would involve specifying the data-access or block-encoding assumptions \cite{Low_2017,Low_2019}, accounting for data-loading and circuit costs, and examining noise in hardware-relevant settings. At the representation level, quantum singular value transformation or eigenvalue processing could provide controlled tests of spectral values while variational operations act on the associated subspaces \cite{Gilyen2019-ry,Low2026-tr}. Investigating how these approaches depend on matrix structure and access cost could connect QSM representation design with scalable quantum implementations.

Taken together, our formal analysis and experiments establish QSMs as an analysable framework for input-adaptive spectral representation learning. The input-derived Hamiltonian supplies sample-conditioned candidate phase carriers through its spectral gaps, while the associated spectral subspaces interact with the initial state, trainable mixer and measurement observable to determine which carriers are realised and with what coefficients. This support--coefficient perspective provides a concrete principle for designing quantum models around matrix-valued data and offers a new lens for studying how data-dependent structure can be embedded directly into model dynamics.

%This package was used to typeset Table~\ref{sample-table}.

%\begin{table}[H]
%  \caption{Sample table caption. Explain what the table shows and add a key take-away message to the caption.}
%  \label{sample-table}
%  \centering
%  \begin{tabular}{lll}
%    \toprule
%    \multicolumn{2}{c}{Part}                   \\
%    \cmidrule(r){1-2}
%    Name     & Description     & Size ($\mu$m) \\
%    \midrule
%    Dendrite & Input terminal  & $\approx$100     \\
%    Axon     & Output terminal & $\approx$10      \\
%    Soma     & Cell body       & up to $10^6$  \\
%    \bottomrule
%  \end{tabular}
%\end{table}

%\section*{References}
\clearpage
%\section*{Acknowledgments}

\section*{Data and Code Availability}

Code available at \url{https://github.com/peiyong-addwater/QuantumSpectralModel}.
\bibliographystyle{plainnat}
\bibliography{Sections/main}

%%%%%%%%%%%%%%%%%%%%%%%%%%%%%%%%%%%%%%%%%%%%%%%%%%%%%%%%%%%%
\newpage
\appendix

\addcontentsline{toc}{section}{Appendix} % Add the appendix text to the document TOC
% \vspace{-5cm}
\part{Appendix} % Start the appendix part
\parttoc % Insert the appendix TOC
\section{Preliminaries on Quantum Computing}\label{sec:appendix-qc-preliminaries}

This appendix introduces the quantum-computing notation and basic linear-algebraic objects needed to follow the paper, assuming no prior background in quantum computing or quantum information.

\subsection{Quantum states and bra-ket notation}
A quantum state is represented by a complex vector. The standard notation for a column vector is a \textbf{ket}, written as $\ket{\psi}$. Its conjugate transpose is a \textbf{bra}, written as $\bra{\psi}$. If
\begin{equation}
    \ket{\psi} = \begin{pmatrix}
        \psi_0 \\ \psi_1 \\ \vdots \\ \psi_{d-1}
    \end{pmatrix} \in \mathbb{C}^d,
\end{equation}
then
\begin{equation}
    \bra{\psi} = \begin{pmatrix}
        \psi_0^{\star} & \psi_1^{\star} & \cdots & \psi_{d-1}^{\star},
    \end{pmatrix}
\end{equation}
where ${}^{\star}$ denotes complex conjugation. The inner product between the two states is written as
\begin{equation}
    \braket{\phi | \psi} = \sum_{j=0}^{d-1} \phi^{\star}_j \psi_j.
\end{equation}
This is the usual complex inner product from linear algebra. The squared norm of a state is therefore
\begin{equation}
    \braket{\psi | \psi} = \sum_{j=0}^{d-1} |\psi_j|^2.
\end{equation}
A valid pure quantum state is normalised:
\begin{equation}
    \braket{\psi | \psi} = 1.
\end{equation}
The entries of $\ket{\psi}$ are called \textbf{amplitudes}, not probabilities. Measurement probabilities are obtained by taking squared magnitudes. For example, if
\begin{equation}
    \ket{\psi} = \alpha \ket{0} + \beta \ket{1},
\end{equation}
then the probability of observing outcome $0$ is $|\alpha|^2$, and the probability of observing outcome $1$ is $|\beta|^2$. Normalisation requires
\begin{equation}
    |\alpha|^2 + |\beta|^2 = 1.
\end{equation}
The notation
\begin{equation}
    \ket{\phi}\bra{\psi}
\end{equation}
denotes an \textbf{outer product}, hence a matrix. In particular,
\begin{equation}
    \ket{\psi}\bra{\psi}
\end{equation}
is the rank-one projector onto the one-dimensional subspace spanned by $\ket{\psi}$. Projectors of this type appear throughout quantum machine learning because measurements and readouts are naturally written as projections onto subspaces.

\subsection{Computational basis states}

A single qubit is a two-dimensional complex vector. The standard basis vectors are
\begin{equation}
    \ket{0} = \begin{pmatrix}
        1 \\ 0
    \end{pmatrix}, \quad \ket{1} = \begin{pmatrix}
        0 \\ 1
    \end{pmatrix}.
\end{equation}
These are called the computational basis states. A general one-qubit pure state is
\begin{equation}
    \ket{\psi} = \alpha \ket{0} + \beta \ket{1} = \begin{pmatrix}
        \alpha \\ \beta
    \end{pmatrix},
\end{equation}
with $|\alpha|^2 + |\beta|^2 = 1$.

For $n$ qubits, the state space has dimension $2^n$. The computational basis consists of all binary strings of length $n$:
\begin{equation}
    \ket{b_1 b_2 \cdots b_n }, \quad b_j \in \{0,1\}.
\end{equation}
For example, for two qubits the computational basis is
\begin{equation}
    \ket{00}, \ket{01}, \ket{10}, \ket{11}.
\end{equation}
A general two-qubit state is 
\begin{equation}
    \ket{\psi} = \alpha_{00} \ket{00} + \alpha_{01} \ket{01} + \alpha_{10} \ket{10} + \alpha_{11}\ket{11},
\end{equation}
where
\begin{equation}
    |\alpha_{00}|^2 + |\alpha_{01}|^2 + |\alpha_{10}|^2 + |\alpha_{11}|^2 = 1.
\end{equation}
More generally, an $n$-qubit state can be written as
\begin{equation}
    \ket{\psi} = \sum_{b\in \{0,1\}^n} \alpha_b \ket{b}, \quad \sum_{b\in \{0,1\}^n} |\alpha_b|^2 = 1.
\end{equation}
Thus an $n$-qubit state is a length-$2^n$ complex vector indexed by binary strings.

A useful single-qubit state used in the paper is
\begin{equation}
    \ket{+} = \frac{\ket{0}+\ket{1}}{\sqrt{2}}.
\end{equation}
The initial state in our experiment is
\begin{equation}
    \ket{+}^{\otimes n},
\end{equation}
which means that every qubit starts in the $\ket{+}$ state. This is the uniform superposition over all computational basis states:
\begin{equation}
    \ket{+}^{\otimes n} = \frac{1}{\sqrt{2^n}} \sum_{b\in \{0,1 \}^n } \ket{b}.
\end{equation}

\subsection{Tensor products}
The tensor product is the operation used to combine quantum systems. If
\begin{equation}
    \ket{a} = \begin{pmatrix}
        a_0 \\ a_1
    \end{pmatrix}, \qquad \ket{b} = \begin{pmatrix}
        b_0 \\ b_1
    \end{pmatrix},
\end{equation}
then
\begin{equation}
    \ket{a}\otimes\ket{b} = \begin{pmatrix}
        a_0 b_0 \\ a_0 b_1 \\ a_1 b_0 \\ a_1 b_1
    \end{pmatrix}.
\end{equation}
For example,
\begin{equation}
    \ket{0}\otimes\ket{1} = \ket{01} = \begin{pmatrix}
        0 \\ 1 \\ 0 \\ 0
    \end{pmatrix}.
\end{equation}
It is common to omit the tensor-product symbol for the basis states, so
\begin{equation}
    \ket{0}\otimes\ket{1}
\end{equation}
is written simply as
\begin{equation}
    \ket{01}.
\end{equation}
Tensor products also combine operators. If $A$ acts on the first qubit and $B$ acts on the second qubit, then
\begin{equation}
    A\otimes B
\end{equation}
acts on the two-qubit system. For example, $X\otimes I$ means ``apply $X$ to the first qubit and do nothing to the second,'' while $I\otimes X$ means ''do nothing to the first qubit and apply $X$ to the second."

For $n$ qubits, a local single-qubit operator acting on qubit $j$ is implicitly tensored with identities on all other qubits, such as the $Y_j$ operator in the main text, which means Pauli $Y$ applied to qubit $j$, with identity operators on the remaining qubits.

\subsection{Pauli matrices}
The Pauli matrices are four $2\times 2$ matrices that form a basic operator basis for one qubit:
\begin{equation}
    \begin{aligned}
        I &= \begin{pmatrix}
            1 & 0 \\ 0 & 1
        \end{pmatrix}, \quad X = \begin{pmatrix}
            0 & 1 \\ 1 & 0
        \end{pmatrix}, \\
        Y &= \begin{pmatrix}
            0 & -i \\ i & 0
        \end{pmatrix}, \quad Z = \begin{pmatrix}
            1 & 0 \\ 0 & -1
        \end{pmatrix}.
    \end{aligned}
\end{equation}
Here $I$ is the identity. The other three are the Pauli $X, Y,$ and $Z$ operators.

The Pauli $X$ matrix flips the computational basis states:
\begin{equation}
    X\ket{0} = \ket{1}, \quad X\ket{1} = \ket{0}.
\end{equation}
The Pauli $Z$ matrix leaves $\ket{0}$ unchanged and changes the sign of $\ket{1}$:
\begin{equation}
    Z\ket{0} = \ket{0}, \quad Z\ket{1} = -\ket{1}.
\end{equation}
The Pauli $Y$ is similar to a bit flip but includes complex phases:
\begin{equation}
    Y\ket{0} = i\ket{1} \quad Y\ket{1} = -i\ket{0}.
\end{equation}
All Pauli matrices are Hermitian and unitary. Hermitian means
\begin{equation}
    P^{\dagger} = P,
\end{equation}
and unitary means
\begin{equation}
    P^{\dagger}P = I.
\end{equation}
Each of $X, Y, Z$ has eigenvalues $+1$ and $-1$. This simple spectrum is why Pauli rotations generate Fourier-like factors in data reuploading circuits.

Multi-qubit Pauli strings are tensor products of one-qubit Pauli operators. For example,
\begin{equation}
    X\otimes Z, \quad Y\otimes I, \quad Z\otimes X \otimes Y
\end{equation}
are Pauli strings. In this paper, the trainable two-qubit SU(4) blocks are parameterised using the 15 non-identity two-qubit Pauli strings. These are all tensor products of $I, X, Y, Z$ on two qubits except $I\otimes I$.

\subsection{Quantum gates and unitary evolution}
A quantum gate is a unitary matrix $U$. Applying a gate to a state gives
\begin{equation}
    \ket{\psi^\prime} = U\ket{\psi}.
\end{equation}
Unitary preserves normalisation:
\begin{equation}
    \braket{\psi^\prime | \psi^\prime} = \bra{\psi}U^\dagger U \ket{\psi} = \braket{\psi | \psi}.
\end{equation}
Many gates are written as matrix exponentials of Hermitian generators. If $H$ is Hermitian, then
\begin{equation}
    U(t) = \exp\big( -i H t \big)
\end{equation}
is unitary. In this paper, the QSM data-encoding unitaries use the convention
\begin{equation}
    U(t;M) = \exp\Big(-\frac{i}{2} H(M) t \Big),
\end{equation}
where $H(M)$ is a Hamiltonian constructed from the input matrix $M$, and $t$ is a trainable upload-time scale.

Single-qubit rotation gates are also matrix exponentials. For example,
\begin{equation}
    R_y (\theta) = \exp\Big(-\frac{i}{2} \theta Y \Big), \quad R_z (\theta) = \exp\Big(-\frac{i}{2} \theta Z \Big).
\end{equation}
These gates rotate a qubit state by an angle $\theta$ around the $Y$ or $Z$ axis of the Bloch sphere. For the purposes of this paper, the important point is that the data value can appear inside the rotation angle. A typical coordinate-wise rotation-gate encoding has the form
\begin{equation}
    R_y (\alpha x_j),
\end{equation}
where $x_j$ is one input feature and $\alpha$ is a fixed scale.

\subsection{Observables, measurement, and projector readout}
An observable is a Hermitian matrix $O$. Given a state $\ket{\psi}$, the expected value of $O$ is 
\begin{equation}
    \braket{O}_{\psi} = \braket{\psi | O | \psi}.
\end{equation}
This is a real number when $O$ is Hermitian.

For a task with $N$ classes, let $m=\lceil\log_2N\rceil$. The leading $m$ qubits form the label register, and class $j$ is associated with the computational-basis projector
\begin{equation}
    P_j=\ket{j}\bra{j}\otimes\mathbf{I}^{\otimes(n-m)},
    \qquad j\in\{0,\ldots,N-1\},
\end{equation}
where $\ket{j}$ is an $m$-qubit basis state and the identity acts on the remaining qubits. The corresponding raw projector mass is
\begin{equation}
    q_j=\bra{\psi}P_j\ket{\psi}.
\end{equation}
If $N=2^m$, the projectors exhaust the label register and $\sum_jq_j=1$. Otherwise, some label states are unused. The experiments condition on the named class outcomes,
\begin{equation}
    p_j=\frac{q_j}{\sum_{k=0}^{N-1}q_k},
    \qquad \sum_{k=0}^{N-1}q_k>0,
\end{equation}
and predict the class with the largest $p_j$, equivalently the largest $q_j$. As noted in \Cref{sec:preliminary}, the Fourier analysis concerns the raw expectation values $q_j$ rather than the normalised ratios $p_j$.

\section{Derivation of the Frequency Support of Common Data-Encoding Unitaries}\label{sec:rot-gate-freq-derivation}

This section derives the maximal one-upload frequency supports stated in \Cref{eqn:ry-freq-support,eqn:ryrz-local-freq-support,eqn:ryrz-freq-support,eqn:trainable-freq-support}. Throughout, the input consists of $d$ scalar features $\boldsymbol{x}=(x_1,\ldots,x_d)$.

\subsection{Fixed single-qubit \texorpdfstring{$R_y$}{Ry} encoding}

For fixed single-qubit $R_y$ rotation-gate encoding, the encoding layer is
\begin{equation}\label{eqn:ry-encoder}
    S_{R_y}(\boldsymbol{x}) = \bigotimes_{j=1}^d R_y(\alpha x_j),
\end{equation}
where $\alpha$ is a constant scaling factor and
\begin{equation}
    R_y(\alpha x_j) = e^{-i\alpha x_jY_j/2},
\end{equation}
with $Y_j$ denoting the Pauli-$Y$ operator on the $j$-th qubit. Its eigenvalues satisfy $s_j\in\{+1,-1\}$, with $Y_j\ket{s_j}_Y=s_j\ket{s_j}_Y$. The corresponding joint $Y$-basis is
\begin{equation}
    \ket{s}_Y = \ket{s_1}_Y\otimes\cdots\otimes\ket{s_d}_Y,
    \qquad s=(s_1,\ldots,s_d)\in\{\pm1\}^d.
\end{equation}
The joint generator is
\begin{equation}
    K(\boldsymbol{x}) = \sum_{j=1}^d \frac{\alpha x_j}{2}Y_j.
\end{equation}
Because the $Y_j$ operators act on different qubits, they commute, and hence $S_{R_y}(\boldsymbol{x})=e^{-iK(\boldsymbol{x})}$. In the joint $Y$-basis,
\begin{equation}
    K(\boldsymbol{x})\ket{s}_Y = \left(\frac{\alpha}{2}\sum_{j=1}^d s_jx_j\right)\ket{s}_Y.
\end{equation}
Because the joint $Y$-eigenstates form a complete orthonormal basis, the spectral theorem gives the operator identity
\begin{equation}\label{eqn:ry-spectral-decomposition}
    S_{R_y}(\boldsymbol{x}) = \sum_{s\in\{\pm1\}^d}\exp\left(-\frac{i\alpha}{2}\sum_{j=1}^d s_jx_j\right)\Pi_s^Y,
\end{equation}
where $\Pi_s^Y$ is the rank-one projector onto $\ket{s}_Y$. In particular, when applied to a joint eigenstate,
\[
    S_{R_y}(\boldsymbol{x})\ket{s}_Y = \exp\left(-\frac{i\alpha}{2}\sum_{j=1}^d s_jx_j\right)\ket{s}_Y.
\]
The phase-carrier vector is
\begin{equation}
    \eta_s = \frac{\alpha}{2}s = \left(\frac{\alpha s_1}{2},\ldots,\frac{\alpha s_d}{2}\right).
\end{equation}
When this encoding is used in a one-upload raw projector output $q_j$ of the form in \Cref{eqn:logits}, the phase factors from the bra and ket combine. Consequently, the candidate frequency vectors in its Fourier expansion are the pairwise differences
\begin{equation}
    \eta_{s^\prime}-\eta_s = \frac{\alpha}{2}(s^\prime-s).
\end{equation}
Since $(s_j^\prime-s_j)/2\in\{-1,0,1\}$, setting $n_j=(s_j^\prime-s_j)/2$ yields the full vector-valued support in \Cref{eqn:ry-freq-support}.

\subsection{Fixed single-qubit \texorpdfstring{$R_yR_z$}{RyRz} encoding}

For fixed single-qubit $R_yR_z$ encoding,
\begin{equation}\label{eqn:ryrz-encoder}
    S_{R_yR_z}(\boldsymbol{x}) = \bigotimes_{j=1}^d R_z(\beta x_j)R_y(\alpha x_j).
\end{equation}
For one feature,
\begin{equation}
    S_j(x_j) = R_z(\beta x_j)R_y(\alpha x_j).
\end{equation}
The individual spectral decompositions are
\begin{equation}
    R_y(\alpha x_j) = \sum_{s=\pm1}\Pi_s^Y e^{-is\alpha x_j/2},
    \qquad
    R_z(\beta x_j) = \sum_{r=\pm1}\Pi_r^Z e^{-ir\beta x_j/2}.
\end{equation}
Therefore,
\begin{equation}
    \begin{aligned}
        S_j(x_j)
        &= \left(\sum_{r=\pm1}\Pi_r^Z e^{-ir\beta x_j/2}\right)
           \left(\sum_{s=\pm1}\Pi_s^Y e^{-is\alpha x_j/2}\right) \\
        &= \sum_{r,s=\pm1}\Pi_r^Z\Pi_s^Y e^{-i(r\beta+s\alpha)x_j/2}.
    \end{aligned}
\end{equation}
This is a finite carrier expansion with carrier rates
\begin{equation}
    \eta_{r,s} = \frac{r\beta+s\alpha}{2}.
\end{equation}
The candidate one-feature frequencies in the model output are the pairwise differences
\begin{equation}
    \eta_{r^\prime,s^\prime}-\eta_{r,s}
    = \frac{(r^\prime-r)\beta+(s^\prime-s)\alpha}{2}.
\end{equation}
Setting $m=(r^\prime-r)/2$ and $n=(s^\prime-s)/2$, where $m,n\in\{-1,0,1\}$, yields the local support in \Cref{eqn:ryrz-local-freq-support}. Taking the Cartesian product over the $d$ independently encoded features gives the full vector-valued support in \Cref{eqn:ryrz-freq-support}.

\subsection{Component-wise trainable-frequency \texorpdfstring{$R_y$}{Ry} encoding}

For component-wise trainable-frequency $R_y$ encoding, generalising the scalar-value version in \cite{Jaderberg2024-ye},
\begin{equation}\label{eqn:trainable-freq-encoder}
    S_{\mathrm{TF}}(\boldsymbol{x},\gamma)
    = \bigotimes_{j=1}^d R_y(\gamma_jx_j)
    = \bigotimes_{j=1}^d e^{-i\gamma_jx_jY_j/2},
\end{equation}
where $\gamma=(\gamma_1,\ldots,\gamma_d)$ contains the trainable scaling parameters. The same joint $Y$-basis derivation gives the carrier vectors
\begin{equation}
    \eta_s(\gamma) = \left(\frac{s_1\gamma_1}{2},\ldots,\frac{s_d\gamma_d}{2}\right).
\end{equation}
For $n_j=(s_j^\prime-s_j)/2\in\{-1,0,1\}$, the pairwise differences are
\begin{equation}
    \eta_{s^\prime}(\gamma)-\eta_s(\gamma)
    = (n_1\gamma_1,\ldots,n_d\gamma_d),
\end{equation}
which yields the full one-upload vector support in \Cref{eqn:trainable-freq-support}.

\section{Derivations for Hamiltonian-Embedding Frequency Support}\label{sec:H-block-derivation-details}

\subsection{Conditioning variable, padding and coefficient convention}

The Hamiltonian-embedding expansions differ from the coordinate-wise Fourier expansions in \Cref{sec:preliminary}. Here the matrix $M$ is fixed and the upload time is the Fourier variable. Let $H^{(0)}(M)\in\mathbb{R}^{d_H\times d_H}$ denote a matrix-derived Hamiltonian before Hilbert-space padding. For a circuit dimension $D=2^n\geq d_H$, the implemented operator is
\begin{equation}\label{eqn:appendix-direct-sum-padding}
    H(M)=H^{(0)}(M)\oplus\boldsymbol{0}_{D-d_H}.
\end{equation}
The main text uses the same symbol for the unpadded and padded operators. Direct-sum padding preserves every nonzero eigenvalue and enlarges the zero eigenspace.

Let $\{\nu_a(M)\}_{a\in\mathcal{A}(M)}$ be the distinct eigenvalues of $H(M)$, and let $\Pi_a(M)$ be the corresponding spectral projectors. The spectral theorem gives
\begin{equation}\label{eqn:appendix-generic-ham-expansion}
    H(M)=\sum_{a\in\mathcal{A}(M)}\nu_a(M)\Pi_a(M),
    \qquad
    S_H(t;M)=\sum_{a\in\mathcal{A}(M)}e^{-i\nu_a(M)t/2}\Pi_a(M).
\end{equation}
For an initial state $\ket{\psi_0}$ and an observable $O_{\boldsymbol{\theta}}$ that includes the trainable mixer, direct substitution gives
\begin{equation}\label{eqn:appendix-generic-output-expansion}
    \begin{aligned}
        f_1^H(M;t,\boldsymbol{\theta})
        &=\bra{\psi_0}S_H(t;M)^{\dagger}O_{\boldsymbol{\theta}}
          S_H(t;M)\ket{\psi_0}\\
        &=\sum_{a,b}A_{ab}^H(M,\boldsymbol{\theta})
          e^{-i[\nu_b(M)-\nu_a(M)]t/2},
    \end{aligned}
\end{equation}
where
\begin{equation}
    A_{ab}^H(M,\boldsymbol{\theta})
    =\bra{\psi_0}\Pi_a(M)O_{\boldsymbol{\theta}}\Pi_b(M)\ket{\psi_0}.
\end{equation}
Several ordered pairs $(a,b)$ can produce the same gap. The Fourier coefficient at frequency $\omega$ is therefore
\begin{equation}\label{eqn:appendix-aggregated-coefficient}
    C_{\omega}^H(M,\boldsymbol{\theta})
    =\sum_{\substack{a,b:\\ (\nu_b(M)-\nu_a(M))/2=\omega}}
      A_{ab}^H(M,\boldsymbol{\theta}).
\end{equation}
Consequently,
\begin{equation}\label{eqn:appendix-generic-candidate-support}
    \operatorname{supp}_{\mathrm{F}}(f_1^H)
    \subseteq
    \Omega_1^H(M)
    =\left\{\frac{\nu_b(M)-\nu_a(M)}{2}:a,b\in\mathcal{A}(M)\right\}.
\end{equation}
The inclusion can be strict because individual amplitudes can vanish or different amplitudes at the same frequency can cancel. Using eigenspace projectors rather than chosen eigenvectors also makes \Cref{eqn:appendix-generic-output-expansion} invariant under basis changes within a degenerate eigenspace.

The implemented global encoders use an independent time $t_{\ell}$ at each reuploading layer. Expanding every upload in a depth-$L$ circuit gives a multivariate trigonometric polynomial of the form
\begin{equation}\label{eqn:appendix-layer-time-expansion}
    f_L^H(M;\boldsymbol{t},\boldsymbol{\Theta})
    =\sum_{\boldsymbol{a},\boldsymbol{b}}
      A_{\boldsymbol{a}\boldsymbol{b}}^H(M,\boldsymbol{\Theta})
      \exp\left[-\frac{i}{2}\sum_{\ell=1}^{L}
      \big(\nu_{b_{\ell}}(M)-\nu_{a_{\ell}}(M)\big)t_{\ell}\right],
\end{equation}
where the amplitudes contain the ordered products of spectral projectors and intervening mixers. Its maximal candidate support in $\boldsymbol{t}=(t_1,\ldots,t_L)$ is the Cartesian product $[\Omega_1^H(M)]^{\times L}$. If all layer times are instead constrained to one scalar variable, the exponents combine and the corresponding scalar candidate set is the $L$-fold Minkowski sum of $\Omega_1^H(M)$.

\subsection{Global symmetric Hamiltonian embedding}

For the symmetric encoder, $M$ is first zero-column padded to a square matrix $\widetilde{M}$ when $p>q$. The natural Hamiltonian is
\begin{equation}
    H_{\mathrm{sym}}^{(0)}(M)
    =\frac{\widetilde{M}+\widetilde{M}^{T}}{2},
\end{equation}
followed by the direct-sum padding in \Cref{eqn:appendix-direct-sum-padding}. Let
\begin{equation}
    H_{\mathrm{sym}}(M)
    =\sum_{a\in\mathcal{A}_{\mathrm{sym}}(M)}
      \lambda_a(M)\Pi_a^{\mathrm{sym}}(M)
\end{equation}
be its decomposition into distinct eigenvalues and their eigenspace projectors. Its upload unitary is
\begin{equation}
    S_{H_{\mathrm{sym}}}(t;M)
    =\sum_a e^{-i\lambda_a(M)t/2}\Pi_a^{\mathrm{sym}}(M).
\end{equation}
Substitution into the observable output yields
\begin{equation}
    \begin{aligned}
        f_1^{H_{\mathrm{sym}}}(M;t,\boldsymbol{\theta})
        &=\sum_{a,b}e^{+i\lambda_a(M)t/2}e^{-i\lambda_b(M)t/2}\\
        &\quad\times
        \bra{\psi_0}\Pi_a^{\mathrm{sym}}(M)O_{\boldsymbol{\theta}}
        \Pi_b^{\mathrm{sym}}(M)\ket{\psi_0},
    \end{aligned}
\end{equation}
which gives the pair amplitudes, aggregated coefficients and eigengap candidate set in \Cref{eqn:sym-ham-one-layer,eqn:sym-ham-coeff,eqn:sym-ham-freq}.

When the eigenspaces indexed by $a$ and $b$ are one-dimensional, one may choose normalised eigenvectors $\ket{v_a(M)}$ and $\ket{v_b(M)}$. Writing $d_a(M)=\braket{v_a(M)|\psi_0}$ gives
\begin{equation}
    A_{ab}^{H_{\mathrm{sym}}}(M,\boldsymbol{\theta})
    =d_a(M)^{\star}d_b(M)
      \bra{v_a(M)}O_{\boldsymbol{\theta}}\ket{v_b(M)}.
\end{equation}
This rank-one expression illustrates the eigenvector dependence but is basis-dependent when an eigenvalue is degenerate. The projector expression in \Cref{eqn:sym-ham-coeff} is the invariant statement used in the main text. Any natural or padding-induced zero eigenspaces are combined into a single projector associated with $\lambda=0$.

\subsection{Global block-Hamiltonian embedding}

For $M\in\mathbb{R}^{p\times q}$, define the natural block Hamiltonian
\begin{equation}\label{eqn:appendix-natural-block-ham}
    H_{\mathrm{block}}^{(0)}(M)
    =\begin{pmatrix}
        \boldsymbol{0}_{p\times p} & M\\
        M^{T} & \boldsymbol{0}_{q\times q}
    \end{pmatrix}
\end{equation}
on $\mathbb{R}^{p}\oplus\mathbb{R}^{q}$. Lloyd \textit{et al.} use the Hermitian Hamiltonian $\left(\begin{smallmatrix}0&A^{\dagger}\\A&0\end{smallmatrix}\right)$ in their quantum polar decomposition algorithm \cite{lloyd2020quantumpolardecompositionalgorithm}. For real $M$, setting $A=M^{T}$ gives \Cref{eqn:appendix-natural-block-ham}. Their construction uses access to this Hamiltonian to implement transformations associated with the polar decomposition; the present model instead uses its time evolution directly as a data-encoding unitary.

Let the compact singular value decomposition of $M$ be
\begin{equation}\label{eqn:appendix-block-svd}
    M=\sum_{j=1}^{r}\sigma_j(M)\ket{u_j(M)}\bra{v_j(M)},
    \qquad r=\operatorname{rank}(M),
\end{equation}
where $\sigma_j(M)>0$, $u_j(M)\in\mathbb{R}^{p}$ and $v_j(M)\in\mathbb{R}^{q}$. The singular-vector equations are
\begin{equation}
    M\ket{v_j}=\sigma_j\ket{u_j},
    \qquad
    M^{T}\ket{u_j}=\sigma_j\ket{v_j}.
\end{equation}
Introduce the direct-sum vectors
\begin{equation}
    \ket{u_j\oplus0}=\begin{pmatrix}u_j\\\boldsymbol{0}_q\end{pmatrix},
    \qquad
    \ket{0\oplus v_j}=\begin{pmatrix}\boldsymbol{0}_p\\v_j\end{pmatrix}.
\end{equation}
Equation~\eqref{eqn:appendix-natural-block-ham} then gives
\begin{equation}
    H_{\mathrm{block}}^{(0)}\ket{u_j\oplus0}
    =\sigma_j\ket{0\oplus v_j},
    \qquad
    H_{\mathrm{block}}^{(0)}\ket{0\oplus v_j}
    =\sigma_j\ket{u_j\oplus0}.
\end{equation}
Hence its restriction to
\begin{equation}
    \mathcal{S}_j=\operatorname{span}\{\ket{u_j\oplus0},\ket{0\oplus v_j}\}
\end{equation}
is $\sigma_jX$. The normalised eigenvectors on this subspace are
\begin{equation}\label{eqn:H-block-eigenvectors}
    \ket{j,+}=\frac{\ket{u_j\oplus0}+\ket{0\oplus v_j}}{\sqrt{2}},
    \qquad
    \ket{j,-}=\frac{\ket{u_j\oplus0}-\ket{0\oplus v_j}}{\sqrt{2}},
\end{equation}
and satisfy
\begin{equation}
    H_{\mathrm{block}}^{(0)}(M)\ket{j,+}=+\sigma_j(M)\ket{j,+},
    \qquad
    H_{\mathrm{block}}^{(0)}(M)\ket{j,-}=-\sigma_j(M)\ket{j,-}.
\end{equation}

The natural zero eigenspace is
\begin{equation}
    \ker H_{\mathrm{block}}^{(0)}
    =\ker(M^{T})\oplus\ker(M).
\end{equation}
Direct-sum Hilbert-space padding adds a further zero subspace of dimension $D-(p+q)$. Let $\Pi_0^{\mathrm{block}}(M)$ project onto their combined span. For a chosen orthonormal singular-vector basis,
\begin{equation}\label{eqn:appendix-block-spectral-decomposition}
    H_{\mathrm{block}}(M)
    =\sum_{j=1}^{r}\sigma_j(M)
      \left(\ket{j,+}\bra{j,+}-\ket{j,-}\bra{j,-}\right)
      +0\,\Pi_0^{\mathrm{block}}(M).
\end{equation}
If singular values are repeated, the rank-one projectors associated with each common eigenvalue are summed to obtain the invariant spectral projectors $\Pi_a^{\mathrm{block}}(M)$ used in the main text. Exponentiating \Cref{eqn:appendix-block-spectral-decomposition} gives
\begin{equation}
    S_{H_{\mathrm{block}}}(t;M)
    =\sum_{j=1}^{r}\left[
      e^{-i\sigma_j(M)t/2}\ket{j,+}\bra{j,+}
      +e^{+i\sigma_j(M)t/2}\ket{j,-}\bra{j,-}\right]
      +\Pi_0^{\mathrm{block}}(M).
\end{equation}

The generic expansion in \Cref{eqn:appendix-generic-output-expansion} now has eigenvalue labels $\mu_a(M)\in\{+\sigma_j(M),-\sigma_j(M),0\}$. Pairwise differences give
\begin{equation}\label{eqn:appendix-explicit-block-gaps}
    \Omega_1^{H_{\mathrm{block}}}(M)
    =\{0\}
    \cup\left\{\pm\frac{\sigma_j(M)-\sigma_k(M)}{2},
                 \ \pm\frac{\sigma_j(M)+\sigma_k(M)}{2}:j,k=1,\ldots,r\right\}
    \cup\Omega_0(M),
\end{equation}
where
\begin{equation}
    \Omega_0(M)=
    \begin{cases}
        \left\{\pm\sigma_j(M)/2:j=1,\ldots,r\right\},
        & \Pi_0^{\mathrm{block}}(M)\neq0,\\
        \emptyset, & \Pi_0^{\mathrm{block}}(M)=0.
    \end{cases}
\end{equation}
The terms with $j=k$ in the half-sum set include $\pm\sigma_j(M)$. Equations~\eqref{eqn:block-ham-coeff} and \eqref{eqn:block-ham-freq} give the invariant coefficient and candidate-support forms; the realised support can again be smaller.

\subsection{Patch-local block-Hamiltonian embedding}

Partition $M$ into $R$ non-overlapping patches $P_1,\ldots,P_R$ as defined in Appendix~\ref{sec:appendix-patch-and-mixer}. For patch $r$, write
\begin{equation}
    H_r\equiv H_{\mathrm{block}}(P_r)
    =\sum_{a_r\in\mathcal{A}_r}
      \mu_{r,a_r}(P_r)\Pi_{r,a_r}(P_r),
\end{equation}
where the distinct eigenvalues consist of the signed singular values of $P_r$ and zero when present. Because the local uploads act on disjoint registers, their joint one-upload unitary is
\begin{equation}\label{eqn:patch-block-joint-upload}
    \begin{aligned}
        S_{\mathrm{patch}\text{-}H_{\mathrm{block}}}(\boldsymbol{t};M)
        &=\bigotimes_{r=1}^{R}\exp\left[-\frac{i}{2}H_rt_r\right]\\
        &=\sum_{\boldsymbol{a}}
          \exp\left[-\frac{i}{2}\sum_{r=1}^{R}
          \mu_{r,a_r}(P_r)t_r\right]\Pi_{\boldsymbol{a}}(M),
    \end{aligned}
\end{equation}
where $\boldsymbol{t}=(t_1,\ldots,t_R)$, $\boldsymbol{a}=(a_1,\ldots,a_R)$ and
\begin{equation}
    \Pi_{\boldsymbol{a}}(M)
    =\bigotimes_{r=1}^{R}\Pi_{r,a_r}(P_r).
\end{equation}
The one-upload observable output is
\begin{equation}\label{eqn:patch-block-one-layer}
    \begin{aligned}
        f_1^{\mathrm{patch}\text{-}H_{\mathrm{block}}}
        (M;\boldsymbol{t},\boldsymbol{\theta})
        &=\sum_{\boldsymbol{a},\boldsymbol{b}}
          A_{\boldsymbol{a}\boldsymbol{b}}^{\mathrm{patch}\text{-}H_{\mathrm{block}}}
          (M,\boldsymbol{\theta})\\
        &\quad\times\exp\left[-\frac{i}{2}\sum_{r=1}^{R}
          \big(\mu_{r,b_r}(P_r)-\mu_{r,a_r}(P_r)\big)t_r\right],
    \end{aligned}
\end{equation}
with pair amplitudes
\begin{equation}\label{eqn:patch-block-coeff}
    A_{\boldsymbol{a}\boldsymbol{b}}^{\mathrm{patch}\text{-}H_{\mathrm{block}}}
    (M,\boldsymbol{\theta})
    =\bra{\psi_0}\Pi_{\boldsymbol{a}}(M)O_{\boldsymbol{\theta}}
      \Pi_{\boldsymbol{b}}(M)\ket{\psi_0}.
\end{equation}
The frequency vector associated with $(\boldsymbol{a},\boldsymbol{b})$ is
\begin{equation}
    \frac{1}{2}\left(
    \mu_{1,b_1}(P_1)-\mu_{1,a_1}(P_1),\ldots,
    \mu_{R,b_R}(P_R)-\mu_{R,a_R}(P_R)\right).
\end{equation}
Taking all index pairs gives the Cartesian-product candidate set in \Cref{eqn:patch-block-freq}. At layer $\ell$, $t_r$ is replaced by the implemented parameter $t_{\ell,r}$. For a depth-$L$ circuit, the maximal support in the $LR$ independent time variables is the Cartesian product of the $R$ local candidate sets over all $L$ layers; particular frequencies can still be absent through zero or cancelling amplitudes.

\section{Additional Information on the Benchmark Datasets}\label{sec:additional-dataset-info}

We evaluate the models on one real-world-data benchmark and two controlled synthetic benchmarks. The real-world data benchmark tests whether the proposed encoders are useful for a small, nontrivial handwritten-digit classification problem with matrix-valued inputs. The synthetic benchmarks are designed to isolate spectral structure: one task is controlled by eigenvalue gaps, and the other by singular values. Together, these datasets let us compare performance on realistic matrix-structured inputs and on tasks where the relevant decision is explicitly spectral.

\subsection{Pendigits benchmark}

The Pendigits dataset \cite{pen-based_recognition_of_handwritten_digits_81} is used as the benchmark for real-world classification tasks. We use two matrix-valued representations of each handwritten digit: the \texttt{DYN} representation, arranged as an $8 \times 2$ rectangular trajectory matrix, and the \texttt{STA4} representation, arranged as a $4 \times 4$ bitmap. These two representations provide complementary views of the same underlying object. \texttt{DYN} retains an ordered pen-trajectory structure, whereas \texttt{STA4} gives a compact square matrix representation. The two representations are useful for testing whether a QFM data-encoding unitary benefits from treating the input as a structured matrix rather than an unstructured feature vector.

For all Pendigits experiments, we use the original train/test split supplied with the dataset. A validation set is formed by holding out 10\% of the original training data. This validation is class-stratified so that each class remains represented in both the training and validation subsets. The final split used in the full $10$-class experiments contains $6744$ training examples, $750$ validation examples, and $3498$ test examples.

Before training, Pendigits features are standardised using statistics computed only from the post-split training set. The resulting training-set mean and standard deviation are then applied to the training, validation and test sets. After standardisation, each sample is reshaped into its corresponding matrix representation for the Hamiltonian-based and patch-local data-encoding unitaries. The visual examples shown in the main text (\Cref{fig:pendigit-data-sample}) display raw feature values for visual interpretation; the models are trained on the standardised inputs described above.

The Pendigits task is included to provide a compact, real-world benchmark whose input dimension is small enough for exact statevector simulation, especially for rotation-gate-based data-encoding unitaries, while still retaining nontrivial matrix structure. The \texttt{DYN} and \texttt{STA4} representations also test different notions of structure: trajectory geometry in the former and spatially arranged pixel-feature geometry in the latter. This makes Pendigits a useful test case for comparing coordinate-wise rotation-gate data-encoding unitaries, patch-local data-encoding unitaries and global QSMs under the same training protocol.

\subsection{Synthetic spectral benchmarks}

In addition to Pendigits, we use two controlled binary classification tasks whose labels are defined by spectral properties of latent clean matrices. These tasks are designed to test whether the model can exploit eigenvalue and singular-value structure rather than only component-value-level features. In both tasks, the classifier observes a noisy matrix, while the label is determined by the spectrum used to construct the corresponding clean matrix. The main experiments use $4096$ samples per task, a data generation seed of $0$, a validation fraction of $0.1$, and a relative noise level of $0.05$.

For the \texttt{SYNTHETIC EIGENGAP} task, each clean sample is generated as a random symmetric matrix with prescribed eigenvalues. For sample index $i$, a Gaussian random matrix is first drawn and factorised to obtain an orthogonal matrix $Q$. Independently, a vector of real eigenvalues is sampled from a standard normal distribution and sorted in ascending order,
\begin{equation}
    \lambda_{i,1} \leq \lambda_{i,2} \leq \cdots \leq \lambda_{i,d}.
\end{equation}
The clean matrix is then
\begin{equation}
    S_i = Q_i \mathbf{diag}(\lambda_{i,1}, \lambda_{i,2}, \cdots, \lambda_{i,d})Q^T.
\end{equation}
In the main experiments, $d=4$. The binary label is determined by the largest eigengap:
\begin{equation}
    y_i = \boldsymbol{1}\Big[ |\lambda_{i,d}-\lambda_{i,d-1}| > \tau_{\mathrm{eig}}   \Big],
\end{equation}
with $\tau_{\mathrm{eig}}=0.75$. 

The observed input matrix is obtained by adding scaled Gaussian noise to the clean matrix. Specifically, for each clean matrix $S_i$, an independent Gaussian noise matrix $ Z_i$ of the same shape is sampled and rescaled as
\begin{equation}
    E_i = \epsilon \frac{||S_i||_F}{\mathbf{max}(||Z_i||_F, 10^{-12})} Z_i,
\end{equation}
where $||\cdot||_F$ denotes for Frobenius norm. Then the input data becomes
\begin{equation}
    X_i = S_i + E_i.
\end{equation}
The main experiments use $\epsilon=0.05$, so the added noise has Frobenius norm equal to $5\%$ of the clean sample norm, up to the numerical denominator guard. The label is still computed from the latent clean eigenvalues, not from the noisy input data matrix.

This construction is aligned with the symmetric Hamiltonian embedding. In the clean limit, the input-derived symmetric Hamiltonian has eigenvalues $\lambda_j (S_i)$, and one upload has phase-difference carriers proportional to
\begin{equation}
    \frac{\lambda_k (S_i) - \lambda_j (S_i) }{2}.
\end{equation}
Thus the label depends directly on one of the spectral gaps that can appear in the sample-conditioned Hamiltonian frequency support. The task is therefore a controlled probe of whether the model can use eigengap information rather than only raw matrix entries.

For the \texttt{SYNTHETIC SINGULAR} value task, each clean sample is generated as a rectangular matrix with prescribed singular values. For sample index $i$, independent Gaussian matrices are factorised to obtain orthogonal factors $U_i$ and $V_i$. A vector of singular values is sampled uniformly on $[0,2]$, sorted in descending order,
\begin{equation}
    \sigma_{i,1} \geq \sigma_{i,2} \geq \cdots \geq \sigma_{i,r}, \quad r = \mathbf{min}(p,q),
\end{equation}
and placed on the diagonal of a rectangular $p\times q$ matrix $\Sigma_i$. The clean matrix is
\begin{equation}
    R_i = U_i \Sigma_i V_i^T.
\end{equation}
In the main experiments, $p=8, q=2$, and hence $r=2$. The label is
\begin{equation}
    y_i = \boldsymbol{1}\Big[\sigma_{i,1} + \sigma_{i,2} > \tau_{\mathrm{sv}}   \Big],
\end{equation}
where $\tau_{\mathrm{sv}} = 2$.

As the eigengap task, the observed \texttt{SYNTHETIC SINGULAR} value task input is a noisy version of the clean latent matrix. The label is determined by the latent singular values, not by the singular values of the noisy observed matrix. The main experiments again use $\epsilon=0.05$.

This task is aligned with the block Hamiltonian embedding. For a rectangular matrix $M$, the block Hamiltonian
\begin{equation}
    H_{\mathrm{block}} (M) = \begin{pmatrix}
        0 & M \\ M^T & 0
    \end{pmatrix}
\end{equation}
has nonzero eigenvalues $\pm \sigma_j (M)$, where $\sigma_j (M)$ are the singular values of $M$. The one-upload phase-difference support therefore contains singular-value differences and sums, including the terms of the form
\begin{equation}
    \pm\frac{\sigma_j (M)-\sigma_k (M)}{2}, \quad \pm\frac{\sigma_j (M)+\sigma_k (M)}{2}.
\end{equation}
The singular-value task uses a decision rule based on $\sigma_1+\sigma_2$, which is one of the spectral combinations naturally exposed by this block construction in the clean setting.

For both synthetic tasks, the train/validation/test split is generated by randomly permuting the 4096 samples with the fixed data seed. The first 20\% of the permuted samples are assigned to the test set. A validation set is then formed from the 10\% of the remaining samples, and the rest are used for training. This yields 819 test examples, 328 validation examples, and 2949 training examples per synthetic task.

The two synthetic benchmarks therefore play complementary diagnostic roles. The eigengap task tests whether a model can exploit eigenvalue-gap information in a matrix whose clean signal is symmetric. The singular-value task tests whether a model can exploit singular-value sums in a rectangular matrix. Since the latent spectral rule is known by construction, these tasks provide controlled evidence about spectral inductive bias, while the added matrix noise prevents the benchmarks from reducing to noiseless spectral lookup.

\section{Additional Information on the Patch-Local Encoders and Mixer Parameterisation}\label{sec:appendix-patch-and-mixer}

\subsection{Patch-local SU(4) and block-Hamiltonian data-encoding unitaries}

The patch-local data-encoding unitaries operate on four non-overlapping $2\times 2$ patches. For a $4\times 4$ matrix
\begin{equation}
    M = \begin{pmatrix}
        M_{1:2, 1:2} & M_{1:2, 3:4} \\
        M_{3:4, 1:2} & M_{3:4, 3:4}
    \end{pmatrix},
\end{equation}
we define
\begin{equation}
    P_1 = M_{1:2, 1:2}, \quad P_2 = M_{1:2, 3:4}, \quad
    P_3 = M_{3:4, 1:2}, \quad P_4 = M_{3:4, 3:4}.
\end{equation}
For an $8\times 2$ matrix, the patches are
\begin{equation}
    P_r = M_{2r-1:2r, 1:2}, \quad r = 1,2,3,4.
\end{equation}
Each patch is represented by a flattened vector
\begin{equation}
    \boldsymbol{p}_r = \operatorname{vec}(P_r) \in \mathbb{R}^4
\end{equation}
and is uploaded to a dedicated two-qubit system. Patch $r$ acts on qubits $2r-1$ and $2r$, so all four patch uploads together act on eight qubits. This applies to fixed patch-$SU(4)$, trainable patch-$SU(4)$, and non-overlap patch block-Hamiltonian embedding.

\subsection{SU(4) convention}

Both patch-$SU(4)$ data-encoding unitaries and the brick-wall mixer use the same two-qubit $SU(4)$ parametrisation. Let
\begin{equation}
    \mathcal{G} = (G_1, \cdots, G_{15})
\end{equation}
be the ordered list of all non-identity two-qubit Pauli strings,
\begin{equation}
    I\otimes X, I\otimes Y, I\otimes Z, X\otimes I, X\otimes X, X\otimes Y, X\otimes Z, Y\otimes I, Y\otimes X, Y\otimes Y, Y\otimes Z, Z\otimes I, Z\otimes X, Z\otimes Y, Z\otimes Z.
\end{equation}
For a coefficient vector $\theta \in \mathbb{R}^{15} $, the corresponding two-qubit unitary is
\begin{equation}
    U_{SU(4)}(\theta) = \exp \Bigg[ i \sum_{a=1}^{15} \theta_a G_a   \Bigg].
\end{equation}
The sign convention is therefore $+i$ for the $SU(4)$ parametrisation, different from the Hamiltonian embedding unitaries.

\subsection{Fixed patch-SU(4) encoder}

The fixed patch-SU(4) encoder maps each flattened patch $\boldsymbol{p}_r\in\mathbb{R}^4$ to an SU(4) coefficient vector by
\begin{equation}
    \theta(P_r) = A_0\boldsymbol{p}_r,
\end{equation}
where $A_0 \in \mathbb{R}^{15\times 4} $ is a fixed matrix. In the main experiments, the fixed map is
\begin{equation}
    A_0 = \frac{1}{2}\begin{pmatrix}
        1 & 0 & 0 & 0 \\
        0 & 1 & 0 & 0 \\
        0 & 0 & 1 & 0 \\
        0 & 0 & 0 & 1 \\
        1 & 1 & 0 & 0 \\
        0 & 0 & 1 & 1 \\
        1 & 0 & 1 & 0 \\
        0 & 1 & 0 & 1 \\
        1 & 0 & 0 & -1 \\
        0 & 1 & -1 & 0 \\
        1 & -1 & 0 & 0 \\
        0 & 0 & 1 & -1 \\
        1 & 1 & -1 & -1 \\
        1 & -1 & 1 & -1 \\
        1 & -1 & -1 & 1
    \end{pmatrix}
\end{equation}
The patch upload is then
\begin{equation}
    U_{\mathrm{patch}}(P_r)
    = \exp \Bigg( i \sum_{a=1}^{15} [A_0\boldsymbol{p}_r]_a G_a \Bigg).
\end{equation}
The same $A_0$ is used for every patch and every reuploading layer.

\subsection{Trainable patch-SU(4) encoder}

The trainable patch-SU(4) encoder uses the same four-patch decomposition and the same local SU(4) exponential, but the linear patch map is made trainable. For layer $\ell$ and patch $r$, let
\begin{equation}
    A_{\ell, r} \in \mathbb{R}^{15\times 4}.
\end{equation}
The patch angles are
\begin{equation}
    \theta_{\ell,r}(P_r)=A_{\ell,r}\boldsymbol{p}_r,
\end{equation}
and the upload is
\begin{equation}
    U_{\mathrm{trainable-patch},\ell,r}(P_r)
    =\exp\Bigg(i\sum_{a=1}^{15}[A_{\ell,r}\boldsymbol{p}_r]_aG_a\Bigg).
\end{equation}
The trainable parameter tensor has shape
\begin{equation}
    \big( A_{\ell, r} \big)^{L, 4}_{\ell = 1, r =1} \in \mathbb{R}^{L\times 4\times 15\times 4}.
\end{equation}
It is initialised near the fixed patch map:
\begin{equation}
    A_{\ell, r} = A_0 + \Delta_{\ell, r},
\end{equation}
where the entries of $\Delta_{\ell, r}$ are initialised as zero-mean Gaussian noise. The main experiments use a noise scale of $0.01$. With zero initialisation noise, the trainable patch encoder exactly reproduces the fixed patch-SU(4) angle map at initialisation.

\subsection{Non-overlapping patch-local block-Hamiltonian data-encoding unitary}

The non-overlapping patch-local block-Hamiltonian data-encoding unitary replaces the local $SU(4)$ patch map with structured Hamiltonian evolution. For each $2\times2$ patch $P_r$, define
\begin{equation}
    H_{\mathrm{patch}}(P_r) = \begin{pmatrix}
        0 & P_r \\ P_r^T &0
    \end{pmatrix} \in \mathbb{R}^{4\times 4}.
\end{equation}
At layer $\ell$, the patch upload is
\begin{equation}
    U_{\mathrm{patch-block},\ell,r}(P_r)
    =\exp\Bigg[-\frac{i}{2}H_{\mathrm{patch}}(P_r)t_{\ell,r}\Bigg].
\end{equation}
This unitary is applied to the two-qubit subsystem assigned to patch $r$. When trainable upload times are enabled, the parameter tensor is
\begin{equation}
    t \in \mathbb{R}^{L\times 4}.
\end{equation}
Thus the model has one scalar Hamiltonian upload time for each layer and each patch. If no fixed time is supplied, all entries are initialised to $1/L$. 

The patch-local block-Hamiltonian QSM differs from its global counterpart. The global QSM constructs one block Hamiltonian from the full matrix $M$, so its spectral structure is determined by the singular values and singular vectors of the whole input matrix. The patch-local QSM instead computes four independent $2\times 2$ block Hamiltonians, preserving local patch singular geometry while discarding cross-patch singular-vector coupling.

\subsection{Brick-wall SU(4) mixer}

After every application of a data-encoding unitary, the model applies a trainable nearest-neighbour brick-wall mixer. For $n$ qubits, define the ordered pair set
\begin{equation}
    \mathcal{B}_n = \big\{ (1,2), (3,4), \cdots \big\} \bigcup \big\{ (2,3), (4,5), \cdots \big\},
\end{equation}
where the even-starting pairs are applied before the odd-starting pairs. For each layer $\ell$ and each pair $b \in \mathcal{B}_n$, the model applies
\begin{equation}
    V_{\ell, b} = \exp \Bigg( i \sum_{a=1}^{15} \theta^{SU(4)}_{\ell, b, a} G_a    \Bigg).
\end{equation}
The full mixer layer is the ordered product of these local blocks:
\begin{equation}
    V_{\ell} = \prod_{b\in \mathcal{B}_n} V_{\ell, b}.
\end{equation}
For the $n$-qubit model used here, $|\mathcal{B}_n|=n-1$, so the mixer contributes $15(n-1)$ trainable parameters per uploading layer. The mixer parameter tensor has shape
\begin{equation}
    \theta^{SU(4)} \in \mathbb{R}^{L\times (n-1)\times 15}.
\end{equation}
It is initialised from a zero-mean Gaussian distribution with scale 0.01.

The number of qubits in the QFM circuit depends on the data-encoding unitary. Matrix-value-wise rotation-gate data-encoding unitaries use one qubit per scalar feature, so both $4\times 4$ and $8\times 2$ inputs use 16 qubits. The three patch-local data-encoding unitaries split the input into four $2\times 2$ patches and assign each patch to two qubits, giving 8 qubits in total. The global QSMs instead use the dimension of the input-derived Hamiltonian: $H_{\mathrm{sym}}$ uses 2 qubits for $4\times 4$ inputs and 3 qubits for $8\times 2$ inputs, while $H_{\mathrm{block}}$ uses 3 and 4 qubits, respectively, before any label-register expansion. For the ten-class Pendigits task, the projector readout requires label qubits, so the global QSMs are promoted to 4 qubits when their natural Hamiltonian dimension is smaller. This promotion is implemented as direct-sum padding $H \mapsto H \oplus \boldsymbol{0}$, equivalently $U\mapsto U \oplus \boldsymbol{I}$, rather than by tensoring extra idle qubits. Binary synthetic tasks require only one label qubit, so no additional expansion is needed for the global QSM data-encoding unitaries.

At $L=32$, the resulting total numbers of trainable scalars, including the mixer and any trainable encoder parameters, are 7200 for fixed-$R_y$, 7200 for fixed-$R_yR_z$, 7712 for trainable-frequency $R_y$, 3360 for fixed patch-$SU(4)$, 11040 for trainable patch-$SU(4)$ and 3488 for the patch-local block-Hamiltonian QSM. For the four-qubit Pendigits global QSMs, both $H_{\mathrm{sym}}$ and $H_{\mathrm{block}}$ contain 1472 trainable scalars. On \texttt{SYNTHETIC EIGENGAP}, the corresponding totals are 512 for the two-qubit symmetric-Hamiltonian QSM and 992 for the three-qubit block-Hamiltonian QSM; on \texttt{SYNTHETIC SINGULAR}, they are 992 for the three-qubit symmetric-Hamiltonian QSM and 1472 for the four-qubit block-Hamiltonian QSM. The projector readout adds no trainable parameters. These totals make the resource mismatch explicit: the main comparison holds the data and optimisation protocol fixed, but it does not normalise circuit width or trainable-parameter count.

\section{Additional Performance Results on the Pendigits and Synthetic Datasets}\label{sec:appendix-additional-main-results}

The main performance archive contains 3840 complete records with finite reported metrics. Of these records, 319 are marked as having been generated from a dirty implementation worktree. However, all 20 records contributing to each of the four depth-32 winner aggregates quoted in the main text, and all 20 records contributing to each of the four global $H_{\mathrm{block}}$ maxima across the tested depths, have \texttt{git\_dirty=False}. We therefore retain those headline aggregates while disclosing the broader archive provenance.

\begin{figure}[ht]
    \centering
    \begin{subfigure}[t]{1\textwidth}
        \includegraphics[width=1\linewidth]{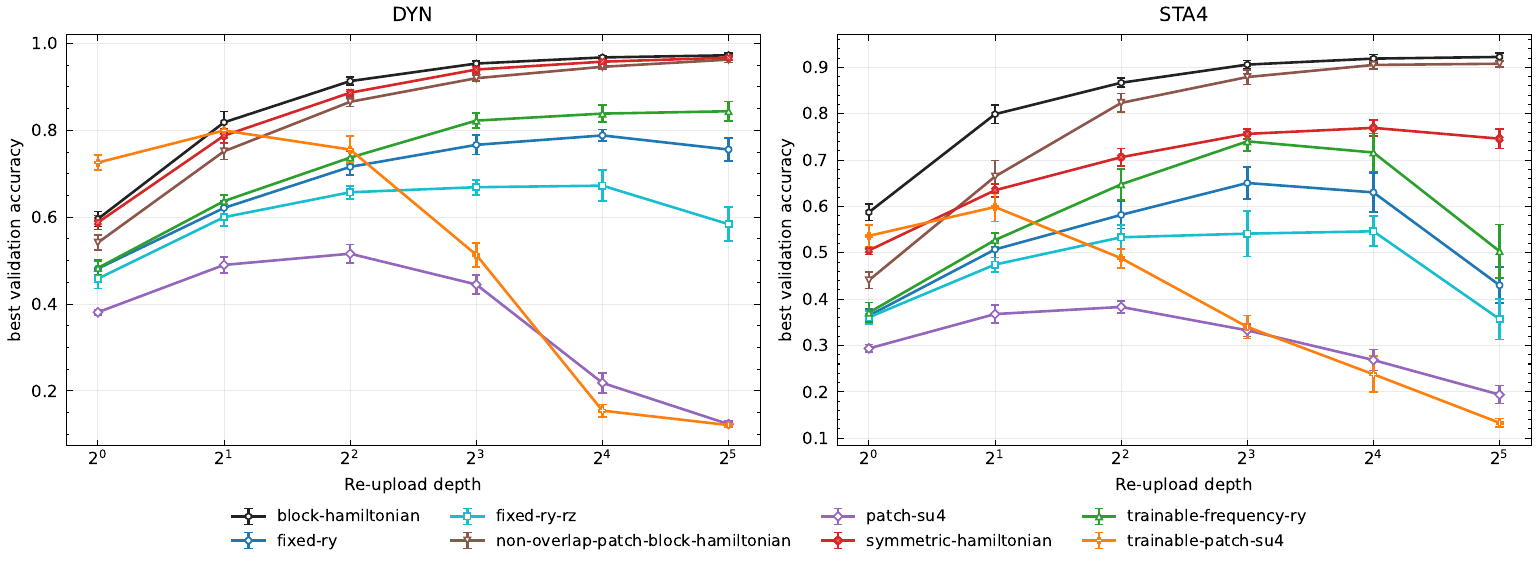}
        \caption{ }\label{fig:appendix-pendigits-best-val}
    \end{subfigure}
    \vfill
    \begin{subfigure}[t]{1\textwidth}
        \includegraphics[width=1\linewidth]{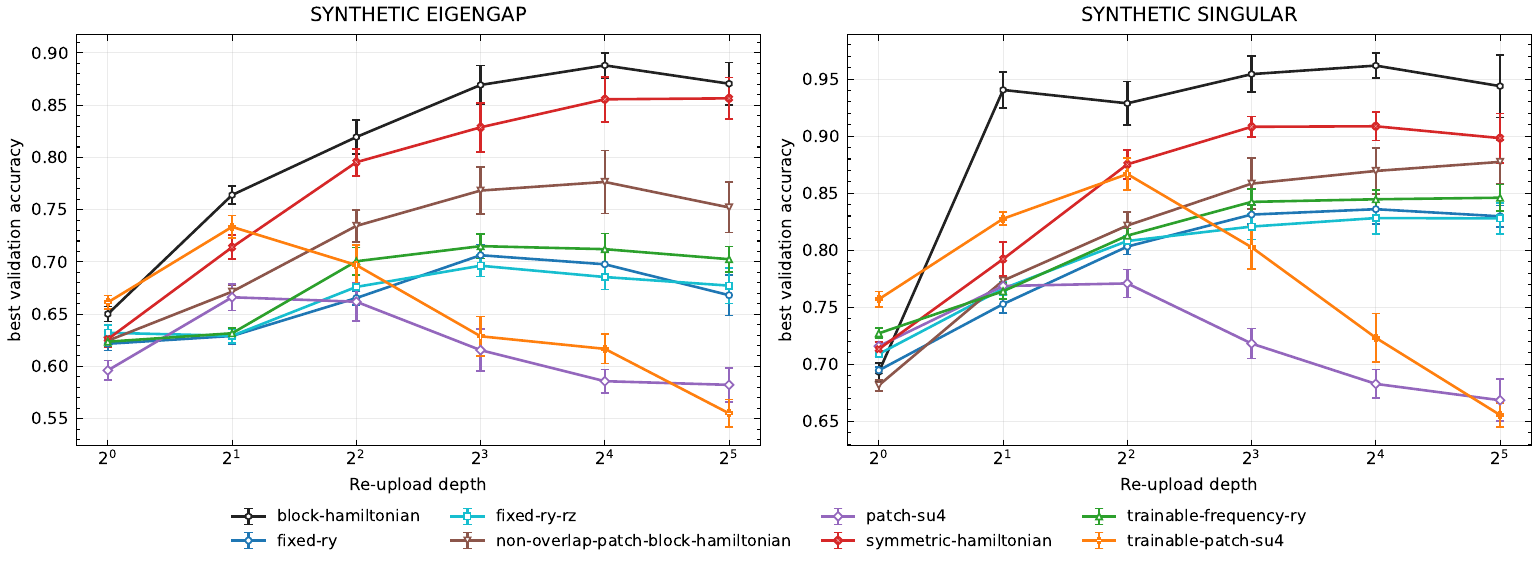}
        \caption{ }\label{fig:appendix-synthetic-best-val}
    \end{subfigure}
    \caption{
    \textbf{(a)} Best validation accuracy on the Pendigits benchmark as a function of reuploading depth. Each point is the mean over 20 seeds, and the error bars show one standard deviation. The metric is the maximum validation accuracy observed over the 2000-step training trajectory for each run. The QSMs achieve the highest best-validation accuracies on both \texttt{DYN} and \texttt{STA4}.
    \textbf{(b)} Best validation accuracy on the two synthetic spectral benchmarks as a function of reuploading depth. Each point is the mean over 20 seeds, and the error bars show one standard deviation. The global QSMs achieve the strongest validation performance on both \texttt{SYNTHETIC EIGENGAP} and \texttt{SYNTHETIC SINGULAR}, with the block-Hamiltonian QSM giving the best overall validation accuracy.
    }
    \label{fig:appendix-best-val}
\end{figure}

\begin{figure}[ht]
    \centering
    \begin{subfigure}[t]{1\textwidth}
        \includegraphics[width=1\linewidth]{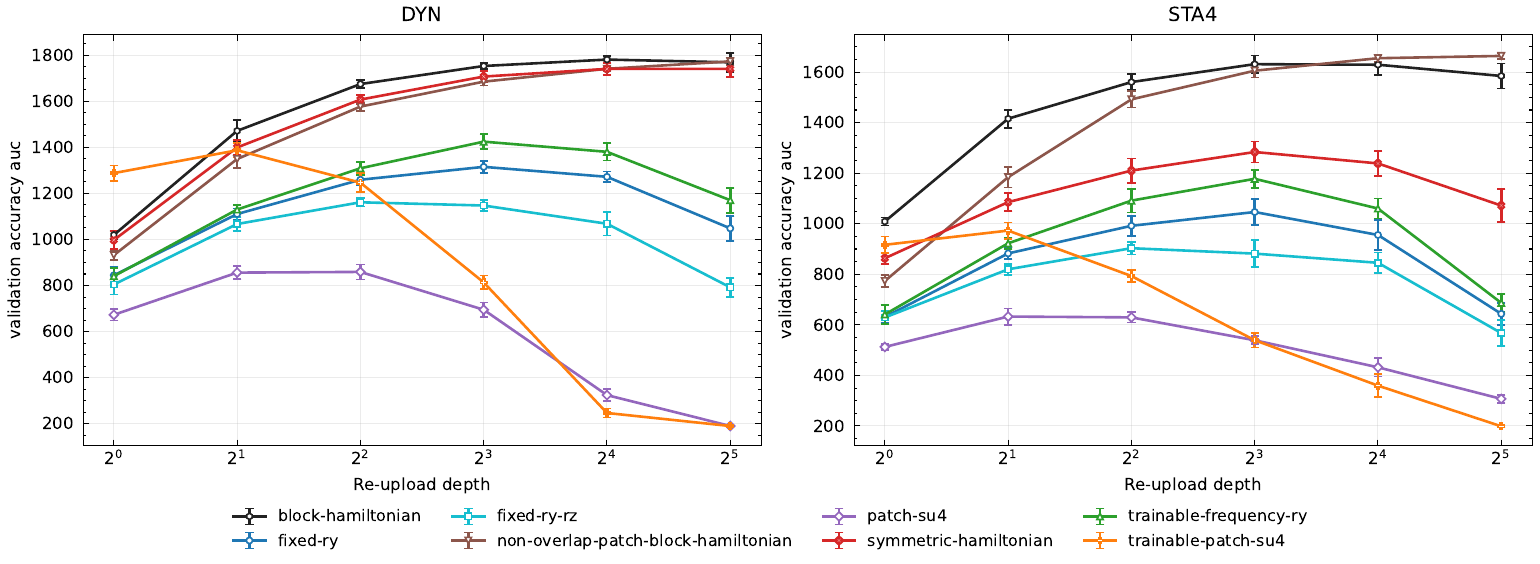}
        \caption{ }\label{fig:appendix-pendigits-val-auc}
    \end{subfigure}
    \vfill
    \begin{subfigure}[t]{1\textwidth}
        \includegraphics[width=1\linewidth]{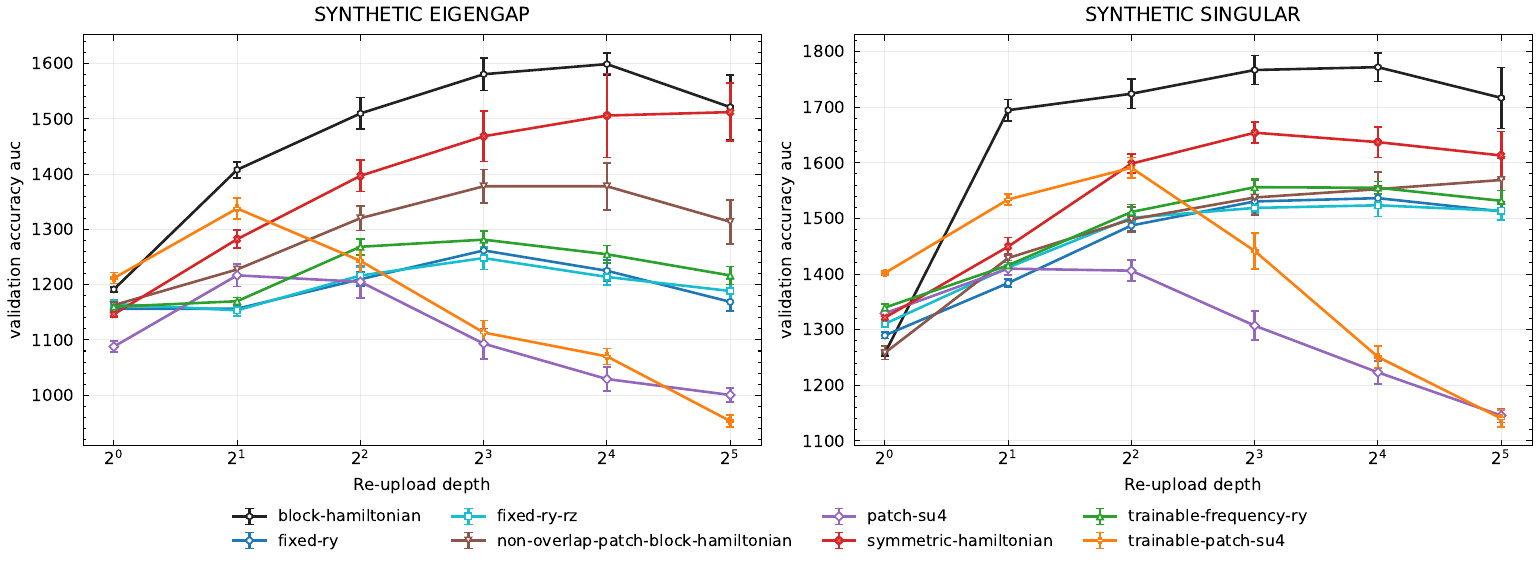}
        \caption{ }\label{fig:appendix-synthetic-val-auc}
    \end{subfigure}
    \caption{
    \textbf{(a)} Area under the validation-accuracy training curve on Pendigits. For each run, AUC is computed by trapezoidal integration of validation accuracy over recorded evaluation steps. The plotted values are means over 20 seeds with one-standard-deviation error bars. The patch-local and global block-Hamiltonian QSMs achieve the highest AUCs, indicating stronger sustained validation performance throughout training.
    \textbf{(b)} Area under the validation-accuracy training curve on \texttt{SYNTHETIC EIGENGAP} and \texttt{SYNTHETIC SINGULAR}. AUC is computed by trapezoidal integration over validation checkpoints and is reported in step units. Each point is the mean over 20 seeds, with error bars showing one standard deviation. The global block-Hamiltonian QSM gives the largest AUC on both synthetic tasks.
    }
    \label{fig:appendix-main-results-val-auc}
\end{figure}

\begin{table*}[htbp!]
  \caption{\textbf{Additional validation-set performance summaries for the Pendigits and synthetic
  benchmark runs.} Best validation accuracy is the largest validation accuracy reached during
  training for each run. Validation AUC is the unnormalised trapezoidal area under the
  validation-accuracy curve over recorded evaluation steps. Values are mean $\pm$ standard
  deviation over 20 random seeds.}
  \label{tab:additional-validation-performance}
  \centering
  \small
  \setlength{\tabcolsep}{3pt}
  \begin{tabular}{llccccc}
    \toprule
    Dataset & Encoder & Best-val depth & Best val. acc. (\%) &
    Best-AUC depth & Val. AUC & Val. AUC / 1900 \\
    \midrule
    Pendigits DYN & fixed-$R_y$ & 16 & $78.84\pm1.34$ & 8 & $1315.0\pm25.3$ & $0.692$ \\
     & fixed-$R_yR_z$ & 16 & $67.27\pm3.54$ & 4 & $1161.2\pm18.2$ & $0.611$ \\
     & TF-$R_y$ & 32 & $84.41\pm2.22$ & 8 & $1424.4\pm31.8$ & $0.750$ \\
     & patch-$SU(4)$ & 4 & $51.61\pm2.16$ & 4 & $859.0\pm32.2$ & $0.452$ \\
     & trainable patch-$SU(4)$ & 2 & $79.97\pm1.85$ & 2 & $1387.0\pm31.0$ & $0.730$ \\
     & patch $H_{\rm block}$ & 32 & $96.30\pm0.66$ & 32 & $1772.6\pm16.5$ & $0.933$ \\
     & $H_{\rm sym}$ & 32 & $96.74\pm0.57$ & 16 & $1740.2\pm26.4$ & $0.916$ \\
     & $H_{\rm block}$ & 32 & $97.29\pm0.53$ & 16 & $1780.7\pm15.8$ & $0.937$ \\
    \midrule
    Pendigits STA4 & fixed-$R_y$ & 8 & $65.05\pm3.43$ & 8 & $1046.4\pm50.9$ & $0.551$ \\
     & fixed-$R_yR_z$ & 16 & $54.63\pm3.23$ & 4 & $903.2\pm25.1$ & $0.475$ \\
     & TF-$R_y$ & 8 & $74.04\pm2.03$ & 8 & $1177.8\pm35.2$ & $0.620$ \\
     & patch-$SU(4)$ & 4 & $38.31\pm1.34$ & 2 & $632.7\pm32.4$ & $0.333$ \\
     & trainable patch-$SU(4)$ & 2 & $59.87\pm3.16$ & 2 & $972.9\pm32.6$ & $0.512$ \\
     & patch $H_{\rm block}$ & 32 & $90.78\pm0.64$ & 32 & $1664.0\pm11.7$ & $0.876$ \\
     & $H_{\rm sym}$ & 16 & $76.95\pm1.71$ & 8 & $1283.8\pm41.7$ & $0.676$ \\
     & $H_{\rm block}$ & 32 & $92.25\pm0.86$ & 8 & $1631.4\pm34.9$ & $0.859$ \\
    \midrule
    Synthetic eigengap & fixed-$R_y$ & 8 & $70.64\pm1.03$ & 8 & $1262.0\pm15.5$ & $0.664$ \\
     & fixed-$R_yR_z$ & 8 & $69.63\pm1.02$ & 8 & $1248.3\pm20.5$ & $0.657$ \\
     & TF-$R_y$ & 8 & $71.51\pm1.18$ & 8 & $1281.6\pm15.8$ & $0.675$ \\
     & patch-$SU(4)$ & 2 & $66.62\pm1.26$ & 2 & $1216.9\pm20.3$ & $0.640$ \\
     & trainable patch-$SU(4)$ & 2 & $73.35\pm1.09$ & 2 & $1338.0\pm19.5$ & $0.704$ \\
     & patch $H_{\rm block}$ & 16 & $77.65\pm3.02$ & 16 & $1378.2\pm42.6$ & $0.725$ \\
     & $H_{\rm sym}$ & 32 & $85.66\pm2.01$ & 32 & $1512.4\pm52.5$ & $0.796$ \\
     & $H_{\rm block}$ & 16 & $88.81\pm1.19$ & 16 & $1599.3\pm20.2$ & $0.842$ \\
    \midrule
    Synthetic singular & fixed-$R_y$ & 16 & $83.60\pm1.28$ & 16 & $1536.5\pm18.6$ & $0.809$ \\
     & fixed-$R_yR_z$ & 16 & $82.82\pm1.38$ & 16 & $1523.6\pm19.9$ & $0.802$ \\
     & TF-$R_y$ & 32 & $84.60\pm1.15$ & 8 & $1556.2\pm14.8$ & $0.819$ \\
     & patch-$SU(4)$ & 4 & $77.09\pm1.23$ & 2 & $1409.7\pm11.0$ & $0.742$ \\
     & trainable patch-$SU(4)$ & 4 & $86.68\pm1.38$ & 4 & $1591.1\pm18.4$ & $0.837$ \\
     & patch $H_{\rm block}$ & 32 & $87.74\pm1.90$ & 32 & $1569.2\pm38.4$ & $0.826$ \\
     & $H_{\rm sym}$ & 16 & $90.87\pm1.24$ & 8 & $1654.2\pm19.0$ & $0.871$ \\
     & $H_{\rm block}$ & 16 & $96.19\pm1.11$ & 16 & $1772.0\pm25.5$ & $0.933$ \\
    \bottomrule
  \end{tabular}
\end{table*}

In this Appendix section, we report two additional validation-set summaries for the same training runs shown in the main text (\Cref{fig:main-res}). The first is the best validation accuracy reached during training, see \Cref{fig:appendix-best-val}. This metric records the largest validation accuracy observed at any evaluation checkpoint for a given run. It is useful for separating the model's best achieved validation performance from the final checkpoint used for the main held-out test evaluation.

The second metric is the validation-accuracy area under the training curve (AUC), as shown in \Cref{fig:appendix-main-results-val-auc}. Let the validation checkpoints for one run be
\begin{equation}
    (t_1, a_1), (t_2, a_2),\cdots ,(t_m, a_m),
\end{equation}
where $t_i$ is the optimisation step and $a_i$ is the validation accuracy recorded at that step. The reported AUC is calculated by the trapezoidal rule,
\begin{equation}
    \mathrm{AUC}_{\mathrm{val}} = \sum_{i=1}^{m-1} \frac{a_i + a_{i+1}}{2} (t_{i+1}-t_i).
\end{equation}
This is the raw area in units of optimisation steps, not a normalised curve. In the main runs, validation is evaluated every 100 steps over 2000 training steps, so the maximum raw AUC is 1900 if validation accuracy is 1 at every recorded interval. A summary of these results can also be found in \Cref{tab:additional-validation-performance}.

The validation curves support the same qualitative conclusion as the final test results. On Pendigits \texttt{DYN}, the three QSMs achieve the highest best validation accuracies, with the global block-Hamiltonian QSM reaching $97.29\% \pm 0.53\%$, the symmetric-Hamiltonian QSM reaching $96.74\%\pm 0.57\%$, and the patch-local block-Hamiltonian QSM reaching $96.30\%\pm 0.66\%$. On Pendigits \texttt{STA4}, the global and patch-local block-Hamiltonian QSMs again lead, reaching $92.25\% \pm 0.86\%$ and $90.78\% \pm 0.64\% $, respectively.

The AUC plots provide a complementary view of training efficiency and sustained validation performance. A model can obtain a high best validation accuracy late in training but a smaller AUC if it learns slowly or is unstable. On \texttt{DYN}, the global block-Hamiltonian QSM achieves the highest mean validation AUC, closely followed by the patch-local block-Hamiltonian QSM. On \texttt{STA4}, the patch-local block-Hamiltonian QSM achieves the highest mean validation AUC, indicating more sustained validation performance throughout training. On \texttt{SYNTHETIC EIGENGAP} and \texttt{SYNTHETIC SINGULAR}, the global block-Hamiltonian QSM gives the largest validation AUC.

\section{Additional Information on the Gradient Analysis}\label{sec:appendix-grad-analysis}

This section defines the gradient statistics used in the main text and reports the supplementary diagnostics.

\subsection{Details for the Gradient Diagnostics Metrics}

Let $B=\big\{ (x_i, y_i) \big\}^m_{i=1}$ be the diagnostic training batch, with $m=32$ in the reported experiments. The batch is selected deterministically using diagnostic seed 0 and held fixed within each run configuration. Let $\mathcal{L}(\theta; B)$ be the mean cross-entropy loss on this batch. For a parameter group $q$, let
\begin{equation}
    g_q (\theta; B) = \nabla_{\theta_q} \mathcal{L}(\theta; B) \in \mathbb{R}^{P_q},
\end{equation}
where $P_q$ is the number of parameters in group $q$. The parameter groups used in the plots are the shared mixer layer parameters $\theta_{SU(4)}$, trainable frequency parameters $\gamma$, Hamiltonian-embedding upload times $t_{\ell}$ or $t_{\ell, r}$, and trainable patch-map parameters.

For each run seed $s$, the initialisation diagnostic evaluates $R=20$ parameter draws, generated with seeds $s+r-1$ for $r=1,\ldots,R$, on the same diagnostic batch. Let
\begin{equation}
    g_{q,r} = g_q (\theta^{(r)};B), \quad r = 1, \cdots, R.
\end{equation}
For each parameter coordinate $p$, the initialisation gradient variance is
\begin{equation}
    v_{q,p} = \frac{1}{R} \sum_{r=1}^R ( g_{q,r,p} - \bar{g}_{q,p}   )^2, \qquad \bar{g}_{q,p} = \frac{1}{R} \sum_{r=1}^R g_{q,r,p}.
\end{equation}
The plotted initialisation-variance statistic is
\begin{equation}
    \mathbf{median}_{p=1,\cdots, P_q} \log_{10} (v_{q,p} + \epsilon),
\end{equation}
with $\epsilon = 10^{-30}$. The RMS-gradient statistic for a single gradient vector is
\begin{equation}
    \mathbf{RMS}(g_q) = \sqrt{\frac{1}{P_q} \sum_{p=1}^{P_q} |g_{q,p}  |^2  }.
\end{equation}
For initialisation, each run first averages the RMS over its $R$ draws. For the final mode, each trained run contributes one final-checkpoint gradient vector. The reported means and error bars then aggregate these run-level statistics over 20 run seeds; the error bars are sample standard deviations across those seeds.

The aggregate initialisation-diagnostic records do not store a \texttt{git\_dirty} field. Their clean-state provenance therefore cannot be checked from the archived aggregate files, although the records used in the reported summaries are present and complete. The final-checkpoint diagnostic records do store this field, allowing the affected comparisons to be qualified below.

For the empirical-Fisher diagnostic, the implementation computes one observed-label loss gradient per diagnostic-batch element:
\begin{equation}
    g_i = \nabla_{\theta} \ell (\theta; x_i, y_i), \quad i = 1,\cdots, m,
\end{equation}
where $\ell$ is the per-sample cross-entropy loss and $\theta$ denotes the full trainable parameter vector. Let
\begin{equation}
    G = \begin{pmatrix}
        g_1^T \\ \vdots \\ g_m^T
    \end{pmatrix} \in \mathbb{R}^{m\times P}.
\end{equation}
The sample-space gradient Gram matrix is
\begin{equation}
    \hat{F} = \frac{1}{m} G G^T.
\end{equation}
This is the common machine-learning empirical-Fisher construction based on ground-truth labels \cite{Kunstner2019-ef}. It has the same nonzero eigenvalues as the parameter-space matrix $G^T G/m$ but is smaller because $m=32$. It is not the true Fisher information, which averages over labels drawn from the model distribution, or the quantum Fisher information used in quantum natural-gradient methods. Nor should it be assumed to represent general loss curvature \cite{Kunstner2019-ef}. We use it only as a finite-batch diagnostic of the sampled gradient directions. At initialisation, each run contributes the first of its 20 parameter draws to this calculation; at the final checkpoint, each run contributes its trained parameters. Its trace is
\begin{equation}
    \mathbf{tr}(\hat{F}) = \sum_j \lambda_j,
\end{equation}
and its participation-ratio effective rank \cite{Gao2017-pr} is
\begin{equation}
    r_{\mathrm{eff}} = \frac{\big(\sum_j \lambda_j \big)^2}{\mathbf{max} \big( \sum_j \lambda_j^2, \delta \big)},
\end{equation}
and the condition number is
\begin{equation}
    \kappa = \frac{\mathbf{max}_{\lambda_j > \delta} \lambda_j}{\mathbf{max} \big(\mathbf{min}_{\lambda_j > \delta}\lambda_j, \delta  \big) },
\end{equation}
with damping threshold $\delta = 10^{-8}$.

\subsection{Additional Gradient Diagnostics Plots and Results}

\begin{figure}[htbp]
    \centering
    \includegraphics[width=1\linewidth]{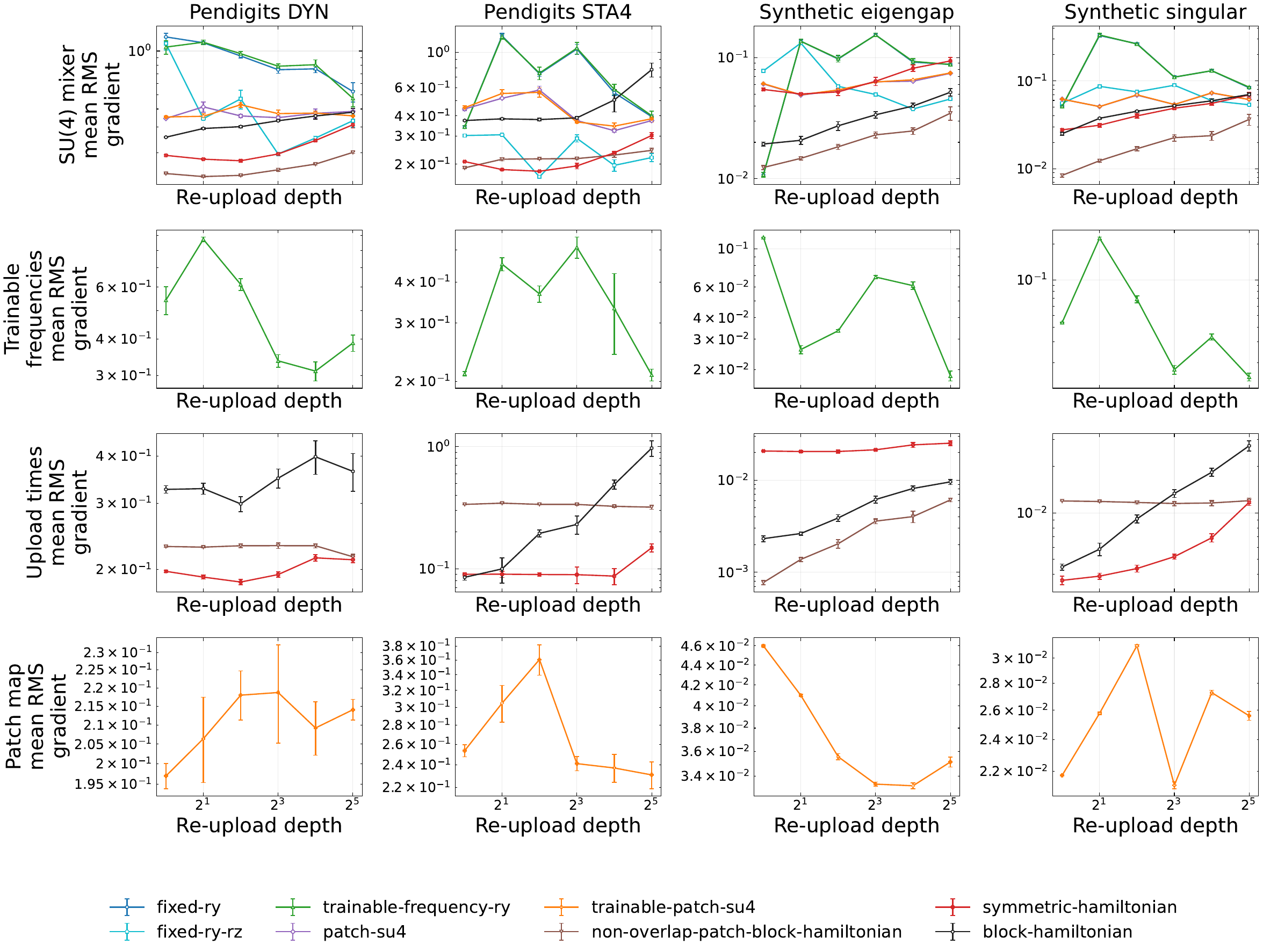}
    \caption{
    Initialisation RMS gradient as a function of reuploading depth. For each run configuration, gradients are computed on a fixed diagnostic training batch for 20 random initialisations. The plotted value is the RMS gradient within each parameter group, averaged over initialisation draws and then over 20 run seeds; error bars show one standard deviation across seeds. This plot complements the initialisation-variance diagnostic (\Cref{fig:gradient-init-var}) by measuring absolute gradient scale rather than parameter-wise variance across initialisations.
    }
    \label{fig:appendix-grad-init-rms}
\end{figure}

\Cref{fig:appendix-grad-init-rms} reports the absolute RMS gradient scale at initialisation. It should be read together with the initialisation-variance plot in \Cref{fig:gradient-init-var}. At depth 32, for example, fixed-$R_y$ on Pendigits \texttt{DYN} has mixer-gradient RMS $0.527\pm0.081$ but median $\log_{10}$ mixer-gradient variance $-11.04\pm0.08$. Nonzero gradient magnitude can therefore coexist with little coordinate-wise variation across initialisations; the two diagnostics measure different properties.

\begin{figure}[htbp]
    \centering
    \includegraphics[width=1\linewidth]{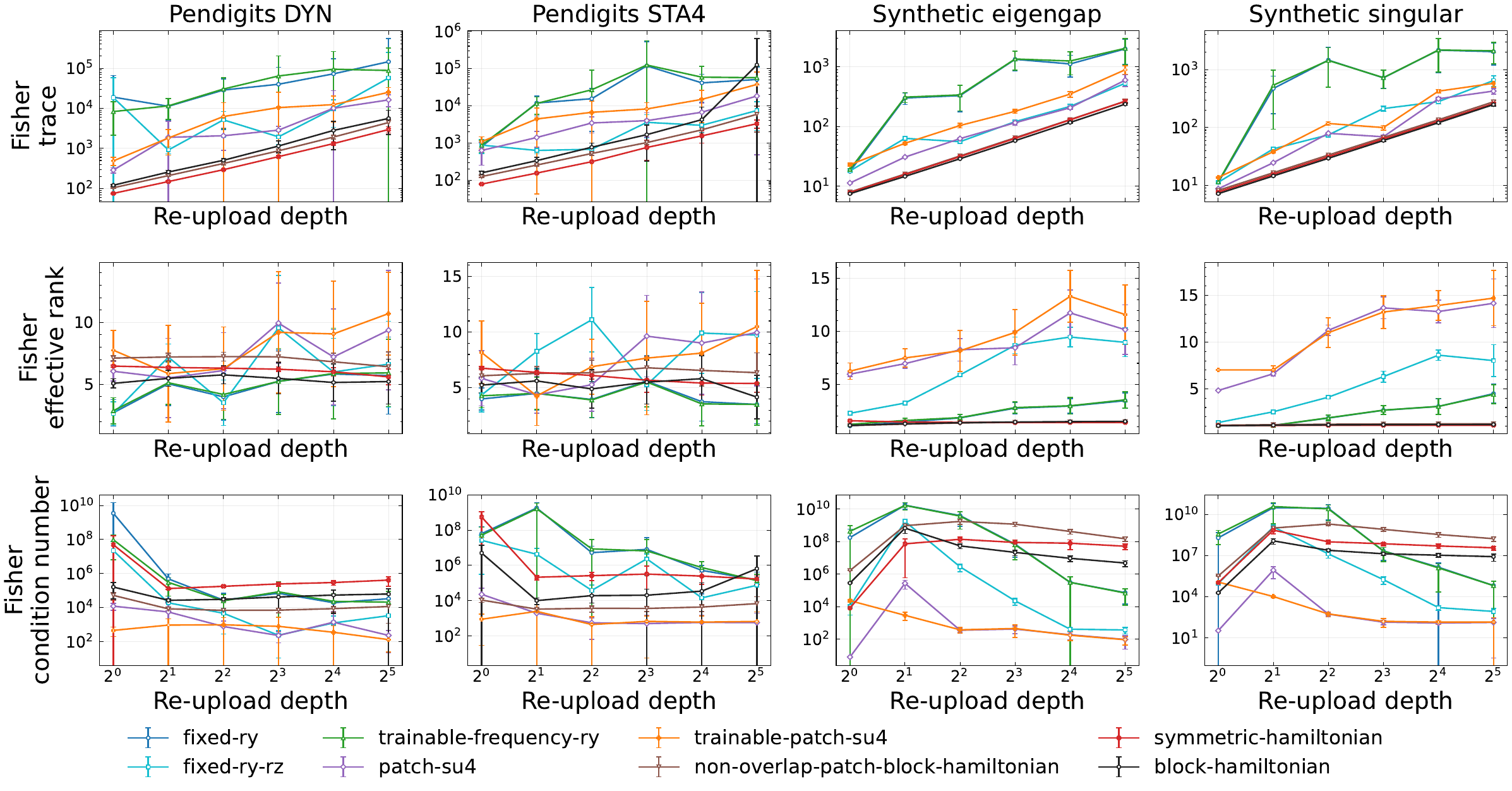}
    \caption{
    Empirical-Fisher diagnostics at initialisation. For each diagnostic batch, per-sample gradients are collected into a gradient matrix, and the empirical Fisher is summarised by its trace, effective rank, and condition number. Each point is the mean over 20 run seeds, and error bars show one standard deviation. The plot characterises the initial tangent-space geometry introduced by each encoder but should not be interpreted as a direct predictor of final test accuracy.
    }
    \label{fig:appendix-grad-init-fisher}
\end{figure}

\Cref{fig:appendix-grad-init-fisher} reports the trace, effective rank and condition number of the empirical-Fisher gradient Gram matrix at initialisation. These quantities describe total squared per-example gradient scale, the participation of sampled gradient directions and their anisotropy, respectively. At depth 32, the initial effective ranks of $H_{\mathrm{sym}}$ and $H_{\mathrm{block}}$ are $1.42\pm0.04$ and $1.52\pm0.02$ on \texttt{SYNTHETIC EIGENGAP}, and $1.10\pm0.02$ and $1.18\pm0.03$ on \texttt{SYNTHETIC SINGULAR}. These low ranks coexist with relatively high test accuracies, so initial empirical-Fisher rank does not track the performance ordering.

\begin{figure}[htbp]
    \centering
    \includegraphics[width=1\linewidth]{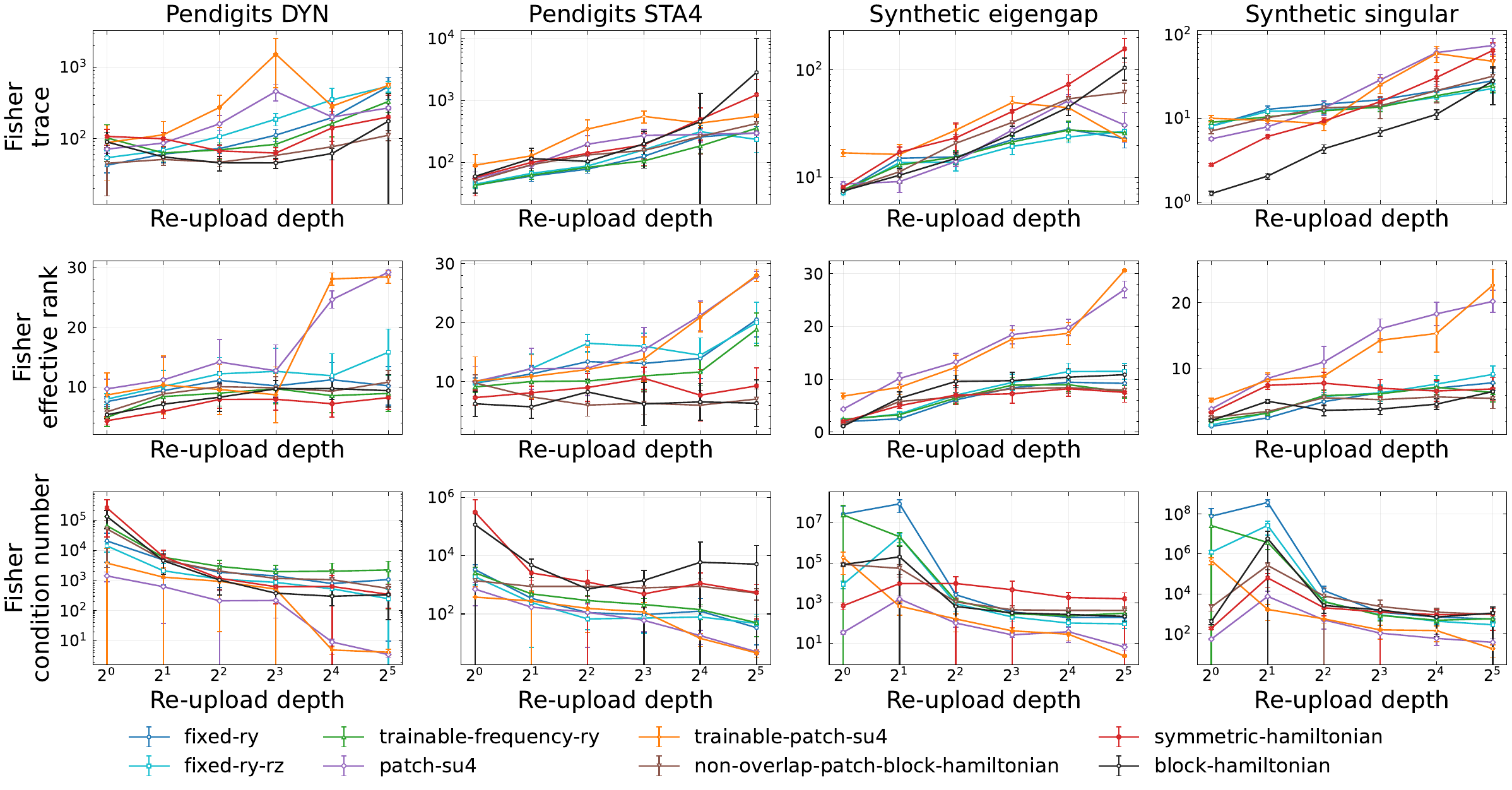}
    \caption{
    Empirical-Fisher diagnostics at the final trained checkpoint. The trace, effective rank, and condition number are computed from per-sample gradients on the fixed diagnostic batch. Each point is the mean over 20 run seeds, and error bars show one standard deviation. High final Fisher effective rank does not by itself imply high predictive performance; several patch-SU(4) models have high final Fisher rank but low depth-32 test accuracy.
    }
    \label{fig:appendix-grad-final-fisher}
\end{figure}

\Cref{fig:appendix-grad-final-fisher} reports the same summaries at the final checkpoint. On \texttt{SYNTHETIC EIGENGAP} at depth 32, trainable patch-$SU(4)$ has effective rank $30.65\pm0.23$ and test accuracy $50.16\%\pm2.36\%$; $H_{\mathrm{block}}$ has effective rank $10.86\pm1.72$ and accuracy $85.15\%\pm2.40\%$. The effective rank therefore does not reproduce the accuracy ordering. This example has a provenance limitation: all 20 trainable patch-$SU(4)$ final-gradient records are marked \texttt{git\_dirty=True}, whereas all 20 $H_{\mathrm{block}}$ records are marked \texttt{git\_dirty=False}. We therefore treat it as a descriptive archived comparison rather than a clean-state reproduction. By contrast, all 20 records in each group used for the depth-32 Pendigits \texttt{DYN} patch-$SU(4)$ versus $H_{\mathrm{block}}$ comparison in the main text have \texttt{git\_dirty=False}.

\begin{table*}[htbp!]
  \caption{\textbf{Summary of the depth-32 performance and gradient analysis.}. Values are mean
  $\pm$ standard deviation over 20 run seeds. Test accuracy is reported in percentages.
  Init logVar is the median $\log_{10}$ variance of the shared $SU(4)$ mixer
  gradients across initialisation draws. Init RMS and final RMS are RMS gradients
  of the shared mixer group. Fisher ranks are empirical-Fisher effective ranks
  computed from per-sample gradients on the diagnostic batch.}
  \label{tab:gradient-depth32-summary}
  \centering
  \scriptsize
  \setlength{\tabcolsep}{3pt}
  \begin{tabular}{llcccccc}
    \toprule
    Dataset & Encoder & Test acc. (\%) & Init logVar & Init RMS &
    Final RMS & Init Fisher rank & Final Fisher rank \\
    \midrule
    Pendigits DYN & fixed-$R_y$ & $72.29\pm2.77$ & $-11.04\pm0.08$ & $0.527\pm0.081$ & $0.054\pm0.010$ & $5.71\pm3.11$ & $10.20\pm3.23$ \\
     & fixed-$R_yR_z$ & $55.84\pm5.30$ & $-11.00\pm0.04$ & $0.333\pm0.033$ & $0.050\pm0.004$ & $6.63\pm3.46$ & $15.83\pm3.88$ \\
     & TF-$R_y$ & $81.37\pm2.36$ & $-11.03\pm0.09$ & $0.473\pm0.060$ & $0.044\pm0.006$ & $5.92\pm2.72$ & $8.92\pm2.98$ \\
     & patch-$SU(4)$ & $9.96\pm0.44$ & $-1.56\pm0.02$ & $0.385\pm0.006$ & $0.050\pm0.002$ & $9.37\pm4.82$ & $29.25\pm0.52$ \\
     & trainable patch-$SU(4)$ & $9.84\pm0.47$ & $-1.50\pm0.02$ & $0.359\pm0.004$ & $0.050\pm0.002$ & $10.71\pm3.34$ & $28.46\pm1.08$ \\
     & patch $H_{\rm block}$ & $92.75\pm0.84$ & $-2.38\pm0.04$ & $0.201\pm0.003$ & $0.048\pm0.005$ & $6.42\pm1.21$ & $10.82\pm2.00$ \\
     & $H_{\rm sym}$ & $91.89\pm3.71$ & $-1.42\pm0.07$ & $0.312\pm0.016$ & $0.079\pm0.034$ & $5.60\pm0.63$ & $8.23\pm2.00$ \\
     & $H_{\rm block}$ & $92.35\pm5.95$ & $-1.05\pm0.05$ & $0.380\pm0.022$ & $0.071\pm0.034$ & $5.22\pm1.81$ & $9.39\pm2.72$ \\
    \midrule
    Pendigits STA4 & fixed-$R_y$ & $40.22\pm4.40$ & $-11.33\pm0.03$ & $0.396\pm0.012$ & $0.040\pm0.006$ & $3.52\pm1.72$ & $20.51\pm2.92$ \\
     & fixed-$R_yR_z$ & $32.79\pm4.28$ & $-11.19\pm0.07$ & $0.220\pm0.013$ & $0.033\pm0.007$ & $9.73\pm3.90$ & $19.98\pm3.49$ \\
     & TF-$R_y$ & $46.85\pm6.39$ & $-11.31\pm0.03$ & $0.401\pm0.028$ & $0.045\pm0.007$ & $3.52\pm1.88$ & $18.86\pm2.81$ \\
     & patch-$SU(4)$ & $16.41\pm1.88$ & $-1.55\pm0.02$ & $0.374\pm0.009$ & $0.054\pm0.002$ & $9.96\pm4.81$ & $27.81\pm0.85$ \\
     & trainable patch-$SU(4)$ & $10.27\pm0.62$ & $-1.55\pm0.03$ & $0.384\pm0.017$ & $0.050\pm0.002$ & $10.50\pm5.02$ & $28.02\pm1.06$ \\
     & patch $H_{\rm block}$ & $86.14\pm1.33$ & $-2.09\pm0.08$ & $0.244\pm0.008$ & $0.079\pm0.013$ & $6.36\pm1.78$ & $7.04\pm1.71$ \\
     & $H_{\rm sym}$ & $59.79\pm14.90$ & $-1.37\pm0.04$ & $0.302\pm0.013$ & $0.164\pm0.053$ & $5.40\pm0.76$ & $9.27\pm3.07$ \\
     & $H_{\rm block}$ & $76.38\pm16.84$ & $0.05\pm0.20$ & $0.779\pm0.078$ & $0.194\pm0.169$ & $4.18\pm1.97$ & $6.34\pm3.96$ \\
    \midrule
    Synthetic eigengap & fixed-$R_y$ & $65.40\pm1.69$ & $-19.73\pm0.06$ & $0.088\pm0.002$ & $0.011\pm0.002$ & $3.47\pm0.73$ & $9.22\pm2.03$ \\
     & fixed-$R_yR_z$ & $65.55\pm1.22$ & $-20.15\pm0.05$ & $0.046\pm0.000$ & $0.011\pm0.001$ & $8.97\pm1.34$ & $11.50\pm1.47$ \\
     & TF-$R_y$ & $67.42\pm1.33$ & $-19.73\pm0.06$ & $0.088\pm0.002$ & $0.010\pm0.001$ & $3.55\pm0.76$ & $7.59\pm1.19$ \\
     & patch-$SU(4)$ & $54.58\pm1.97$ & $-4.91\pm0.04$ & $0.074\pm0.000$ & $0.017\pm0.003$ & $10.17\pm2.34$ & $27.02\pm1.58$ \\
     & trainable patch-$SU(4)$ & $50.16\pm2.36$ & $-4.92\pm0.04$ & $0.075\pm0.001$ & $0.012\pm0.000$ & $11.59\pm2.81$ & $30.65\pm0.23$ \\
     & patch $H_{\rm block}$ & $71.44\pm2.11$ & $-9.44\pm0.03$ & $0.035\pm0.004$ & $0.023\pm0.004$ & $1.50\pm0.03$ & $7.93\pm1.23$ \\
     & $H_{\rm sym}$ & $82.81\pm3.16$ & $-3.01\pm0.06$ & $0.094\pm0.006$ & $0.108\pm0.019$ & $1.42\pm0.04$ & $7.52\pm1.87$ \\
     & $H_{\rm block}$ & $85.15\pm2.40$ & $-3.72\pm0.05$ & $0.052\pm0.004$ & $0.062\pm0.011$ & $1.52\pm0.02$ & $10.86\pm1.72$ \\
    \midrule
    Synthetic singular & fixed-$R_y$ & $84.34\pm1.24$ & $-19.69\pm0.09$ & $0.084\pm0.002$ & $0.012\pm0.002$ & $4.48\pm1.01$ & $7.86\pm1.20$ \\
     & fixed-$R_yR_z$ & $84.54\pm1.01$ & $-20.04\pm0.05$ & $0.053\pm0.001$ & $0.011\pm0.002$ & $8.02\pm1.71$ & $9.13\pm1.28$ \\
     & TF-$R_y$ & $85.40\pm0.63$ & $-19.69\pm0.09$ & $0.085\pm0.002$ & $0.013\pm0.003$ & $4.40\pm1.02$ & $6.38\pm1.44$ \\
     & patch-$SU(4)$ & $63.41\pm2.03$ & $-5.07\pm0.02$ & $0.061\pm0.000$ & $0.026\pm0.004$ & $14.15\pm2.61$ & $20.22\pm1.67$ \\
     & trainable patch-$SU(4)$ & $61.75\pm1.91$ & $-5.06\pm0.02$ & $0.061\pm0.001$ & $0.018\pm0.001$ & $14.71\pm2.95$ & $22.66\pm2.47$ \\
     & patch $H_{\rm block}$ & $87.50\pm2.06$ & $-9.29\pm0.04$ & $0.036\pm0.005$ & $0.023\pm0.009$ & $1.26\pm0.03$ & $5.49\pm1.55$ \\
     & $H_{\rm sym}$ & $88.75\pm2.69$ & $-3.62\pm0.05$ & $0.069\pm0.005$ & $0.061\pm0.015$ & $1.10\pm0.02$ & $6.92\pm1.72$ \\
     & $H_{\rm block}$ & $92.72\pm4.04$ & $-3.64\pm0.04$ & $0.070\pm0.003$ & $0.036\pm0.009$ & $1.18\pm0.03$ & $6.53\pm1.17$ \\
    \bottomrule
  \end{tabular}
\end{table*}

\section{Additional Information on Latent Diagnostics}\label{sec-appendix-latent}

\subsection{Additional details on the latent diagnostic metrics}

Except for the 20-seed test accuracies in \Cref{tab:latent-depth32-summary}, the latent diagnostics use the final checkpoint from seed 0 and a fixed 32-example validation batch. They do not estimate variation across training seeds and should therefore be interpreted as illustrative within-run diagnostics rather than seed-robust estimates. The batched latent archive contains 864 complete records, of which 63 are marked \texttt{git\_dirty=True}. Among the final-checkpoint depth-32 records, only two are dirty: Pendigits \texttt{DYN} with trainable patch-$SU(4)$ and \texttt{SYNTHETIC EIGENGAP} with trainable patch-$SU(4)$. The depth-32 QSM examples quoted in the main text---Pendigits \texttt{DYN} with $H_{\mathrm{sym}}$ and $H_{\mathrm{block}}$, Pendigits \texttt{STA4} with patch $H_{\mathrm{block}}$, and \texttt{SYNTHETIC SINGULAR} with $H_{\mathrm{sym}}$ and $H_{\mathrm{block}}$---all have \texttt{git\_dirty=False}. Let
\begin{equation}
    \big\{ \ket{\psi_i^{(s)}} \big\}^m_{i=1}
\end{equation}
denote the quantum states of the $m=32$ diagnostic data examples at a particular stage $s$ of the circuit. In the final-state summary plots, $s$ is the final post-mixer state. In the layerwise plots, $s$ indexes the post-mixer state after each reuploading layer. We then compute the pure-state fidelity kernel
\begin{equation}
    K_{ij}^{(s)} = \big| \braket{\psi_i^{(s)} | \psi_j^{(s)} } \big|^2.
\end{equation}
Let $y_i$ be the class label of example $i$. The label-equality kernel is
\begin{equation}
    Y_{ij} = \boldsymbol{1}[y_i = y_j].
\end{equation}
This is the target kernel used in the kernel-target alignment calculation.

For a square matrix $A\in \mathbb{R}^{m\times m}$, define the centering operator
\begin{equation}
    C = I_m - \frac{1}{m} \boldsymbol{1}\boldsymbol{1}^T, \qquad \boldsymbol{1}=\begin{pmatrix}
        1 \\ 1 \\ \vdots \\ 1
    \end{pmatrix} \in \mathbb{R}^m,
\end{equation}
so the centred Gram matrix is
\begin{equation}
    \tilde{A} = CAC.
\end{equation}
The centred kernel-target alignment \cite{Kornblith2019-vz} used in the plots is 
\begin{equation}
    \mathbf{KTA}(K, Y) = \frac{\braket{CKC, CYC}_F}{ || CKC ||_F ||CYC ||_F},
\end{equation}
with a small numerical guard in the denominator. This is the centred-kernel alignment (CKA) between the fidelity and label kernels.

The within-minus-between fidelity gap uses only off-diagonal pairs. Let
\begin{equation}
    \mathcal{W} = \big\{ (i,j): i<j, y_i = y_j  \big\}, \quad \mathcal{B} = \big\{ (i,j): i<j, y_i \neq y_j  \big\}.
\end{equation}
Then
\begin{equation}\label{eqn:within-minus-between-fidelity-gap}
    \Delta_{\mathrm{fid}} = \frac{1}{|\mathcal{W}|} \sum_{(i,j)\in \mathcal{W} } K_{ij} - \frac{1}{| \mathcal{B} |} \sum_{(i,j)\in\mathcal{B}} K_{ij}.
\end{equation}
Positive values mean same-label examples are closer in quantum-state fidelity than different-label data examples. 

The participation-ratio effective rank \cite{Gao2017-pr} of the fidelity kernel is computed from the non-negative eigenvalues $\lambda_1,\ldots,\lambda_m$ of the symmetrised kernel:
\begin{equation}
    r_{\mathrm{eff}} (K) = \frac{(\sum_j \lambda_j )^2}{\sum_j \lambda_j^2}.
\end{equation}
This scale-invariant participation ratio summarises how broadly the kernel eigenvalue mass is spread: it equals the number of equally weighted nonzero eigenmodes in the special case where those eigenvalues are equal. It does not use the class labels and, by itself, does not measure class separation or predictive accuracy. The centred effective rank uses the same expression after centring the kernel.

For adjacent-layer CKA, we apply the same centred-alignment formula to fidelity kernels from consecutive post-mixer layers:
\begin{equation}
     \mathbf{CKA}_{\ell, \ell+1} = \frac{\braket{CK^{(\ell)}C, CK^{(\ell + 1)}C}_F}{ || CK^{(\ell)}C ||_F ||C K^{(\ell+1)} C ||_F},
\end{equation}
This measures similarity between consecutive batch geometries; values near one indicate little layer-to-layer change.

The trajectory plots evaluate the same final-state metrics at saved checkpoints rather than only at the final checkpoint. If $c$ indexes training checkpoints, the trajectory target alignment is
\begin{equation}
    \mathbf{KTA}(K^{(c)}_{\mathrm{final}}, Y),
\end{equation}
where $K^{(c)}_{\mathrm{final}}$ is the fidelity kernel of the final post-mixer state at checkpoint $c$.

The diagnostic-batch projector accuracy reported in \Cref{tab:latent-depth32-summary} is computed by applying the same projector readout used for the classifier to the final quantum states and comparing the argmax prediction with the diagnostic labels.

\subsection{Additional figures and tables}

\begin{figure}[htbp]
    \centering
    \includegraphics[width=1\linewidth]{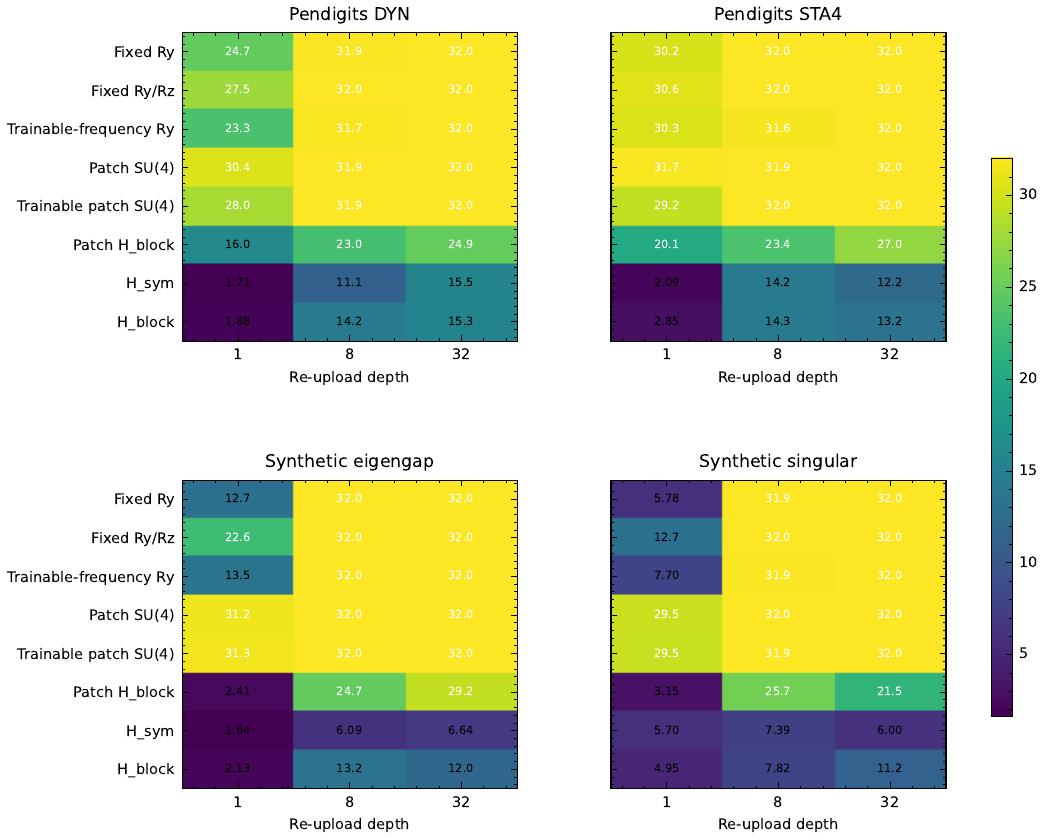}
    \caption{
    \textbf{Effective rank of the final quantum-state fidelity kernel.} Effective rank is computed as $\mathbf{tr}(K)^2/\mathbf{tr}(K^2)$ from the eigenvalues of the final fidelity kernel. A high effective rank indicates that the kernel spectrum is spread across many eigenmodes rather than concentrated in a few; it does not, by itself, imply label alignment or high predictive performance.
    }
    \label{fig:appendix-latent-summary-kernel-effective-rank}
\end{figure}

The final kernel effective-rank plot in \Cref{fig:appendix-latent-summary-kernel-effective-rank} measures spectral spread in the final quantum-state fidelity kernel. This is useful as a control because it separates broad participation across kernel eigenmodes from class-aligned kernel geometry. In \Cref{fig:appendix-latent-summary-kernel-effective-rank}, rotation-gate data-encoding unitaries often have an effective rank of approximately 32 (the diagnostic batch size), but their fidelity gap is approximately zero. Patch-$SU(4)$ data-encoding unitaries also often have high effective rank while performing poorly at depth 32. High effective rank is therefore insufficient for good classification in these runs. The label-aware evidence instead comes from the fidelity gap and kernel-target alignment, which must still be interpreted as associations rather than causes.

\begin{figure}[htbp]
    \centering
    \includegraphics[width=1\linewidth]{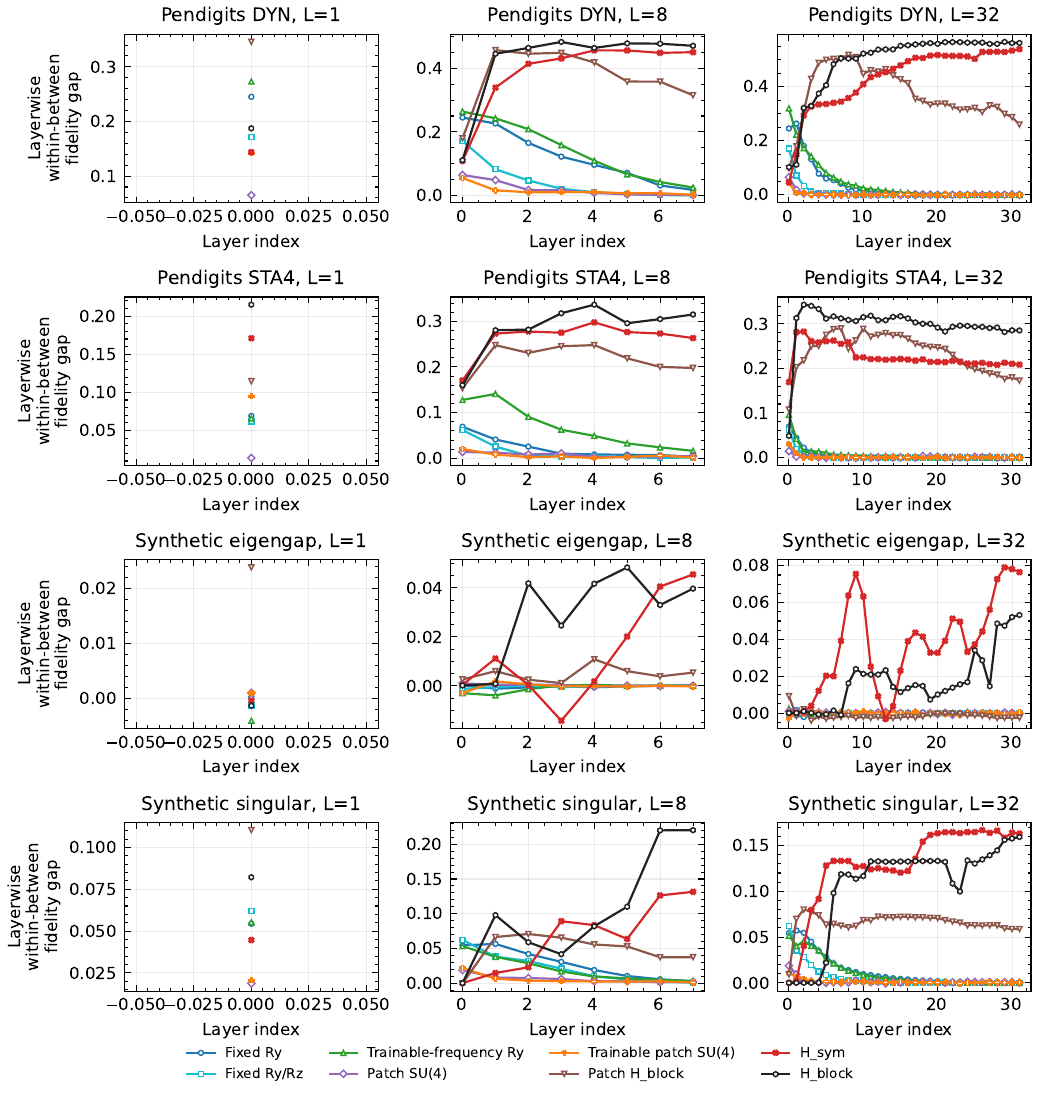}
    \caption{
    \textbf{Layerwise within-minus-between fidelity gap.} For each post-mixer layer, we compute the pure-state fidelity kernel on the diagnostic validation batch and report the mean same-label off-diagonal fidelity minus the mean different-label off-diagonal fidelity. The curves show where class separation emerges inside the reuploading circuit. QSMs generally build larger class-separating fidelity gaps than the coordinate-wise rotation-gate and patch-$SU(4)$ baselines.
    }
    \label{fig:appendix-latent-layerwise-kernel-gap}
\end{figure}

The layerwise fidelity-gap plot (\Cref{fig:appendix-latent-layerwise-kernel-gap}) shows how class separation evolves across the circuit layers at the final trained checkpoint. For each layer, the diagnostic computes the same within-minus-between fidelity gap used in the main text, but on the post-mixer state at that layer. This plot is useful because it shows whether class separation appears early or is progressively built by the reuploading architecture. The QSMs tend to produce larger layerwise gaps than the rotation-gate baselines, especially at depth 32. The patch-local block-Hamiltonian QSM is particularly relevant on STA4, where its local patch structure produces strong class separation in later layers.

\begin{figure}[htbp]
    \centering
    \includegraphics[width=1\linewidth]{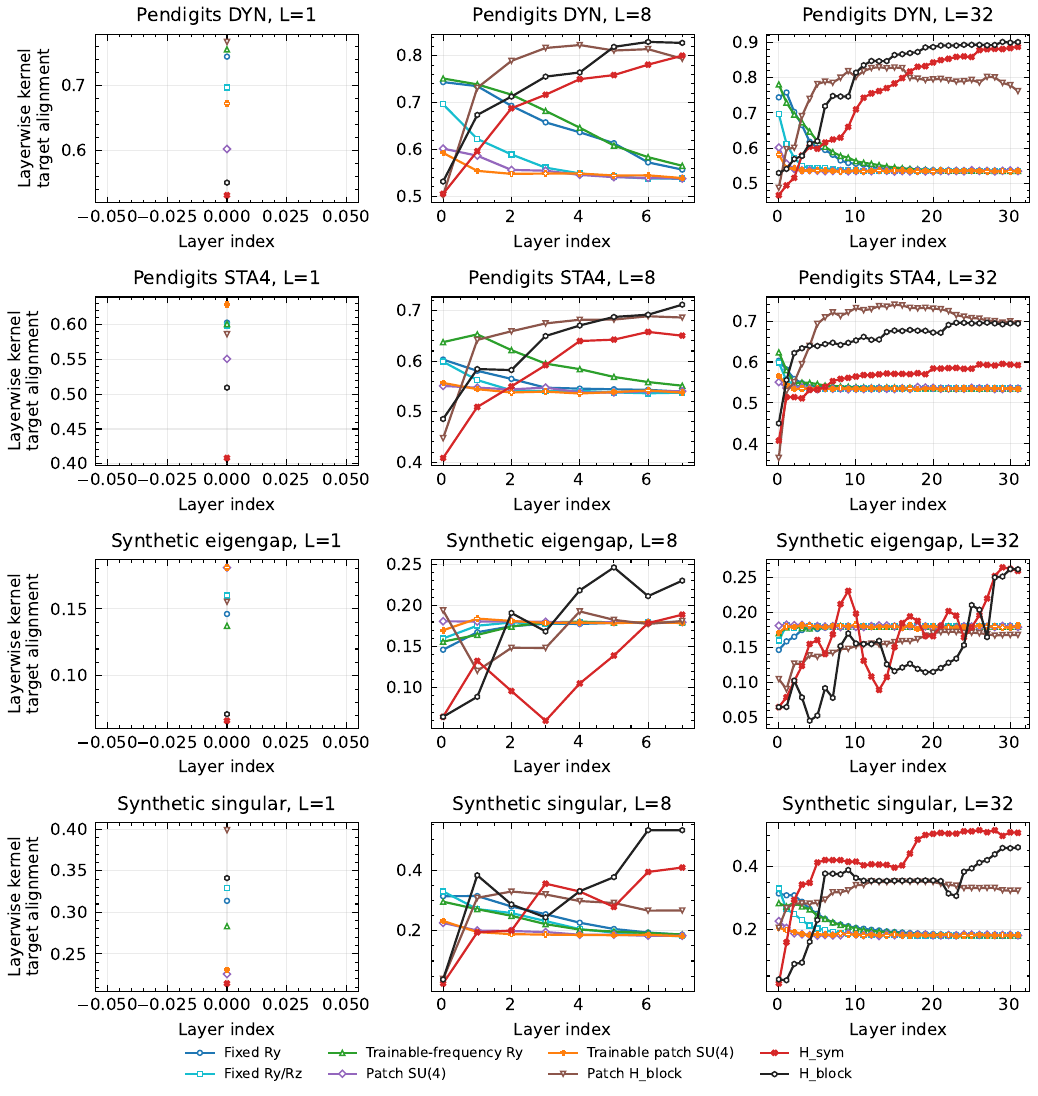}
    \caption{
    \textbf{Layerwise kernel-target alignment.} At each post-mixer layer, the fidelity kernel of the diagnostic validation states is centred and aligned with the label-equality kernel. The resulting curves show how class-aligned quantum-state geometry develops across depth. This diagnostic complements the layerwise fidelity-gap plot by comparing the full kernel structure to the label kernel.
    }
    \label{fig:appendix-latent-layerwise-kernel-target-alignment}
\end{figure}

The layerwise target-alignment plot \Cref{fig:appendix-latent-layerwise-kernel-target-alignment} gives a complementary view of representation formation. Instead of measuring only the within-minus-between fidelity gap, it compares the full layerwise fidelity kernel with the label-equality kernel using centred kernel alignment. This plot is useful when the label structure is not captured solely by a mean gap. The main caveat is the same as for the final target-alignment plot: target alignment can be nonzero for high-rank kernels even when off-diagonal class separation is weak, so it should be interpreted together with the fidelity-gap and effective-rank plots.

\begin{figure}[htbp]
    \centering
    \includegraphics[width=1\linewidth]{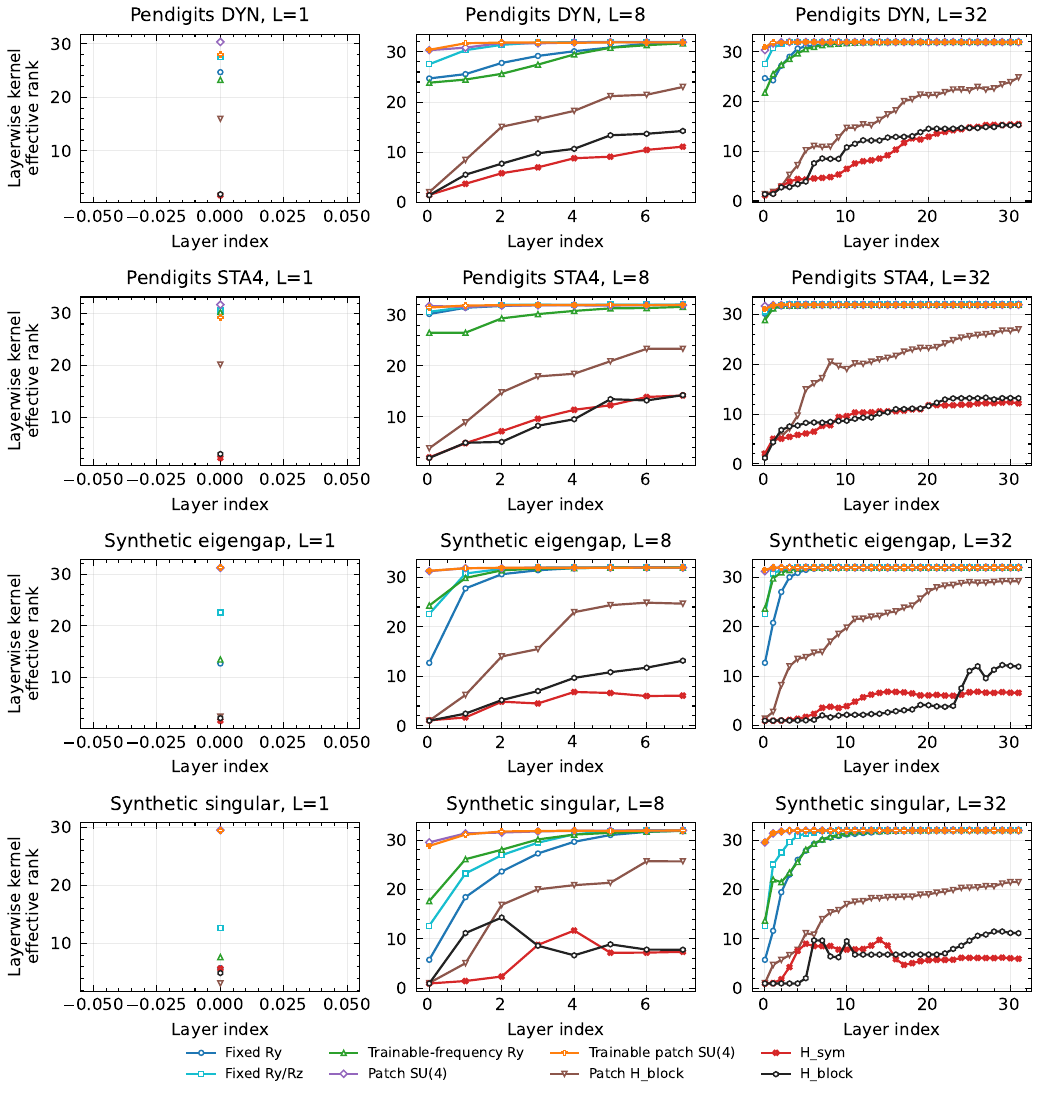}
    \caption{
    \textbf{Layerwise effective rank of the fidelity kernel.} Effective rank is computed from the post-mixer fidelity kernel at each layer. The plot measures spectral spread in the kernel, not label alignment. High effective rank alone is not sufficient for high test accuracy in these runs, as several rotation-gate and patch-$SU(4)$ models produce high-rank kernels without strong class-separated fidelity geometry.
    }
    \label{fig:appendix-latent-layerwise-kernel-effective-rank}
\end{figure}

The layerwise effective-rank plot shown in \Cref{fig:appendix-latent-layerwise-kernel-effective-rank} tracks how the fidelity-kernel eigenvalue mass is distributed across modes as the circuit depth increases. It is most useful as a negative control. Rotation-gate data-encoding unitaries can maintain very high effective rank across layers, but this does not imply strong class separation or high test accuracy. Conversely, QSMs can have a lower effective rank while achieving larger fidelity gaps and higher test accuracy. The layerwise result therefore shows that broad kernel spectral spread alone does not reproduce the observed accuracy pattern.

\begin{figure}[htbp]
    \centering
    \includegraphics[width=1\linewidth]{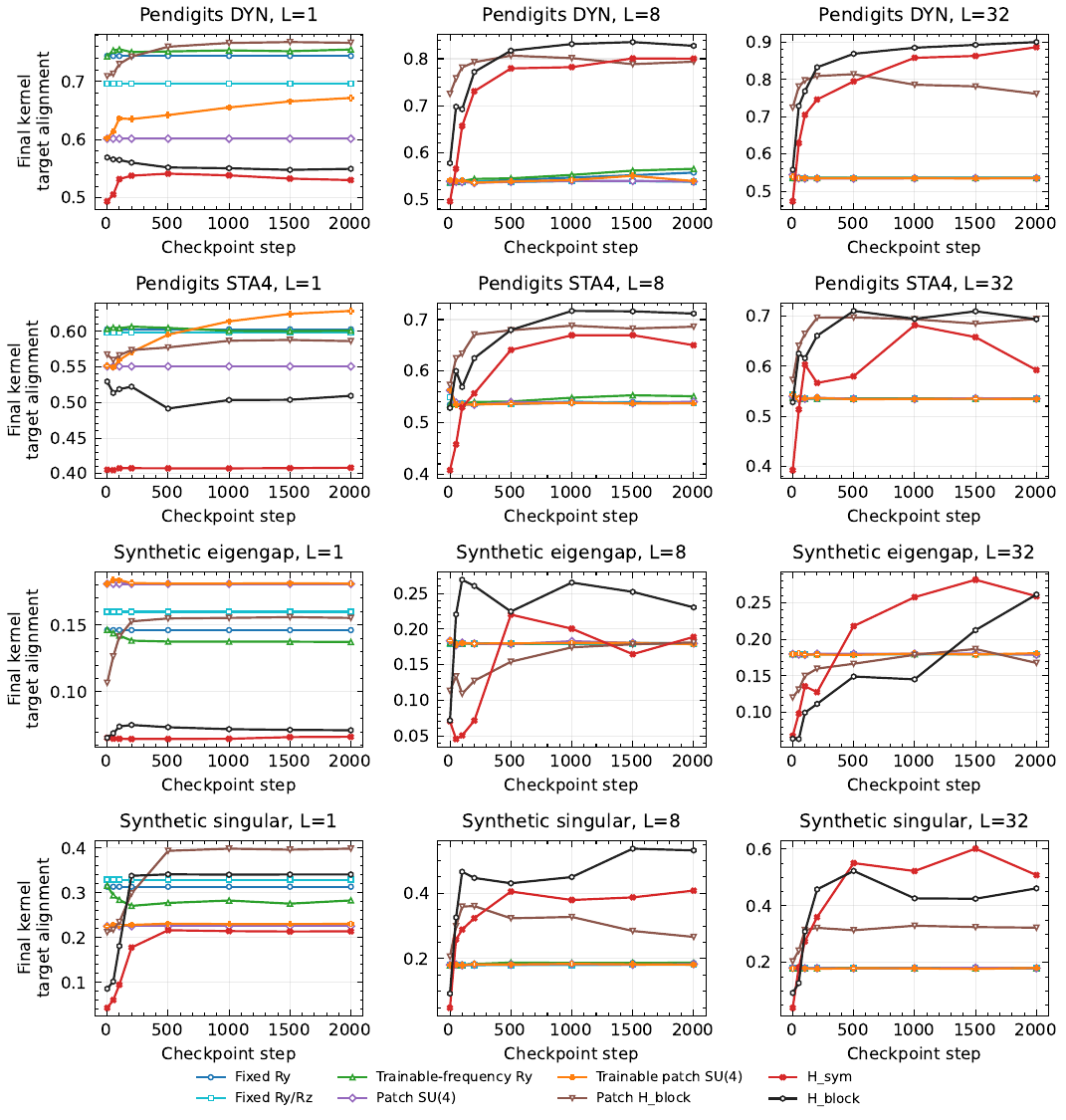}
    \caption{
    \textbf{Training trajectory of kernel-target alignment.} The curves show centred alignment between the final-state fidelity kernel and the label-equality kernel at saved checkpoints. The diagnostic is evaluated on the same validation batch used in the final latent-state summaries. QSMs tend to increase target alignment during optimisation, whereas several coordinate-wise rotation-gate baselines remain comparatively flat.
    }
    \label{fig:appendix-latent-trajectory-kernel-target-alignment}
\end{figure}

The trajectory plot in \Cref{fig:appendix-latent-trajectory-kernel-target-alignment} shows how final-state kernel-target alignment changes during training, using the stored sequence of checkpoints. This helps distinguish a model that starts with label-aligned geometry from one that learns it during optimisation. In \Cref{fig:appendix-latent-trajectory-kernel-target-alignment}, the global block-Hamiltonian QSM on Pendigits \texttt{DYN} increases target alignment from 0.558 at initialisation to 0.900 at the final checkpoint, while fixed-$R_y$ remains flat at 0.535. On \texttt{SYNTHETIC SINGULAR}, the global block-Hamiltonian QSM increases from 0.093 at initialisation to 0.461 at the final checkpoint, while fixed-$R_y$ remains at 0.180. This supports the interpretation that the QSMs learn class-aligned quantum-state geometry over training, rather than merely starting with it.

\begin{figure}[htbp]
    \centering
    \includegraphics[width=1\linewidth]{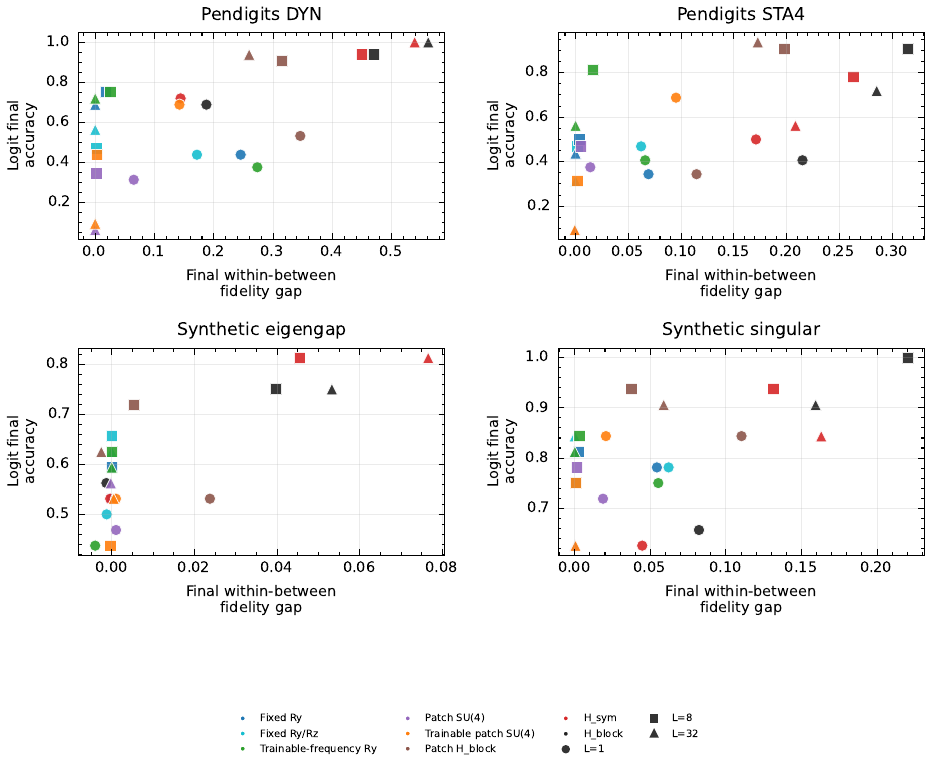}
    \caption{
    \textbf{Relationship between fidelity gap and diagnostic-batch accuracy.} Each point corresponds to one encoder-depth configuration on the diagnostic validation batch. The horizontal axis is the final within-minus-between fidelity gap, and the vertical axis is final projector accuracy on the same batch. Larger positive fidelity gaps generally correspond to stronger diagnostic-batch classification.
    }
    \label{fig:appendix-latent-scatter-kernel-gap}
\end{figure}

The scatter plot in \Cref{fig:appendix-latent-scatter-kernel-gap} relates the fidelity-gap diagnostic to diagnostic-batch accuracy. It directly connects the latent-state geometry to prediction quality on the diagnostic batch. Points with larger within-minus-between fidelity gap generally correspond to better diagnostic-batch accuracy, especially in the Pendigits panels. The plot also makes the negative-control cases visible: some encoders may have a high effective rank, but without a positive fidelity gap, they do not form a label-separated quantum-state geometry.

\begin{figure}[htbp]
    \centering
    \includegraphics[width=0.7\linewidth]{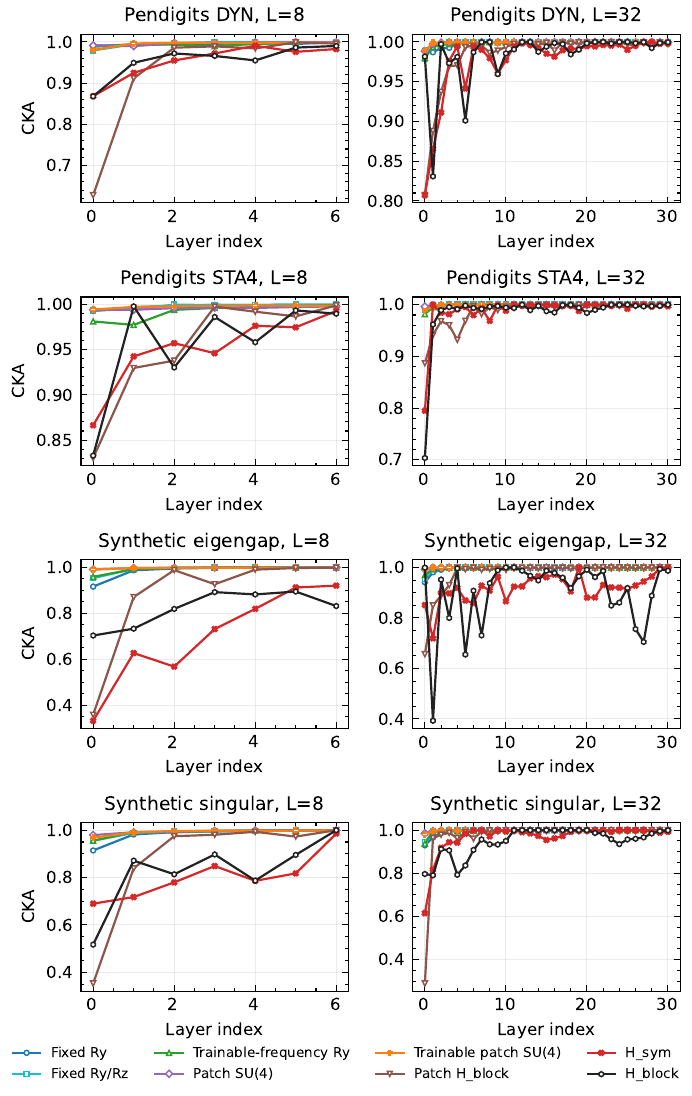}
    \caption{
    \textbf{Adjacent-layer CKA of post-mixer fidelity kernels.} For each trained model, we compute centred kernel alignment between the fidelity kernels of consecutive post-mixer layers. High values indicate that adjacent layers preserve a similar batch geometry. These diagnostic measures representational stability across depth and are included as a supplementary control rather than as a direct predictor of test accuracy.
    }
    \label{fig:appendix-latent-cka-post-mixer-adjacent-layers}
\end{figure}

The adjacent-layer CKA plot in \Cref{fig:appendix-latent-cka-post-mixer-adjacent-layers} measures representational shifts between consecutive post-mixer states. A high value means that the fidelity kernel changes little from one layer to the next. This diagnostic could assess whether deeper circuits are repeatedly reshaping the batch geometry or have entered a stable representation regime. In \Cref{fig:appendix-latent-cka-post-mixer-adjacent-layers}, adjacent-layer CKA is often high at large depths, with final depth-32 values averaging about 0.986 across complete records, but this high stability does not, by itself, distinguish successful from unsuccessful encoders.

\begin{table*}[htbp!]
  \caption{\textbf{Summary of the latent diagnostic results.} Test accuracy is the mean
  $\pm$ standard deviation over 20 random seeds, reported in percentage. The remaining
  columns are final-checkpoint diagnostics computed on the fixed 32-example
  validation diagnostic batch for seed 0. Diagnostic accuracy is the projector
  accuracy on this batch. KTA is centred kernel-target alignment between the
  final fidelity kernel and the label-equality kernel. Gap is the mean within-class
  off-diagonal fidelity minus the mean between-class off-diagonal fidelity.
  Effective rank is computed from the final fidelity kernel.}
  \label{tab:latent-depth32-summary}
  \centering
  \scriptsize
  \setlength{\tabcolsep}{3pt}
  \begin{tabular}{llccccc}
    \toprule
    Dataset & Encoder & Test acc. (\%) & Diag. acc. & KTA & Gap & Eff. rank \\
    \midrule
    Pendigits DYN & fixed-$R_y$ & $72.29\pm2.77$ & 0.688 & 0.535 & 0.000 & 32.00 \\
     & fixed-$R_yR_z$ & $55.84\pm5.30$ & 0.562 & 0.535 & 0.000 & 32.00 \\
     & TF-$R_y$ & $81.37\pm2.36$ & 0.719 & 0.535 & 0.000 & 32.00 \\
     & patch-$SU(4)$ & $9.96\pm0.44$ & 0.062 & 0.535 & $-0.000$ & 31.97 \\
     & trainable patch-$SU(4)$ & $9.84\pm0.47$ & 0.094 & 0.535 & $-0.000$ & 31.97 \\
     & patch $H_{\rm block}$ & $92.75\pm0.84$ & 0.938 & 0.761 & 0.260 & 24.87 \\
     & $H_{\rm sym}$ & $91.89\pm3.71$ & 1.000 & 0.887 & 0.538 & 15.46 \\
     & $H_{\rm block}$ & $92.35\pm5.95$ & 1.000 & 0.900 & 0.561 & 15.29 \\
    \midrule
    Pendigits STA4 & fixed-$R_y$ & $40.22\pm4.40$ & 0.438 & 0.535 & 0.000 & 32.00 \\
     & fixed-$R_yR_z$ & $32.79\pm4.28$ & 0.312 & 0.535 & $-0.000$ & 32.00 \\
     & TF-$R_y$ & $46.85\pm6.39$ & 0.562 & 0.535 & 0.000 & 32.00 \\
     & patch-$SU(4)$ & $16.41\pm1.88$ & 0.094 & 0.535 & $-0.000$ & 31.97 \\
     & trainable patch-$SU(4)$ & $10.27\pm0.62$ & 0.094 & 0.534 & $-0.001$ & 31.97 \\
     & patch $H_{\rm block}$ & $86.14\pm1.33$ & 0.938 & 0.694 & 0.173 & 27.03 \\
     & $H_{\rm sym}$ & $59.79\pm14.90$ & 0.562 & 0.592 & 0.208 & 12.18 \\
     & $H_{\rm block}$ & $76.38\pm16.84$ & 0.719 & 0.694 & 0.285 & 13.20 \\
    \midrule
    Synthetic eigengap & fixed-$R_y$ & $65.40\pm1.69$ & 0.594 & 0.180 & $-0.000$ & 32.00 \\
     & fixed-$R_yR_z$ & $65.55\pm1.22$ & 0.625 & 0.180 & $-0.000$ & 32.00 \\
     & TF-$R_y$ & $67.42\pm1.33$ & 0.594 & 0.180 & $-0.000$ & 32.00 \\
     & patch-$SU(4)$ & $54.58\pm1.97$ & 0.562 & 0.179 & $-0.000$ & 31.97 \\
     & trainable patch-$SU(4)$ & $50.16\pm2.36$ & 0.531 & 0.181 & 0.000 & 31.97 \\
     & patch $H_{\rm block}$ & $71.44\pm2.11$ & 0.625 & 0.168 & $-0.003$ & 29.21 \\
     & $H_{\rm sym}$ & $82.81\pm3.16$ & 0.812 & 0.259 & 0.077 & 6.64 \\
     & $H_{\rm block}$ & $85.15\pm2.40$ & 0.750 & 0.261 & 0.053 & 11.95 \\
    \midrule
    Synthetic singular & fixed-$R_y$ & $84.34\pm1.24$ & 0.812 & 0.180 & 0.000 & 32.00 \\
     & fixed-$R_yR_z$ & $84.54\pm1.01$ & 0.844 & 0.180 & 0.000 & 32.00 \\
     & TF-$R_y$ & $85.40\pm0.63$ & 0.812 & 0.180 & 0.000 & 32.00 \\
     & patch-$SU(4)$ & $63.41\pm2.03$ & 0.625 & 0.180 & 0.000 & 31.95 \\
     & trainable patch-$SU(4)$ & $61.75\pm1.91$ & 0.625 & 0.181 & 0.000 & 31.97 \\
     & patch $H_{\rm block}$ & $87.50\pm2.06$ & 0.906 & 0.323 & 0.059 & 21.50 \\
     & $H_{\rm sym}$ & $88.75\pm2.69$ & 0.844 & 0.508 & 0.163 & 6.00 \\
     & $H_{\rm block}$ & $92.72\pm4.04$ & 0.906 & 0.461 & 0.159 & 11.17 \\
    \bottomrule
  \end{tabular}
\end{table*}

\section{Additional Information on Classical Baseline and Ablation Study}

\Cref{tab:classical-baselines} reports the classical controls described in the main text. For every quantum ablation in \Cref{tab:ablation-summary}, the transformation is constructed before training, applied to the training, validation and test splits, and followed by retraining. Labels are unchanged. Any reference basis or spectrum is computed from the training split only.

The entry-permutation control generates one fixed permutation of the flattened matrix entries and applies it to every sample in all three splits. The row/column control similarly generates one fixed row permutation and one fixed column permutation. The permutations are therefore dataset-level transformations, not independently sampled perturbations of individual examples.

For the symmetric-Hamiltonian controls, let $Q_0$ be the eigenvector matrix of the mean training-set $H_{\mathrm{sym}}$ and let $\boldsymbol{\lambda}_0$ be the coordinatewise median of the ordered training-set eigenvalues. The spectrum-only control retains each sample's eigenvalues and replaces its eigenvectors by $Q_0$. The eigenvectors-only control retains each sample's eigenvectors and replaces its eigenvalues by $\boldsymbol{\lambda}_0$.

For the block-Hamiltonian controls, let $U_0$ and $V_0$ be the left and right singular-vector matrices of the mean training matrix, and let $\boldsymbol{\sigma}_0$ be the coordinatewise median of the ordered training-set singular values. The singular-values-only control combines each sample's singular values with $U_0$ and $V_0$. The singular-vectors-only control retains each sample's left and right singular vectors and replaces its singular values by $\boldsymbol{\sigma}_0$. Because bases within degenerate spectral subspaces are not unique, the vector-only controls are implementation-dependent in those subspaces.

\begin{table*}[htbp!]
  \caption{\textbf{Classical baseline results.} Each baseline is a one-hidden-layer MLP
  trained on either flattened raw entries or spectral-value descriptors. Values
  are mean $\pm$ standard deviation over 20 random seeds, reported in percentage.
  $\Delta$ is the test-accuracy difference, in percentage points, between the
  classical baseline and the largest observed QSM mean on the same dataset
  panel.}
  \label{tab:classical-baselines}
  \centering
  \small
  \setlength{\tabcolsep}{4pt}
  \begin{tabular}{llccccc}
    \toprule
    Dataset & Feature source & Feature dim. & Params & Val. acc. (\%) &
    Test acc. (\%) & \shortstack{$\Delta$ vs. largest\\QSM mean} \\
    \midrule
    Pendigits DYN & raw entries & 16 & 1468 & $99.31\pm0.22$ & $96.88\pm0.24$ & $+3.07$ \\
     & $H_{\rm sym}$ eig. & 8 & 1473 & $66.97\pm1.12$ & $59.98\pm0.60$ & $-33.83$ \\
     & $H_{\rm block}$ sing. & 2 & 1466 & $42.75\pm0.70$ & $39.67\pm0.48$ & $-54.14$ \\
    \midrule
    Pendigits STA4 & raw entries & 16 & 1468 & $95.46\pm0.47$ & $91.32\pm0.46$ & $+4.66$ \\
     & $H_{\rm sym}$ eig. & 4 & 1465 & $38.53\pm0.95$ & $39.80\pm0.50$ & $-46.86$ \\
     & $H_{\rm block}$ sing. & 4 & 1465 & $31.39\pm1.00$ & $30.62\pm0.53$ & $-56.04$ \\
    \midrule
    Synthetic eigengap & raw entries & 16 & 515 & $75.20\pm2.51$ & $74.88\pm1.27$ & $-11.76$ \\
     & $H_{\rm sym}$ eig. & 4 & 513 & $98.28\pm0.28$ & $98.01\pm0.35$ & $+11.36$ \\
     & $H_{\rm block}$ sing. & 4 & 989 & $67.21\pm1.03$ & $67.56\pm0.60$ & $-19.08$ \\
    \midrule
    Synthetic singular & raw entries & 16 & 1465 & $90.37\pm0.80$ & $91.32\pm0.60$ & $-3.92$ \\
     & $H_{\rm sym}$ eig. & 8 & 992 & $98.29\pm0.51$ & $97.28\pm0.23$ & $+2.04$ \\
     & $H_{\rm block}$ sing. & 2 & 1472 & $98.57\pm0.36$ & $98.50\pm0.20$ & $+3.25$ \\
    \bottomrule
  \end{tabular}
\end{table*}

\begin{table*}[htbp!]
  \caption{\textbf{Ablation-study summary for Hamiltonian-based QSMs.} Values are
  mean $\pm$ standard deviation over 20 random seeds, reported in percentage.
  ``Original'' gives the depth and largest observed mean test accuracy of the
  unablated QSM variant on the same dataset panel; ``Ablated'' gives the
  corresponding descriptive maximum after retraining under the specified
  ablation. These maxima are computed across the tested depths rather than at
  depths selected by independent validation. $\Delta$ is the
  ablated-minus-original difference in percentage points. Spectrum/eigenvector
  controls are run for $H_{\rm sym}$; singular-value/vector controls are run for
  $H_{\rm block}$; permutation controls are shown for the global QSMs and the
  patch-local block-Hamiltonian QSM.}
  \label{tab:ablation-summary}
  \centering
  \scriptsize
  \setlength{\tabcolsep}{3pt}
  \begin{tabular}{lllccc}
    \toprule
    Dataset & Encoder & Original & Ablation & Ablated & $\Delta$ \\
    \midrule
    Pendigits DYN & $H_{\rm sym}$ & L32, $91.89\pm3.71$ & spectrum only & L8, $48.87\pm3.90$ & $-43.02$ \\
     & $H_{\rm sym}$ & L32, $91.89\pm3.71$ & eigenvectors only & L16, $89.80\pm4.56$ & $-2.08$ \\
     & $H_{\rm sym}$ & L32, $91.89\pm3.71$ & entry permutation & L16, $89.49\pm10.42$ & $-2.39$ \\
     & $H_{\rm sym}$ & L32, $91.89\pm3.71$ & row/column permutation & L32, $92.69\pm3.79$ & $+0.80$ \\
     & $H_{\rm block}$ & L16, $93.81\pm1.06$ & singular values only & L4, $31.65\pm2.44$ & $-62.16$ \\
     & $H_{\rm block}$ & L16, $93.81\pm1.06$ & singular vectors only & L16, $93.56\pm0.84$ & $-0.24$ \\
     & $H_{\rm block}$ & L16, $93.81\pm1.06$ & entry permutation & L32, $93.94\pm2.34$ & $+0.13$ \\
     & $H_{\rm block}$ & L16, $93.81\pm1.06$ & row/column permutation & L32, $94.82\pm1.59$ & $+1.01$ \\
     & patch $H_{\rm block}$ & L32, $92.75\pm0.84$ & entry permutation & L32, $92.77\pm1.31$ & $+0.03$ \\
     & patch $H_{\rm block}$ & L32, $92.75\pm0.84$ & row/column permutation & L32, $92.09\pm1.40$ & $-0.66$ \\
    \midrule
    Pendigits STA4 & $H_{\rm sym}$ & L16, $70.14\pm8.29$ & spectrum only & L4, $28.35\pm2.11$ & $-41.79$ \\
     & $H_{\rm sym}$ & L16, $70.14\pm8.29$ & eigenvectors only & L4, $64.30\pm2.02$ & $-5.84$ \\
     & $H_{\rm sym}$ & L16, $70.14\pm8.29$ & entry permutation & L32, $72.45\pm6.58$ & $+2.31$ \\
     & $H_{\rm sym}$ & L16, $70.14\pm8.29$ & row/column permutation & L16, $70.73\pm4.05$ & $+0.59$ \\
     & $H_{\rm block}$ & L16, $86.66\pm2.17$ & singular values only & L4, $25.25\pm2.02$ & $-61.41$ \\
     & $H_{\rm block}$ & L16, $86.66\pm2.17$ & singular vectors only & L16, $84.74\pm4.51$ & $-1.91$ \\
     & $H_{\rm block}$ & L16, $86.66\pm2.17$ & entry permutation & L16, $84.59\pm4.05$ & $-2.07$ \\
     & $H_{\rm block}$ & L16, $86.66\pm2.17$ & row/column permutation & L16, $86.16\pm2.85$ & $-0.50$ \\
     & patch $H_{\rm block}$ & L32, $86.14\pm1.33$ & entry permutation & L16, $83.10\pm1.03$ & $-3.04$ \\
     & patch $H_{\rm block}$ & L32, $86.14\pm1.33$ & row/column permutation & L16, $83.92\pm1.38$ & $-2.22$ \\
    \midrule
    Synthetic eigengap & $H_{\rm sym}$ & L16, $83.42\pm1.37$ & spectrum only & L8, $97.92\pm0.38$ & $+14.50$ \\
     & $H_{\rm sym}$ & L16, $83.42\pm1.37$ & eigenvectors only & L1, $61.28\pm0.07$ & $-22.14$ \\
     & $H_{\rm sym}$ & L16, $83.42\pm1.37$ & entry permutation & L8, $67.83\pm1.41$ & $-15.59$ \\
     & $H_{\rm sym}$ & L16, $83.42\pm1.37$ & row/column permutation & L4, $69.80\pm1.39$ & $-13.62$ \\
     & $H_{\rm block}$ & L16, $86.65\pm1.18$ & singular values only & L16, $65.54\pm1.07$ & $-21.11$ \\
     & $H_{\rm block}$ & L16, $86.65\pm1.18$ & singular vectors only & L8, $68.66\pm1.65$ & $-17.99$ \\
     & $H_{\rm block}$ & L16, $86.65\pm1.18$ & entry permutation & L16, $74.46\pm1.35$ & $-12.19$ \\
     & $H_{\rm block}$ & L16, $86.65\pm1.18$ & row/column permutation & L16, $83.47\pm1.18$ & $-3.18$ \\
     & patch $H_{\rm block}$ & L16, $74.37\pm2.59$ & entry permutation & L16, $72.85\pm1.46$ & $-1.52$ \\
     & patch $H_{\rm block}$ & L16, $74.37\pm2.59$ & row/column permutation & L16, $74.74\pm1.33$ & $+0.37$ \\
    \midrule
    Synthetic singular & $H_{\rm sym}$ & L8, $90.16\pm1.08$ & spectrum only & L8, $97.25\pm0.48$ & $+7.08$ \\
     & $H_{\rm sym}$ & L8, $90.16\pm1.08$ & eigenvectors only & L2, $51.29\pm1.04$ & $-38.87$ \\
     & $H_{\rm sym}$ & L8, $90.16\pm1.08$ & entry permutation & L8, $88.35\pm1.00$ & $-1.81$ \\
     & $H_{\rm sym}$ & L8, $90.16\pm1.08$ & row/column permutation & L16, $88.64\pm1.27$ & $-1.53$ \\
     & $H_{\rm block}$ & L16, $95.24\pm1.53$ & singular values only & L32, $98.42\pm0.30$ & $+3.17$ \\
     & $H_{\rm block}$ & L16, $95.24\pm1.53$ & singular vectors only & L16, $49.88\pm1.96$ & $-45.36$ \\
     & $H_{\rm block}$ & L16, $95.24\pm1.53$ & entry permutation & L8, $90.55\pm0.58$ & $-4.69$ \\
     & $H_{\rm block}$ & L16, $95.24\pm1.53$ & row/column permutation & L16, $95.06\pm2.30$ & $-0.18$ \\
     & patch $H_{\rm block}$ & L32, $87.50\pm2.06$ & entry permutation & L32, $87.04\pm2.04$ & $-0.46$ \\
     & patch $H_{\rm block}$ & L32, $87.50\pm2.06$ & row/column permutation & L32, $86.43\pm3.37$ & $-1.06$ \\
    \bottomrule
  \end{tabular}
\end{table*}

%%%%%%%%%%%%%%%%%%%%%%%%%%%%%%%%%%%%%%%%%%%%%%%%%%%%%%%%%%%%

%\newpage
%\input{checklist.tex}

\end{document}